\documentclass[11pt, reprint, aps, prc, notitlepage, tightenlines,nofootinbib,superscriptaddress]{revtex4-1}
\usepackage[utf8]{inputenc}
\usepackage{mathtools}
\usepackage{amsfonts}
\usepackage{amssymb,hyperref, amsmath}
\usepackage{color}
\usepackage{graphicx}
\usepackage{subfigure}
\graphicspath{ {image/} }
\usepackage{multirow}
\newcommand*\diff{\mathop{}\!\mathrm{d}}
\usepackage{braket}
\usepackage{bm}
\usepackage{txfonts}
\usepackage{mathrsfs} 
\usepackage{slashed}
\usepackage{cleveref}
\usepackage{hepunits}

\begin{document}

\bibliographystyle{apsrev4-1}

\title{Ultrarelativistic quark-nucleus scattering in a light-front Hamiltonian approach}
\author{Meijian Li}
\email{meijianl@iastate.edu}
\affiliation{Department of Physics and Astronomy, Iowa State University, Ames, IA, 50011, USA}
\affiliation{Department of Physics, P.O. Box 35, FI-40014 University of Jyv\"{a}skyl\"{a},
Finland}
\affiliation{
Helsinki Institute of Physics, P.O. Box 64, FI-00014 University of Helsinki,
Finland
}
\author{Xingbo Zhao}
\email{corresponding author: xbzhao@impcas.ac.cn}
\affiliation{Institute of Modern Physics, Chinese Academy of Sciences, Lanzhou 730000, China}
\affiliation{University of Chinese Academy of Sciences, Beijing 100049, China}
\author{Pieter Maris}
\email{pmaris@iastate.edu}
\affiliation{Department of Physics and Astronomy, Iowa State University, Ames, IA, 50011, USA}
 \author{Guangyao Chen}
 \email{gchen@highlands.edu}
 \affiliation{Department of Physics and Astronomy, Iowa State University, Ames, IA, 50011, USA}
 \affiliation{Division of Natural Sciences, Georgia Highlands College, Marietta, Georgia 30067, USA
 }
\author{Yang Li}
\email{leeyoung@iastate.edu}
\affiliation{Department of Physics and Astronomy, Iowa State University, Ames, IA, 50011, USA}
\author{Kirill Tuchin}
\email{tuchin@iastate.edu}
\affiliation{Department of Physics and Astronomy, Iowa State University, Ames, IA, 50011, USA}
\author{James P. Vary}
\email{vary@iastate.edu}
\affiliation{Department of Physics and Astronomy, Iowa State University, Ames, IA, 50011, USA}
\begin{abstract}
  We investigate the scattering of a quark on a heavy nucleus at high energies using the time-dependent basis light-front quantization (tBLFQ) formalism, which is the first application of the tBLFQ formalism in QCD. 
  We present the real-time evolution of the quark wave function in a strong classical color field of the relativistic nucleus, described as the Color Glass Condensate. 
  The quark and the nucleus color field are simulated in the QCD SU(3) color space. 
  We calculate the total and the differential cross sections, and the quark distribution in coordinate and color spaces using the tBLFQ approach. We recover the eikonal cross sections in the eikonal limit. 
  We find that the differential cross section from the tBLFQ simulation is in agreement with a perturbative calculation at large $p_\perp$, and it deviates from the perturbative calculation at small $p_\perp$ due to higher-order contributions. 
  In particular, we relax the eikonal limit by letting the quark carry realistic finite longitudinal momenta. 
  We study the sub-eikonal effect on the quark through the transverse coordinate distribution of the quark with different longitudinal momentum, and we find the sub-eikonal effect to be sizable. 
  Our results can significantly reduce the theoretical uncertainties in small $p_\perp$ region which has important implications to the phenomenology of the hadron-nucleus and deep inelastic scattering at high energies.
\end{abstract}
\maketitle

\section{Introduction}
Scattering of an ultrarelativistic quark off a heavy nucleus is one of the most direct ways to study the structure of the cold nuclear matter at low values of Bjorken’s $x$. The perturbative calculations involve resummation of the multiple scatterings of the quark in the nucleus~\cite{Dumitru:2002qt} and of the radiative processes~\cite{Tuchin:2004rb,Kovchegov:2006qn} all in the eikonal limit. Due to the gluon saturation at small $x$, the typical transverse momentum scale in this process is the semi-hard saturation momentum, which makes the perturbative approach possible~\cite{Gribov:1984tu, Mueller:1985wy}. The resummation can be very efficiently performed by means of the Color Glass Condensate (CGC) theory that treats heavy nucleus as a random color field~\cite{Gelis:2010nm, Weigert:2005us}. 

While the perturbative eikonal approach yielded essential insights into the structure and dynamics of the cold nuclear matter at small $x$~\cite{Kovchegov:2012mbw}, the corresponding phenomenological approaches often suffer from uncertainties that arise from the infrared and sub-eikonal corrections. Motivated by the future experimental program at the Electron-Ion Collider~\cite{Accardi:2012qut} we initiate in this paper investigation of the sub-eikonal non-perturbative corrections to the quark-nucleus scattering using the computational formalism the time-dependent basis light-front quantization (tBLFQ)~\cite{Zhao:2013cma}. We ignore the radiative effects that contribute to the quantum evolution of the quark wave function with energy.

The tBLFQ formalism is a natural extension of the basis light-front quantization (BLFQ) approach \cite{1stBLFQ}, that has been developed based on the light-front quantum field theory and the Hamiltonian formalism to tackle bound state problems.
The implementation of the basis function representation allows us to choose a basis with the same symmetries of the system under investigation, and is therefore advantageous for carrying out efficient numerical calculations.
This method has been applied to study the QED bound state system of positronium ~\cite{positronium}, the QCD bound states of heavy and light mesons~\cite{Yang_fix, Yang_run, Shuo_Bc, Tang:2019gvn, Jia:2018ary}, and the bound states of the nucleon-pion system with a simple chiral model~\cite{Du:2019ips}.
It has been shown that tBFLQ is particularly well-suited for calculating non-perturbative effects through the applications of the nonlinear Compton scattering~\cite{Zhao:2013cma, Hu:2019hjx}, and the interaction of an electron with intense electromagnetic fields~\cite{Chen:2017uuq}. 
Its counterpart in quantum mechanics, the time-dependent basis function approach, has been applied to investigate deuteron heavy ion scatterings under the Coulomb interaction~\cite{Du:2018tce, Yin:2019kqv}.

In this paper we apply the tBLFQ formalism to investigate the quark-nucleus scattering, by treating the nucleus as a classical SU(3) color field using the CGC theory~\cite{McLerran:1993ni,McLerran:1993ka,McLerran:1994vd}. 
In particular, we solve for the time evolution of the quark as a quantum state inside the CGC. 
We calculate the quark-nucleus elastic and total scattering cross sections, and we study the evolution of the quark's distribution in the transverse coordinate space as well as in the color space. The intrinsic non-perturbative feature of the tBLFQ formalism provides us an opportunity to study the sub-eikonal effects. 
At high energy, the propagation time of the quark through the target nucleus is short and its transverse position does not change substantially during the propagation. Neglecting such change is usually implemented as the eikonal limit in many studies~\cite{Kovchegov:2012mbw}. 
However, in reality the quark carries a finite longitudinal momentum and therefore admits sub-eikonal effects. A variety of works using pQCD has studied sub-eikonal effects from different aspects, including helicity change of the quark, longitudinal momentum exchange, and finite length of the background field~\cite{Jalilian-Marian:2017ttv, Jalilian-Marian:2019kaf, Jalilian-Marian:2018iui, Kovchegov:2018znm, Kovchegov:2018zeq, Altinoluk:2015gia, Chirilli:2018kkw}.
In this work, we treat the quark with finite energy and keep its interaction time with the nucleus finite, and we reveal a sub-eikonal effect through the evolution of the quark's transverse coordinate distribution from the non-perturbative aspect.



The results presented in this paper can be used to calculate particle production in $p$A collisions in the proton fragmentation region by convoluting the quark total cross section with the quark distribution function of the proton and with the quark-hadron (jet) fragmentation function. Generalization to dipole-nucleus scattering is also straightforward and will allow us in the future to investigate deep inelastic scattering and exclusive vector meson production using light-front wavefunction obtained in the BLFQ formalism. This work also provides the foundation for the study of particle production and evolution in the glasma field created by heavy-ion collisions, where the initial gluon field can be solved analytically~\cite{Chen:2015wia}.

The layout of this paper is as follows. We introduce the formalism of tBLFQ in the application to the quark-nucleus scattering problem in Sec.~\ref{sec:background}. The numerical results are presented and discussed in Sec.~\ref{sec:results}. We conclude the work in Sec.~\ref{sec:summary}.

\section{Methodology: time-dependent basis light-front quantization}
\label{sec:background}

\begin{figure}[t]
  \centering
  \includegraphics[width=0.4\textwidth]{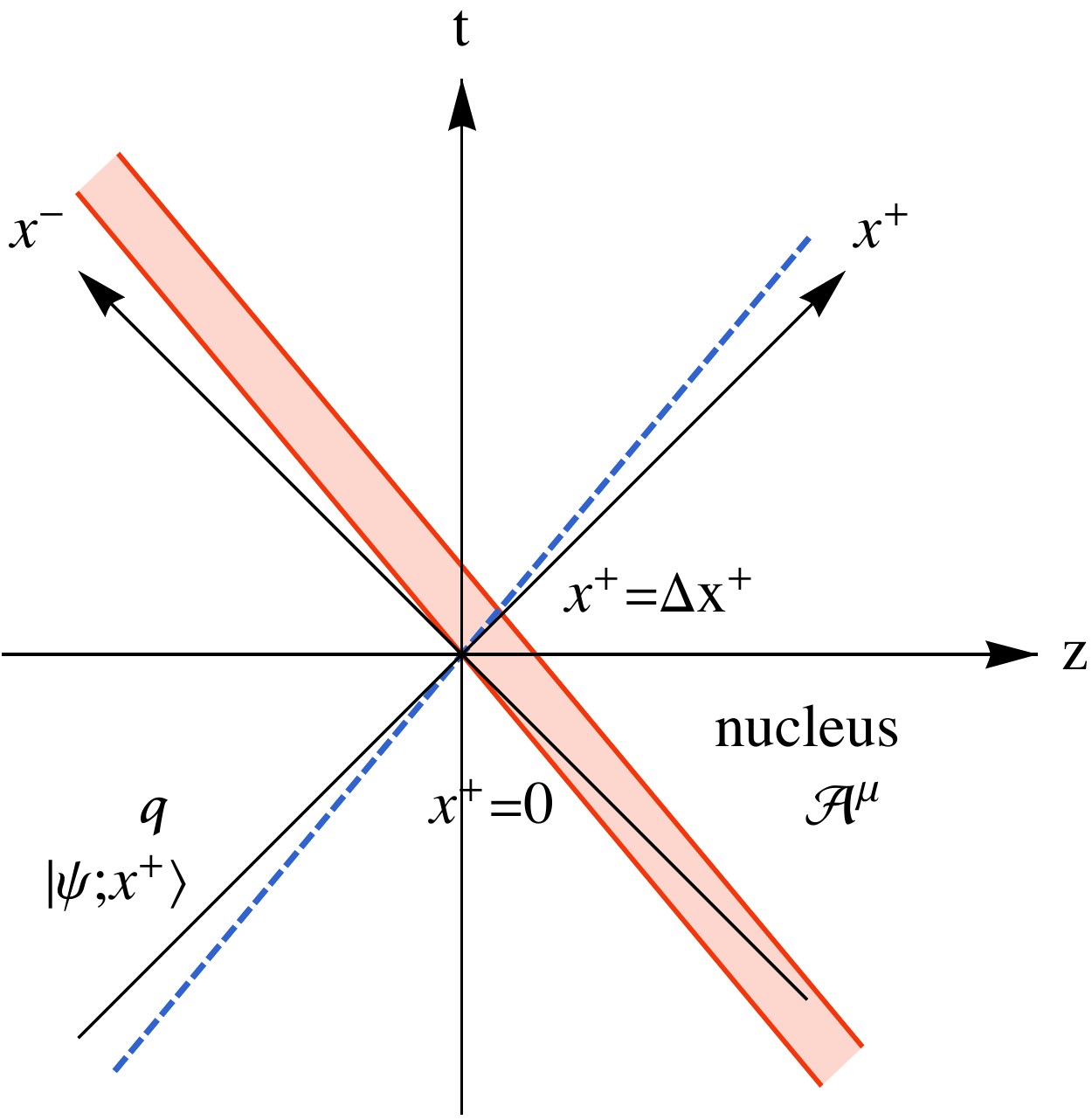}
  \caption{
  The quark is moving along the positive-z direction scatters on the nucleus along the negative-z direction. The dashed line is the worldline of the quark, $z=\beta_q t$ with $\beta_q$ the speed of the quark. The band represents worldlines of the nucleus, $z=-\beta_A t$ for one end and $z=-\beta_A t + d'$ for the other end. $\beta_A$ is the speed of the nucleus and $d'=d\sqrt{1-\beta_A^2}$ with $d$ the width of the nucleus in its rest frame. In the ultra-relativistic limit of $\beta_A\to 1$, the red band in the diagram shrinks to a single line aligned with $x^+=0$.}
 \label{fig:dis_zt}
\end{figure}

We start by considering scattering of a high-energy quark moving in the positive $z$ direction, on a high-energy nucleus moving in the negative $z$ direction, as shown in Fig.~\ref{fig:dis_zt}. The quark has momentum $p^\mu$ and $p^+>>p^-, p_\perp$ whereas the nucleus has momentum $P^\mu$ and $P^->>P^+, P_\perp$ (see definitions of the light-front variables in Appendix A). We treat the quark state at the amplitude level and the nucleus as an external background field. The quark interacts with the nuclear field at $0\le x^+\le \Delta x^+$. 

\subsection{Time evolution under a background field}
We consider a quark interacting with the background field generated by the heavy nucleus. To start with, we truncate the Fock space of the quark to the leading  sector as $\ket{q}$. Consequently, the QCD Lagrangian reduces to 
\begin{align}
  \label{eq:Lag}
  \mathcal{L}_{q}=\overline{\Psi}(i\gamma^\mu \bm D_\mu - \bm m)\Psi\;,
\end{align}
 where $D^\mu\equiv \partial_\mu \bm I+ig \bm{\mathcal{A}}^\mu$ and $\bm m = m\bm I$. $\bm I$ is the 3 by 3 unit matrix in color space, and $\bm{\mathcal{A}}^\mu = \mathcal{A}^{a \mu} T^a$ is the background gluon field.  The light-front Hamiltonian is derived from the Lagrangian through the standard Legendre transformation~\cite{Brodsky:1997de},
 \begin{align}
  \begin{split}
    P^-=&\int\diff x^-\diff^2 x_\perp
    \bigg \{
    \frac{1}{2}\bar{\Psi}\gamma^+\frac{m^2-\nabla_\perp^2}{i\partial^+}\Psi\\
    &+g \bar\Psi\gamma^\mu T^a\Psi \mathcal{A}_\mu^a
    +\frac{g^2}{2} \bar\Psi \gamma^\mu T^a \mathcal{A}^a_\mu\frac{\gamma^+}{i\partial^+}\gamma^\nu T^b \mathcal{A}^b_\nu\Psi
    \bigg \}\;.
  \end{split}
\end{align}
The standard QCD light-front Hamiltonian is formulated in the light cone gauge of $A^+=0$. Here the dynamical gauge field is absent due to Fock sector truncation, and we apply the condition $\mathcal{A}^+=0$ to the background field.
The first term is the light-front QCD Hamiltonian in the $\ket{q}$ sector, which is the kinetic energy of the quark, denoted as $P_{QCD}^-$. The two terms in the second line are the interactions introduced by the external field, and together they are signified by $V$. 
The interaction term, in general, could have a time dependence arising from the external field, such that $P^-(x^+) = P_{QCD}^- + V(x^+)$.

We are interested in how the quark, as an eigenstate of the QCD Hamiltonian, $P^-_{QCD}$, evolves due to interactions with the background field. It is therefore natural to use an interaction picture to solve the evolution equation on the light front,
\begin{align}
  \label{eq:ShrodingerEq}
  i\frac{\partial}{\partial x^+}\ket{\psi;x^+}_I=\frac{1}{2}V_I(x^+)\ket{\psi;x^+}_I\;.
\end{align}
$V_I(x^+)=e^{i\frac{1}{2}P^-_{QCD}x^+}V(x^+)e^{-i\frac{1}{2}P^-_{QCD}x^+}$ is the interaction Hamiltonian in the interaction picture. The solution of Eq.~\eqref{eq:ShrodingerEq} describes the state of the investigated system at any given light-front time $x^+$,
\begin{align}\label{eq:ShrodingerEqSol}
  \ket{\psi;x^+}_I=\mathcal{T}_+\exp\left[-\frac{i}{2}\int_0^{x^+}\diff z^+V_I(z^+)\right]\ket{\psi;0}_I\;,
\end{align}
where $\mathcal{T}_+$ is the light-front time ordering. In the perturbative calculations, the time-ordered exponential is written as a
Taylor series expansion, and only the leading terms are retained. However, in cases where the external fields are strong, the perturbative treatment may not be sufficient. Our aim is to solve the problem through a non-perturbative treatment. We
decompose the time-evolution operator into many small steps of the light-front time $x^+$,
\begin{align}
  \begin{split}
    \mathcal{T}_+ & \exp\left[-\frac{i}{2}\int_0^{x^+}\diff z^+V_I(z^+)\right] \\
    &=\mathcal{T}_+ \lim_{n\to\infty}\prod^n_{k=1}
    \left[1-\frac{i}{2}V_I(x_k^+) \frac{x^+}{n}\right]
    \\
  &=\lim_{n\to\infty}\left[1-\frac{i}{2}V_I(x_n^+)\delta x^+\right]\ldots \left[1-\frac{i}{2}V_I(x_1^+)\delta x^+\right]
  \;,
  \end{split}
\end{align}
The step size is $\delta x^+ \equiv x^+/n$, and the intermediate times are $x_k^+=k\delta x^+ (k=1,2,\ldots,n)$. This product expansion is exact in the continuum limit of $n\to\infty$. In practical calculations, the value of $n$ could be determined so as to achieve a desiring convergence of the final state. Observables could then be evaluated from the evolved state.

\subsection{Gluon field as the Color Glass Condensate}
The CGC formalism provides a description of gluon dynamics in the small-x region~\cite{Jalilian-Marian:2017ttv}.  
The underlying approximation involved in the CGC theory of high energy scattering is the eikonal approximation, i.e. small angle deflection of a high energy projectile traversing a medium. 
The classical gluon field is found from the Yang-Mills equation,
\begin{equation}
  D_{\mu}\mathcal{F}^{\mu\nu}=J^\nu \;.
\end{equation}
$J^\nu = J^\nu_a T_a$ ($a = 1, 2, \ldots, 8$) is the color current, and $T_a$ is the color generator. The current generated by the high-energy nucleus moving along the negative z direction has only one nonzero component, $J^\nu_a=\delta^{\nu-}\rho_a$, and it is independent of its time $x^-$~\cite{Kovchegov:2012mbw}. 

Due to  Lorentz contraction, the $x^+$ dependence of the nucleus is peaked around $x^+=0$, and in the extreme limit it is usually taken to be a delta function. Here we keep the $x^+$ dependence to allow for an extended target. The valence charges are treated as stochastic variables satisfying the correlation relation,
\begin{equation}\label{eq:chgcor}
  \Braket{\rho_a(\vec{x}_\perp,x^+)\rho_b(\vec{y}_\perp,y^+)}=g^2\mu^2\delta_{ab}\delta^2(\vec{x}_\perp-\vec{y}_\perp)\delta(x^+-y^+)\;.
\end{equation}
This correlation relation could be achieved by taking the color charge density $\rho_a(\vec{x}_\perp,x^+)$ to be a stochastic random variable with
a local Gaussian distribution~\cite{McLerran:1993ni, McLerran:1993ka},
\[f[\rho^2_a(\vec{x}_\perp,x^+)]=\exp \bigg[-\frac{\delta x^+ \delta^2 x_\perp}{g^2 \mu^2}  \rho^2_a(\vec{x}_\perp,x^+)\bigg]\;.\]
$\delta x^+$ and $\delta^2 x_\perp$ are the unit lengths in the $x^+$ and $\vec x_\perp$ directions respectively.
The Gaussian form is reasonable when the color charges at high rapidity are uncorrelated and random~\cite{McLerran:1998nk,JalilianMarian:1996xn}. 

The field in the covariant gauge of $\partial^\mu \mathcal{A}_\mu = 0$ has only one nonzero component $\mathcal{A}^-$,
\begin{align}\label{eq:poisson}
  (m_g^2-\nabla^2_\perp )  \mathcal{A}^-_a(\vec{x}_\perp,x^+)=\rho_a(\vec{x}_\perp,x^+)\;.
\end{align}
The gluon mass $m_g$ is introduced to regularize the infrared (IR) divergence in the field, which simulates color neutrality on the source distribution~\cite{krasnitz2003gluon}. 
The field solved from this regularized Poisson equation can be expressed in terms of the Green's function
\begin{align}
  \mathcal{A}^-_a(\vec{x}_\perp,x^+)=\int\diff^2 y_\perp G_0(\vec x_\perp-\vec y_\perp)\rho_a(\vec y_\perp, x^+) \;,
\end{align}
where
\begin{align}\label{eq:Green}
  G_0(\vec x_\perp-\vec y_\perp)=-\int \frac{\diff^2 k_\perp}{{(2\pi)}^2}\frac{e^{-i \vec k_\perp \cdot (\vec x_\perp-\vec y_\perp)}}{m_g^2+\vec k_\perp^2}\;.
\end{align}
The field is logarithmically ultraviolet (UV) divergent. The divergence corresponds to the large momentum modes in the nuclear wavefunction, which are the degrees of freedom not meant to be included in the classical field. It is then natural to introduce a UV regulator~\cite{Lappi:2006hq}. In the numerical calculations of this work, the discretization of the transverse space automatically introduces a UV cutoff. Alternatively, one can introduce an additional parameter $\Lambda_{UV}$ when solving the gluon field as in Eq.~\eqref{eq:Green}, and the integral measure becomes $\int^{\Lambda_{UV}} \diff k_\perp$~\cite{Muller:2019bwd}. 

In this work, we follow the McLerran-Venugopalan (MV) model, where a quantum correction to saturation scale $Q_s$ is not implemented. Consequently, the saturation scale is a constant for a fixed charge density $g^2\mu$, $L_\eta$, and the extension of the field along $x^+$~\cite{Dumitru:2002qt,fukushima2007light},
\begin{align}\label{eq:Qs}
  Q_s^2=\frac{(g^2\mu)^2L_\eta}{2\pi^2}\;.
\end{align}
This differs from more sophisticated methods where the saturation scale is related to the gluon structure function of the nucleus and depends on $x$~\cite{Kovchegov:2012mbw}. In our numerical analysis we vary the density parameter $g^2\mu$ at fixed $L_\eta=50~\GeV^{-1}\approx 10~\text{fm}$ (see Appendix~\ref{app:Leta} for more discussion of $L_\eta$).

Note that since the background field has only one nonzero component $\mathcal{A}^-$, the quark instantaneous interaction vanishes and only one term remains in $V$, 
\begin{align}
  V=\int\diff x^-\diff^2 x_\perp
    g \bar\Psi\gamma^+ T^a\Psi \mathcal{A}_+^a
  \;.
\end{align}
This interaction changes the transverse dependence and the color distribution of the quark, but leaves the longitudinal distribution and the spin of the quark unchanged.

\subsection{Basis construction}
To solve the time evolution equation of Eq.~\eqref{eq:ShrodingerEq}, one could select a basis and then work with the matrix form of the equation. An optimal basis should preserve the symmetries of the system and approximate the eigenfunctions of the Hamiltonian. 



To begin, let us first identify the eigenstates $\ket{\beta}$ and eigenvalues $P^-_\beta$ of $P^-_{QCD}$, such that
\begin{align} 
    P^-_{QCD}\ket{\beta}=P^-_\beta\ket{\beta}
  \;.
\end{align}
Since $P^-_{QCD}$ only contains the kinetic energy of the quark, its eigenstates are therefore the momentum states of the quark. Considering that the background field interacts with the quark in the transverse space and the color space, we construct the basis state as $\ket{\beta} = \ket{k_x, k_y, c}$, where $k_x$ and $k_y$ are the transverse momentum of the quark and $c$ is the color of the quark. This choice of basis is very similar to the discretized momentum representation~\cite{Pauli:1985pv,Eller:1986nt}.

We then expand the quark state as a sum over the QCD eigenstates:
\begin{align} 
  \ket{\psi;x^+}_I=\sum_\beta c_\beta (x^+)\ket{\beta}
  \;,
\end{align}
where $c_\beta(x^+)\equiv\braket{\beta|\psi;x^+}_I$ are the basis coefficients. The initial state at $x^+=0$ can be specified by $c_\beta(0)$ as a vector $c(0)$. The solution of Eq.~\eqref{eq:ShrodingerEqSol} can be written in the QCD eigenstate basis in the matrix form as
\begin{align}\label{eq:ShrodingerEqSolBasis}
  c(x^+) = \mathcal{T}_+\exp{\left(-i\int_0^{x^+}\diff z^+ \mathcal{M}(z^+)\right)} c(0)\;,
\end{align}
where the matrix elements of $\mathcal{M}(x^+)$ are defined as $\mathcal M_{\beta\beta'}(x^+)\equiv \braket{\beta|V_I(x^+)/2|\beta'}$. Once we know the wavefunction of the state via $c(x^+)$, it is straightforward to evaluate observables from it.

\subsection{Numerical scheme}
In the numerical calculation, the fields are color SU(3) matrices on the sites of a 3-dimensional discrete space.
The 2-dimensional transverse space is a lattice extending from $-L$ to $L$ for each side. The number of transverse lattice sites is
$2N$, giving the lattice spacing $a=L/N$. As such, a vector $\vec r_\perp=(r_x,r_y)$ would read as,
\[
  r_i= n_i a ~(i = x,y) , \quad n_i=-N,-N+1,\ldots,N-1 .
\]
This space satisfies periodic boundary conditions. It follows that in the
momentum space, for any vector, $\vec p_\perp=(p_x,p_y)$,
\[
  p_i=k_i d_p~ (i = x,y), \quad k_i=-N,-N+1,\ldots,N-1,
\]
where $d_p\equiv \pi/L$ is the resolution in momentum space. The momentum space extends from $-\pi/a$ to $\pi/a$. Therefore, the transverse lattice introduces a pair of IR and UV cutoffs, $\lambda_{IR}=\pi/L$ and $\lambda_{UV}=N\pi/L$. We include details of the conventions and relations in the discrete space in Appendix~\ref{app:dis}.

The longitudinal dimension of the field $x^+$ (note that this is the
light-front time of the incident quark) is discretized into a number of $N_\eta$ layers~\cite{lappi2008wilson}. If the field extends $L_\eta$ along $x^+$,
each layer would have an expansion of $\tau=L_\eta/N_\eta$. For example, the $k$-th ($k=1,2,\ldots,N_\eta$) layer extends as $x^+ = [(k-1)\tau,k\tau]$.

To summarize, our calculation depends on those numerical parameters.
\begin{itemize}
\item $g^2\mu$, color charge density parameter. We take different values for it and study how observables depend on it.
\item $m_g$, screening mass, the IR regulator. We will take $m_g = 0.1~\GeV$ and use a range of values when investigating its role.
\item The transverse lattice: size $L$, number $N$ and spacing $a=L/N$. In most cases, we take $L=50~\GeV^{-1}(=9.87~\text{fm})$ as estimated from the radius of gold nucleus. Exceptions will be separately noted.
\item The $x^+$ direction: duration $L_\eta$, the number of layers $N_\eta$ and interval $\tau=L_\eta/N_\eta$. We take $L_\eta=50~\GeV^{-1}$ and study the convergence on $N_\eta$. 
\end{itemize}
In this discretized space, the correlation relation of the color charge as defined in Eq.~\eqref{eq:chgcor} also takes a discrete form as,
\begin{equation}\label{eq:chgcor_dis}
  \begin{split}
  \Braket{\rho_a(n_x,n_y,k)\rho_b(n'_x,n'_y,k')} &\\
  =g^2\mu^2\delta_{ab}
  &\frac{\delta_{n_x,n'_x}\delta_{n_y,n'_y}}{a^2}
  \frac{\delta_{k,k'}}{\tau}\;.
\end{split}
\end{equation}
Note that the Kronecker delta dividing the discrete resolution replaces the Dirac delta in Eq.~\eqref{eq:chgcor}, and they converge in the continuous limit of $a\to 0$ and $\tau\to 0$.
\section{Numerical Results}
\label{sec:results}
In this section, we discuss various observables obtained from the tBLFQ formalism. We first study the total and elastic cross sections, and support our approach by comparing to predictions in the eikonal limit. We then study the differential cross section and look into the time evolution of the quark's distribution in transverse coordinate space and color space. We further relax the eikonal condition and explore sub-eikonal effects with finite $p^+$.
\subsection{The cross sections}
The cross section is calculated as the sum of the squares of the transition amplitudes.~\cite{Peskin:1995ev}, 
\begin{align}
  \begin{split}
    \frac{\diff\sigma}{\diff^2 b}
    =&\sum_{\phi_f}
    {|  M(\phi_f; \psi_i)|}^2
    =\sum_{\phi_f}
    {|\braket{\phi_f|S|\psi_i}-\braket{\phi_f|\psi_i}|}^2
    \;.
  \end{split}
\end{align}
$\psi_i$ stands for the initial state, and $\phi_f$ is the final state; $\sum_{\phi_f}$ sums over the phase space of the final state. The $S$ in the equation is the evolution operator from the initial state to the final through a finite time transition, and is different from the case where one takes the infinite time limit. In evaluating the cross section, one should average over the color charge density $\rho$ as in Eq.~\eqref{eq:chgcor}. This would give us the total cross section by definition,
\begin{align}
  \begin{split}
    \frac{\diff\sigma_{\mathrm{tot}}}{\diff^2 b}
    =&\langle\sum_{\phi_f}{|  M(\phi_f; \psi_i)|}^2\rangle
    \;.\\
  \end{split}
\end{align}
The total cross section is the summation of the elastic $2\to 2$ contribution and the inelastic contributions ($2\to 3$, $2\to 4$, etc.),
\(\sigma_{tot}=\sigma_{el} + \sigma_{inel}\)~\cite{Kovchegov:2012mbw}. To calculate the elastic cross section, we carry out the configuration average on the amplitude level to get the elastic scattering amplitude first, and afterwards square it~\cite{Mueller:1997ik, Kovchegov:1999kx, Dumitru:2002qt}. 
\begin{align}
    \frac{\diff\sigma_{\mathrm{el}}}{\diff^2 b}
    =\sum_{\phi_f}{|  \langle M(\phi_f; \psi_i) \rangle|}^2
    \;.
\end{align}

In the eikonal limit, the longitudinal momentum of the quark is infinite, $p^+=\infty$, thus the phase factor $e^{\pm ip^-x^+}$ (see text associated with Eq.~\eqref{eq:ShrodingerEq}) is 1 and $V_I(x^+)$ reduces to $V(x^+)$. 
In this limit, the cross sections can be expressed analytically in terms of the charge density $g^2\mu$, the interaction duration $L_\eta$, the IR cutoff $\Lambda_{IR}$ and the UV cutoff $\Lambda_{UV}$~\cite{Dumitru:2002qt}. A detailed derivation of the interaction matrix is included in Appendix~\ref{app:Wilsonline}.
\begin{align}\label{eq:crss_eikonal}
  \begin{split}
    \frac{\diff\sigma_{\mathrm{tot}}}{\diff^2 b}&\bigg|_{p^+=\infty}\\
    =&2\bigg\{
    1-
    \exp\bigg[
    -\frac{(N_c^2-1)(g^2\mu)^2L_\eta}{16\pi N_c}(\frac{1}{\Lambda_{IR}^2}-\frac{1}{\Lambda^2_{UV}})
    \bigg]
    \bigg\}
    \;,\\
    \frac{\diff\sigma_{\mathrm{el}}}{\diff^2 b}&\bigg|_{p^+=\infty}\\
    =&\bigg\{1-
    \exp\bigg[
    -\frac{(N_c^2-1)(g^2\mu)^2L_\eta}{16\pi N_c}(\frac{1}{\Lambda_{IR}^2}-\frac{1}{\Lambda^2_{UV}})
    \bigg]
    \bigg\}^2
    \;.
  \end{split}
\end{align}
We have introduced the screening mass as the IR regulator, therefore $\Lambda_{IR}=m_g$. Though the gluon field has a UV divergence, the cross sections have finite $\Lambda_{UV}\to\infty$ limits. We therefore do not implement such a cutoff in calculating the cross sections unless specified.

We first calculate the total and elastic cross sections in the eikonal limit and compare our results with the eikonal expectations in Eq.~\eqref{eq:crss_eikonal}. We also study the sensitivity of
the cross sections to the parameters, $N, L, N_\eta$, and $m_g$. We then relax the eikonal condition to allow a finite $p^+$, and explore potential effects. The light-front kinetic energy of the quark is calculated as $p^-=(\vec p_\perp^2 + m_q^2)/p^+$, we use $m_q=0.15~\GeV$ in the presented results. We have checked that using quark mass in the range of $m_q=0.05-4.50~\GeV$ does not make noticeable change in the results.

We then check the dependence of the cross sections on the lattice by varying $N$ and $L$. Note that a reasonable numerical grid should cover the physical range of interest. In this case, we should make sure that the numerical IR cutoff $\lambda_{IR}=\pi/L$ is much smaller than the physical IR cutoff $\Lambda_{IR}=m_g$, and the numerical UV cutoff $\lambda_{UV}=N\pi/L$ much higher than that.
Thus a suitable grid for our investigation should satisfy:
\begin{align}\label{eq:grid_cutoffs}
\frac{\pi}{L} \ll m_g\ll N\frac{\pi}{L}\;.
  \end{align}

\begin{figure*}[htp!]
  \centering
  \subfigure[\ The total cross section \label{fig:L50N_eikonal_tot}]
  {\includegraphics[width=0.45\textwidth]{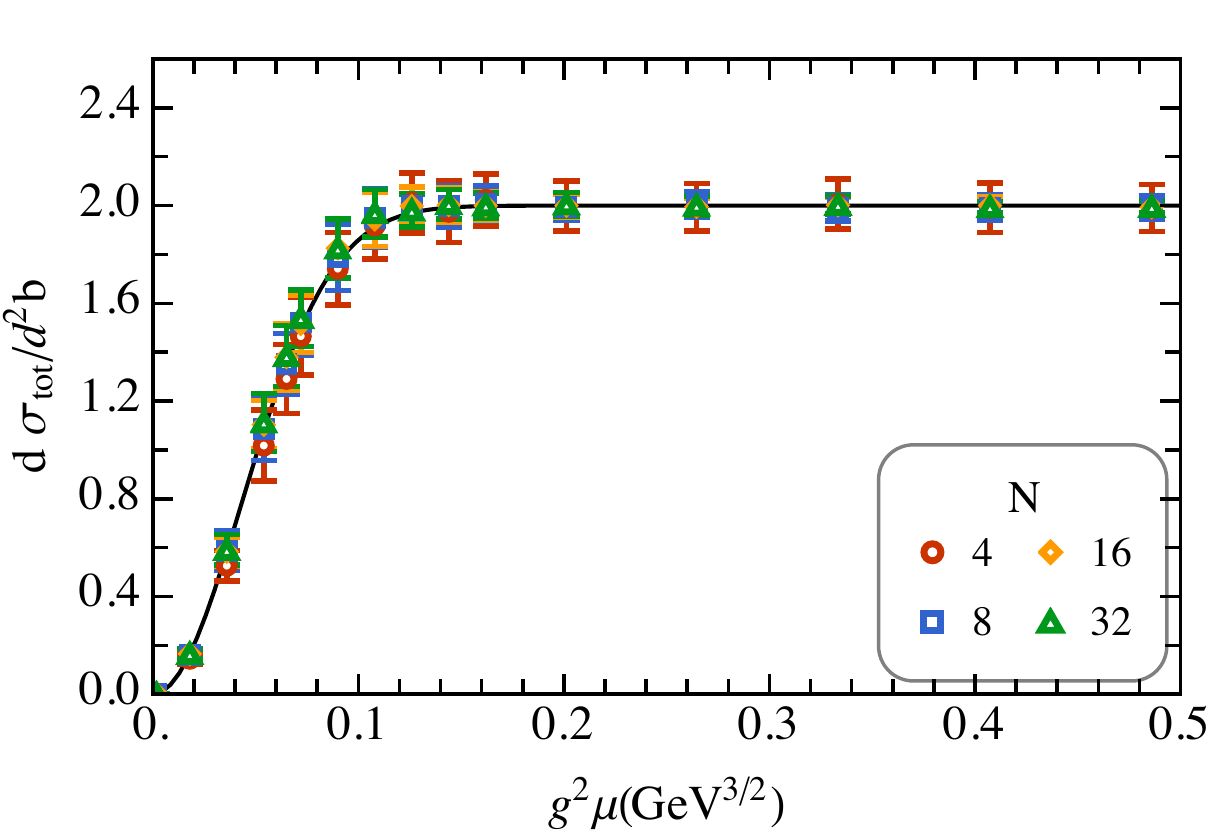}
  } 
  \subfigure[\ The elastic cross section \label{fig:L50N_eikonal_els}]
  {\includegraphics[width=0.45\textwidth]{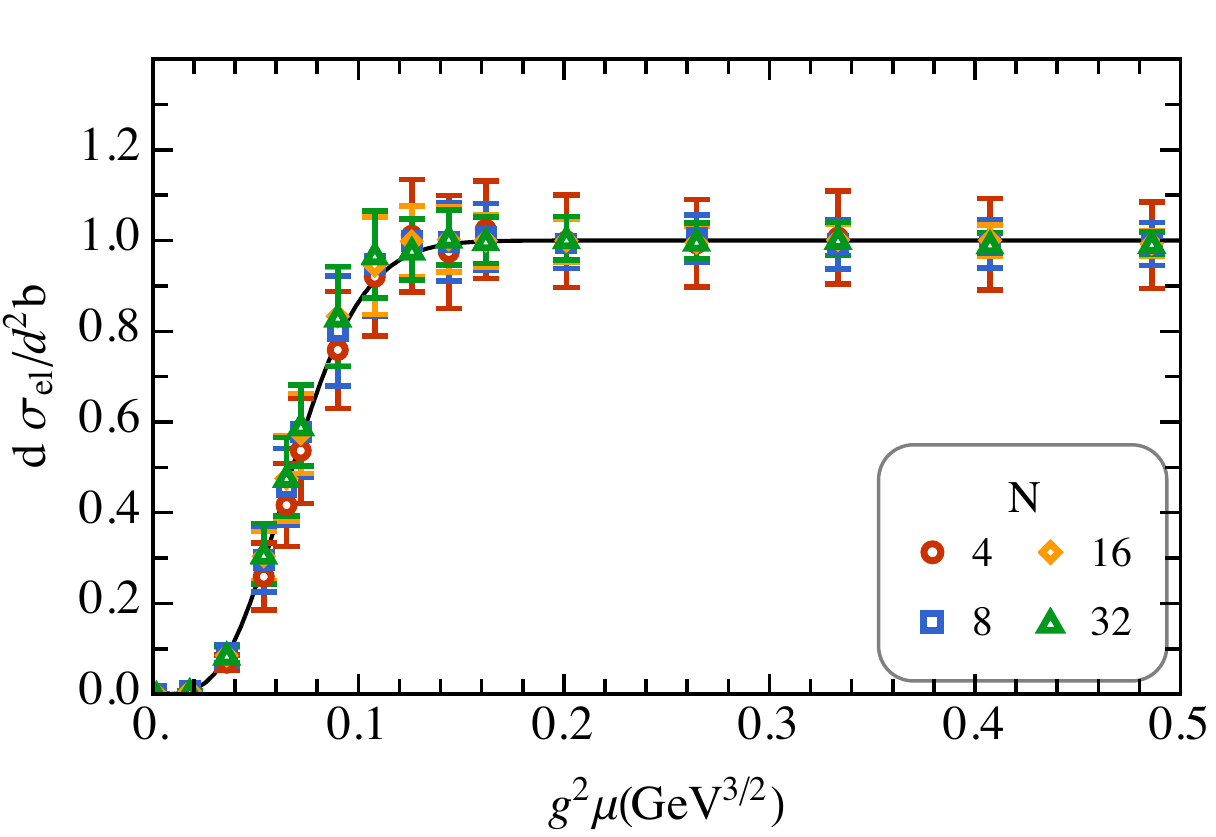}
  } 
  \caption{The dependence on the transverse grid number $N$ of (a) the total and (b) the elastic cross sections at $L=50~\GeV^{-1}$. The cross sections are calculated as functions of $g^2\mu$ with $L_\eta=50~\GeV^{-1}$, $N_\eta=4$ and $p^+=\infty$.
  The solid lines are the eikonal predictions as calculated from Eq.~\eqref{eq:crss_eikonal}. Each data point results from an average over 100 configurations,
  and the standard deviation is taken as the uncertainty bar. }
  \label{fig:crss_N}
\end{figure*}

Figure~\ref{fig:crss_N} represents the total and elastic cross sections as functions of $g^2\mu$ at different $N$ for a fixed $L$. 
The results show a convergence with increasing $N$. We take the standard deviation of the 100 averaged configurations as the uncertainty. Such uncertainty is smaller at larger $N$. This is not hard to imagine, since with more sites on the lattice, the fluctuation of each configuration is more likely to smooth out when averaged over equal number of events.

\begin{figure*}[htp!]
  \centering
  \subfigure[\ The total cross section ]
  {\includegraphics[width=0.45\textwidth]{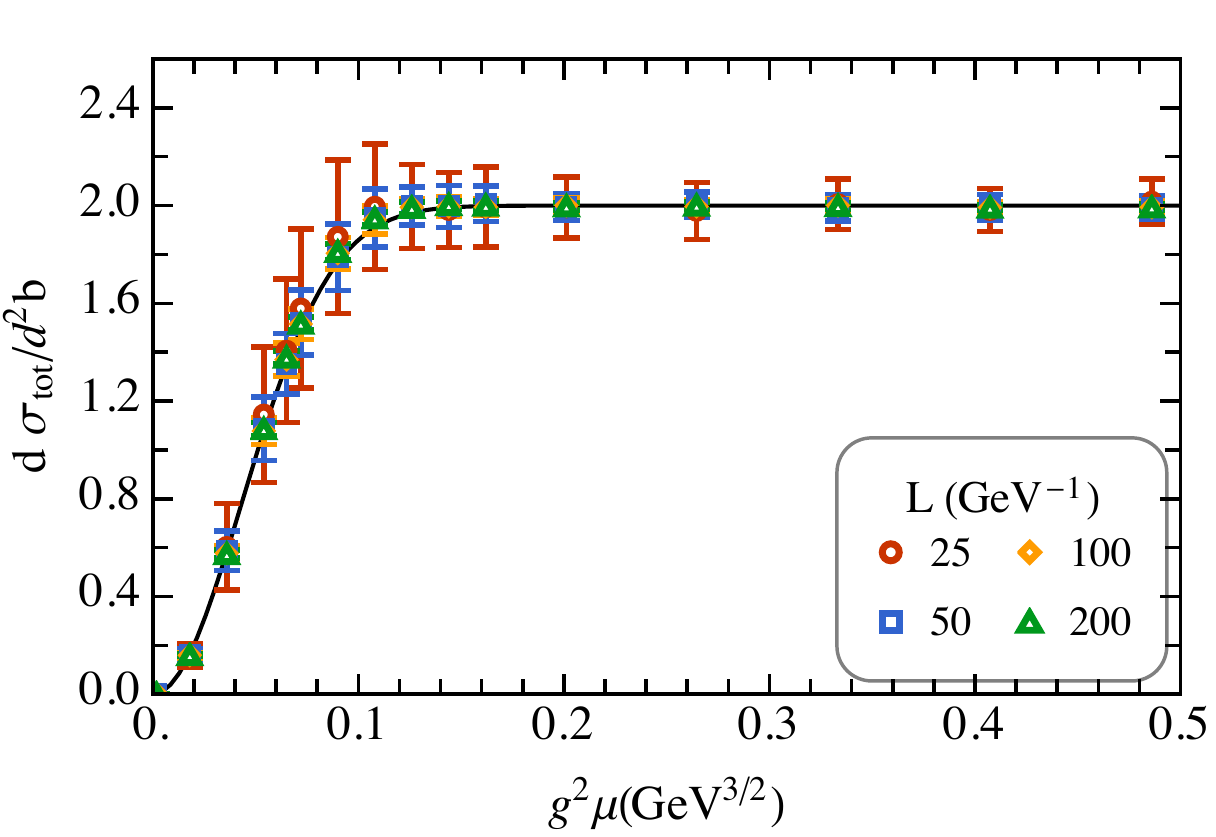}
  } 
  \subfigure[\ The elastic cross section]
  {\includegraphics[width=0.45\textwidth]{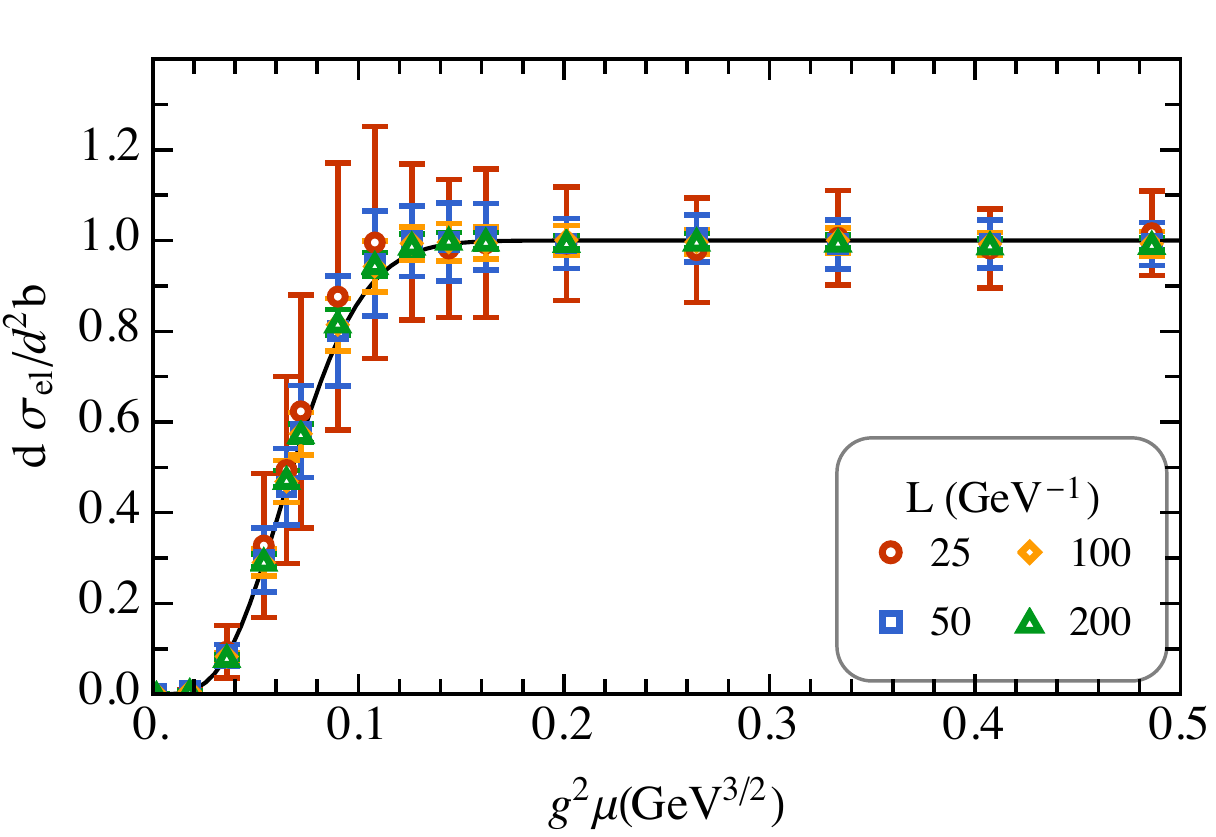}
  }
  \caption{The dependence on the transverse grid length $L$ of (a) the total and (b) the elastic cross sections. 
  The lattice spacing is fixed as $a=L/N=6.25~\GeV^{-1}$ for these results.
  The cross sections of the quark are plotted as functions of $g^2\mu$ at $L_\eta=50~\GeV^{-1}$, $N_\eta=4$ and $p^+=\infty$. The solid lines are the eikonal predictions as calculated from Eq.~\eqref{eq:crss_eikonal}. 
  Each data point is averaged over 100 configurations, and the standard deviation is taken as the uncertainty bar. }
  \label{fig:crss_L}
\end{figure*}

Most importantly, there is a good agreement between the tBLFQ results and the eikonal analytical expectations calculated from Eq.~\eqref{eq:crss_eikonal}. This agreement helps verify our formalism.
The dependence of the cross sections on the grid size $L$ is also checked and shown in Figure~\ref{fig:crss_L}. The total and elastic cross sections are calculated as functions of $g^2\mu$ at different $L$ for a fixed lattice spacing $a=L/N=6.25~\GeV^{-1}$. 
The results show agreement with the eikonal analytical expectations from Eq.~\eqref{eq:crss_eikonal}. We again observe that the lattice with a larger number of grids has smaller uncertainties. The cross sections are not
sensitive to the grid size.

\begin{figure*}[t]
  \centering
  \subfigure[\ The total cross section ]
  {\includegraphics[width=0.45\textwidth]{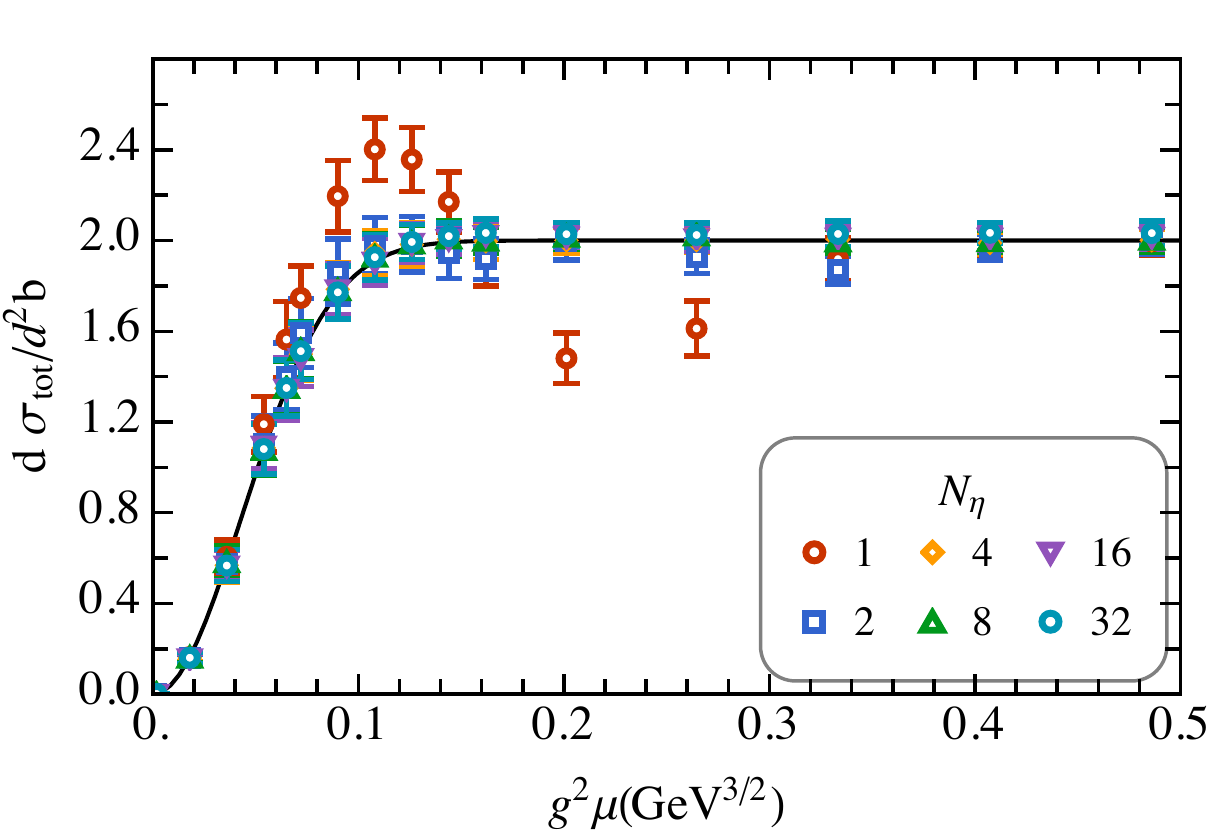}
  } 
  \subfigure[\ The elastic cross section]
  {\includegraphics[width=0.45\textwidth]{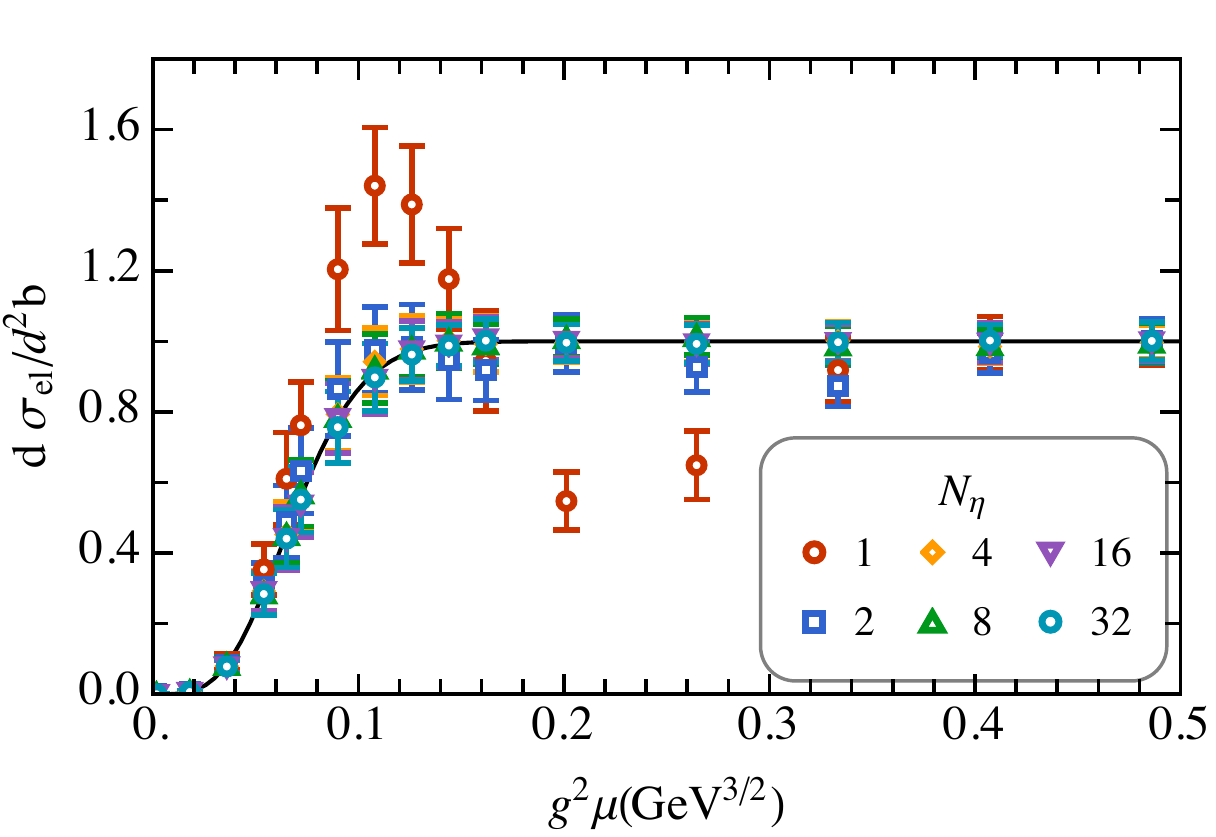}
  }
  \caption{The dependence on $N_\eta$ of (a) the total and (b) the elastic cross sections. 
  Parameters for those results: $L=50 ~\GeV^{-1}$, $N=8$, $L_\eta=50 ~\GeV^{-1}$ and $p^+=\infty$.
  The solid lines are the eikonal predictions as calculated from Eq.~\eqref{eq:crss_eikonal}. 
  Each data point is averaged over 100 configurations, and the standard deviation is taken as the uncertainty bar. }
  \label{fig:crss_Ny}
\end{figure*}

We next show in Fig.~\ref{fig:crss_Ny} the dependence of the cross sections on the number of layers in the longitudinal direction, $N_\eta$. An interesting
``oscillation'' pattern is observed when $N_\eta=1$. At $N_\eta=1$, the source hence the gluon field along $x^+$ is
constant, this breaks one necessary ingredient for the CGC field: sources are uncorrelated along $x^+$, as in Eq.~\eqref{eq:chgcor}.
It follows that in deriving the analytical expression of the cross section, the contraction of multiple sources is no longer
preserved, causing a nontrivial ``oscillation''. In our calculation, the $x^+=[0,L_\eta]$ duration is divided into $N_\eta$ layers, each lasting equally for
$\tau=L_\eta/N_\eta$. The color charges from different layers belong to different nucleons, so they are uncorrelated with each
other, as in Eq.~\eqref{eq:chgcor_dis}. Within each layer, the field is constant along $x^+$. The
continuum limit is restored at $N_\eta\to\infty$, as in Eq.~\eqref{eq:Ny_limit}.

This ``oscillation'' gets strongly suppressed when $N_\eta=2$, and for larger
$N_\eta(\ge 4)$, the physical results converge to the analytical expectation and depend very little on $N_\eta$, as shown in Fig.~\ref{fig:crss_Ny}.

\begin{align}\label{eq:Ny_limit}
  \begin{split}
    \int_{-\infty}^{+\infty} &\diff x^+\int_{-\infty}^{+\infty}\diff y^+\braket{\rho_a(x^+, x_\perp)\rho_b(y^+, y_\perp)}\\
    &=g^2\mu^2\delta_{ab}\delta^2(\vec{x}_\perp-\vec{y}_\perp)\int_{-\infty}^{+\infty}\diff x^+\int_{-\infty}^{+\infty}\diff y^+
    \frac{\delta_{k,k'}}{\tau}\\
    &=g^2\mu^2\delta_{ab}\delta^2(\vec{x}_\perp-\vec{y}_\perp)
    (\sum_{k=1}^{N_\eta}\tau^2)
    \frac{1}{\tau}\\
    &=g^2\mu^2\delta_{ab}\delta^2(\vec{x}_\perp-\vec{y}_\perp)L_\eta
    \;.
  \end{split}
\end{align}

\begin{figure*}[t]
  \centering
  \subfigure[\ The total cross section ]
  {\includegraphics[width=0.45\textwidth]{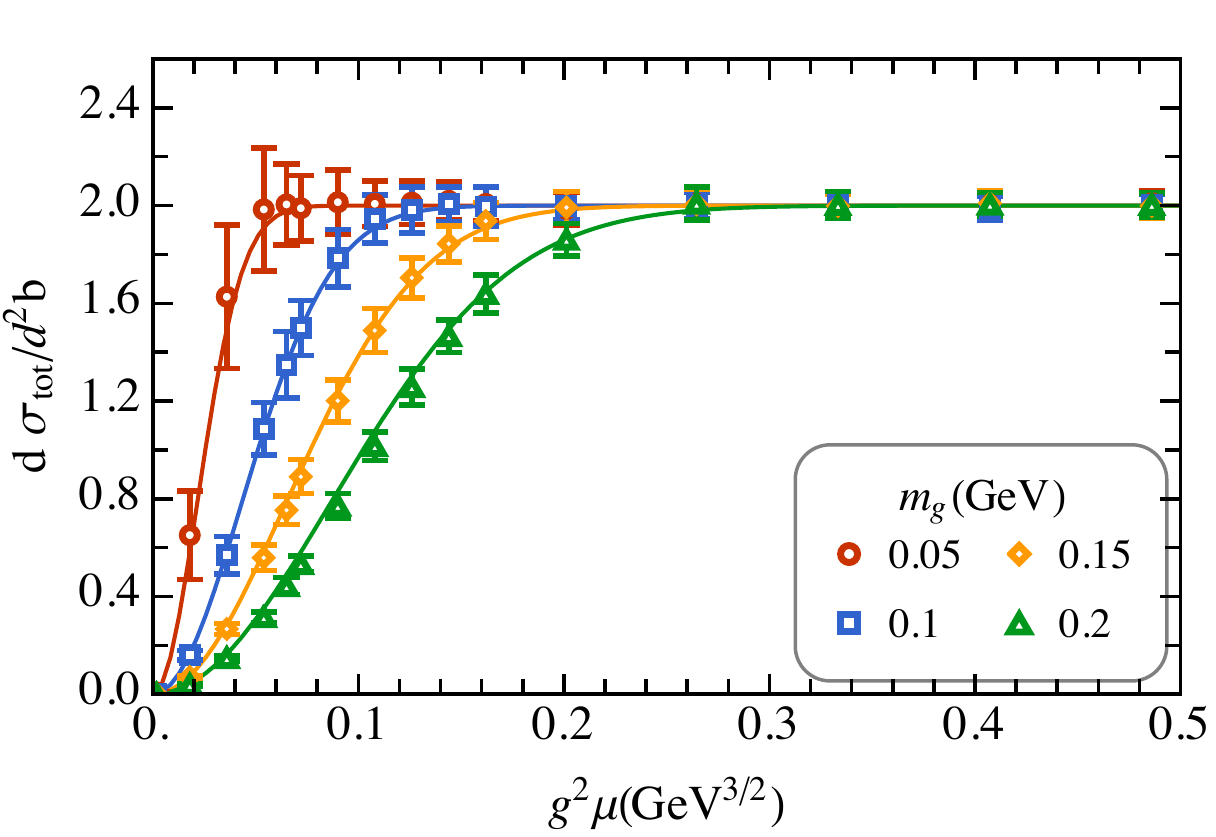}
  } 
  \subfigure[\ The elastic cross section]
  {\includegraphics[width=0.45\textwidth]{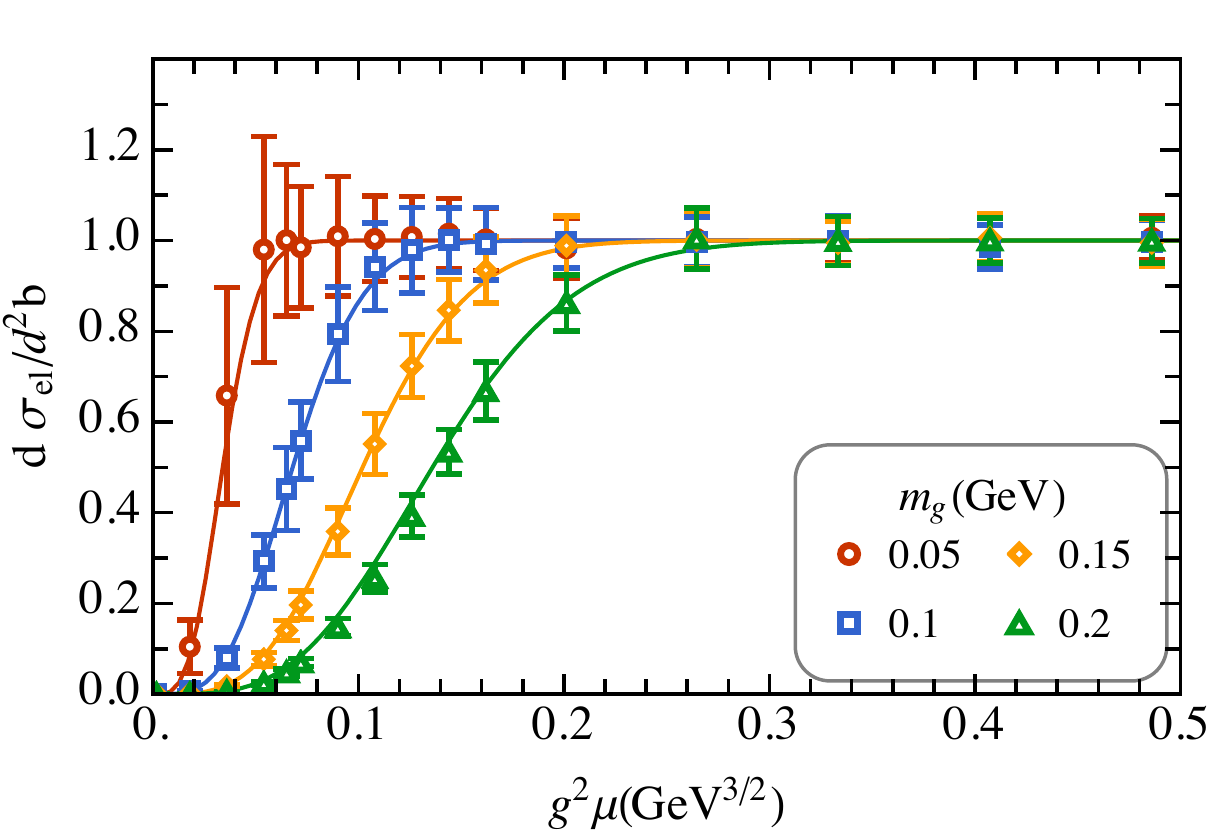}
  }
  \caption{The dependence of (a) the total and (b) the elastic cross sections on $m_g$. Parameters for those panels, $N=8$, $L=50 ~\GeV^{-1}$, $L_\eta=50 ~\GeV^{-1}$, $N_\eta=4$ and $p^+=\infty$. 
  The transverse grid parameters introduce numerical IR cutoff $\lambda_{IR}=\pi/L\approx 0.06~\GeV$ and UV cutoff $\lambda_{UV}=N\pi/L\approx 0.5~\GeV$ to the momentum space. 
  The physical IR cutoff $m_g$ should be inside the numerical range to obtain a valid result.
  The solid lines are the eikonal predictions as calculated from Eq.~\eqref{eq:crss_eikonal}.  Each data point is averaged over 100 configurations, and the standard deviation is taken as the uncertainty bar. }
  \label{fig:crss_mg}
\end{figure*}

Another dependence of the cross sections comes from the IR cutoff $\Lambda_{IR}=m_g$. Fig.~\ref{fig:crss_mg} presents the cross sections evaluated at different $m_g$ on the same grid. These $m_g$ values are covered by the grid range (see Eq.~\eqref{eq:grid_cutoffs} and the associated discussion), and the cross sections agree well with the analytical eikonal expectation. Though not shown in the figure, we found that when the $m_g$ value is not between the numerical IR and UV cutoffs $[\lambda_{IR},\lambda_{UV}]$, the results would start to deviate from the expectations.

\begin{figure*}[ht]
  \centering
  \subfigure[\ The total cross section ]
  {\includegraphics[width=0.45\textwidth]{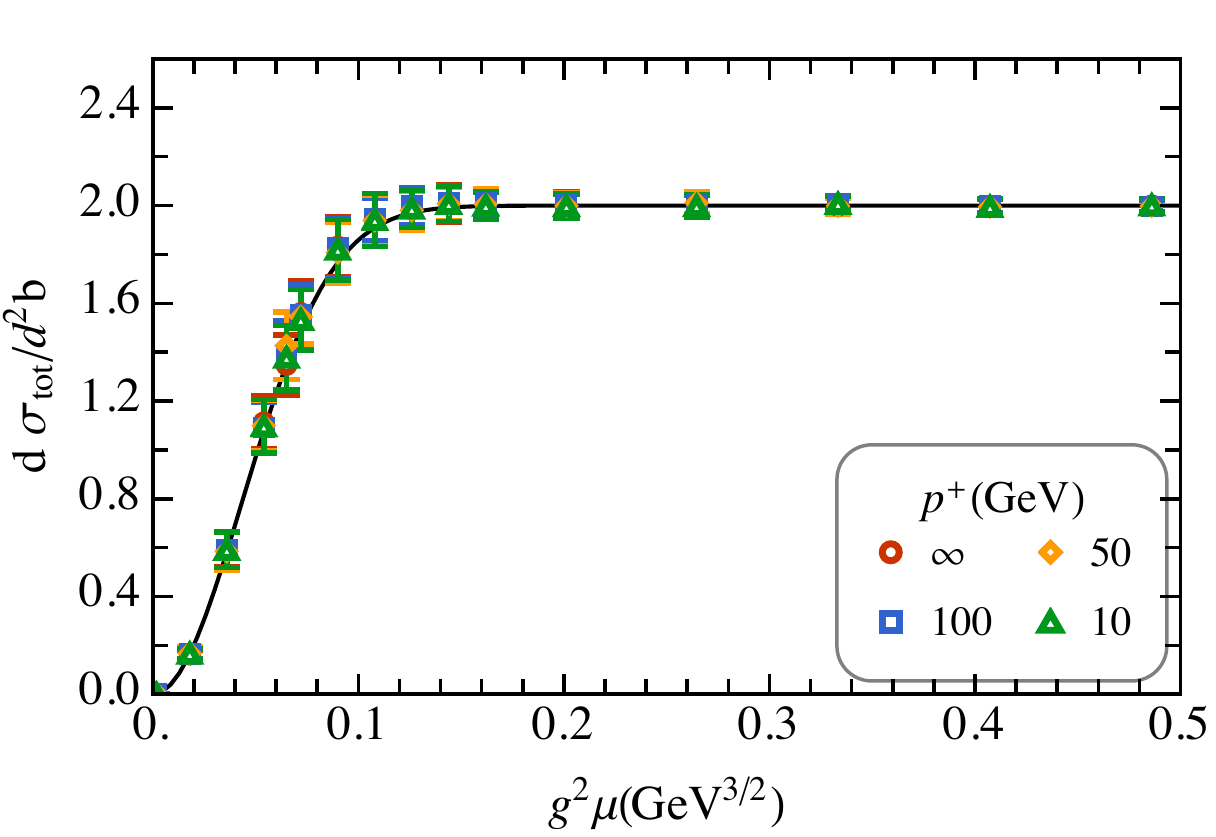}
  } 
  \subfigure[\ The elastic cross section]
  {\includegraphics[width=0.45\textwidth]{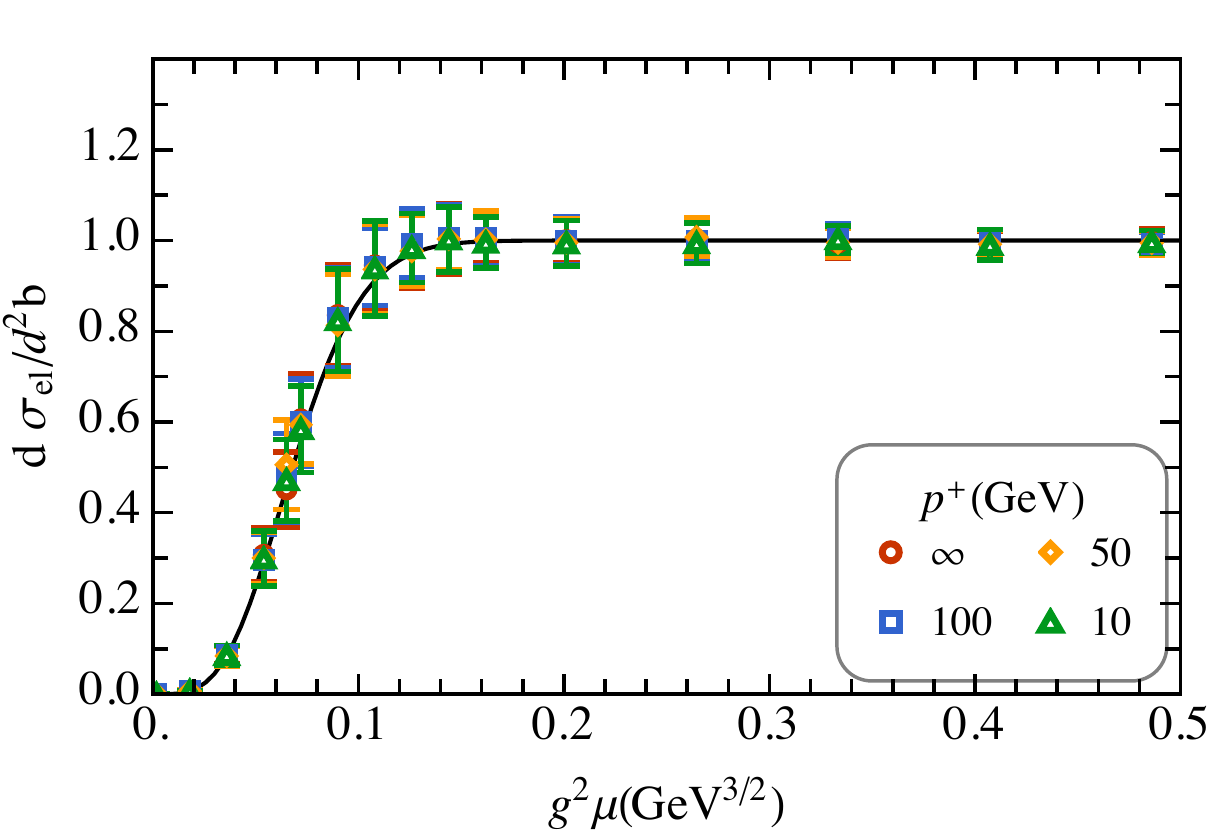}
  }
  \caption{The dependence on $p^+$ of (a) the total and (b) the elastic cross sections at $L=50 ~\GeV^{-1}$ and $N=18$. 
  The cross sections of the quark as functions of $g^2\mu$ for $L_\eta=50 ~\GeV^{-1}$ with $N_\eta=4$. 
  The solid lines are the eikonal predictions ($p^+=\infty$). 
  Each data point is averaged over 100 configurations, and the standard deviation is taken as the uncertainty bar. }
  \label{fig:crss_ppl}
\end{figure*}

We have seen that the cross sections in the eikonal limit agree with the analytical expectations. We now relax the condition so that we have finite $p^+$ and see if this could affect the cross section.
Figure~\ref{fig:crss_ppl} presents the cross sections at different $p^+$ values. 
It turns out that even for very small $p^+$, the cross section does not show noticeable differences from the $p^+=\infty$ case.

The tBLFQ results of the cross sections agree well with the analytical eikonal expectations in the eikonal limit. It also shows good numerical convergences on the various parameters. To study differences from the eikonal limit, we will investigate other observables that depend on additional kinematic variables in what follows.

\subsection{The differential cross sections}
The differential cross section $\diff \sigma/(\diff^2 b\diff^2p_\perp)$ is also of great interest. 
Convoluted with the quark distribution function of the proton at the factorization scale, the $p+A\to h+X$ cross section can be obtained from the $qA$ cross section.

\begin{figure*}[ht]
  \centering
  \hspace{-1mm}
  \includegraphics[width=0.24\textwidth]{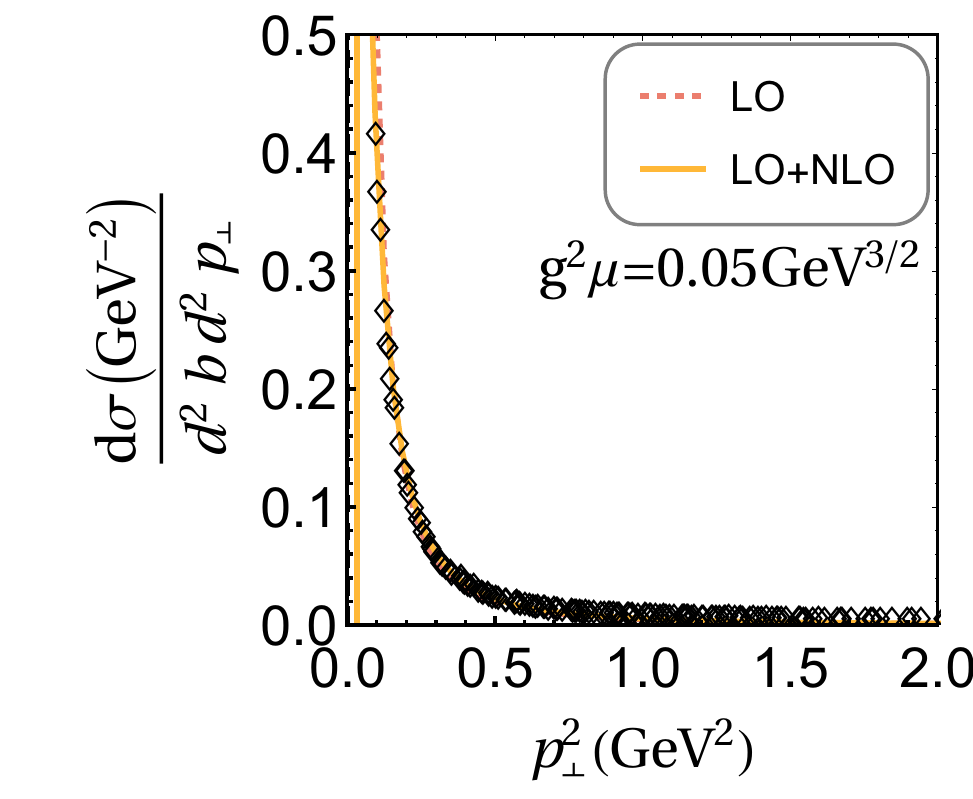}
\hspace{-1mm}
  \includegraphics[width=0.24\textwidth]{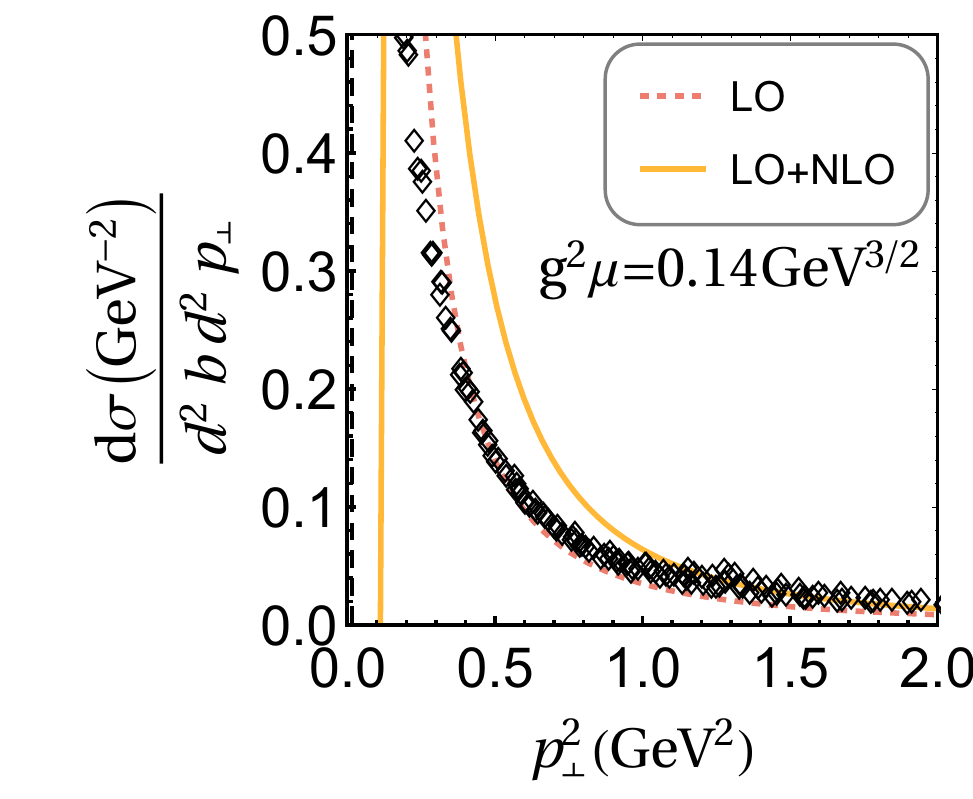}
\hspace{-1mm}
  \includegraphics[width=0.24\textwidth]{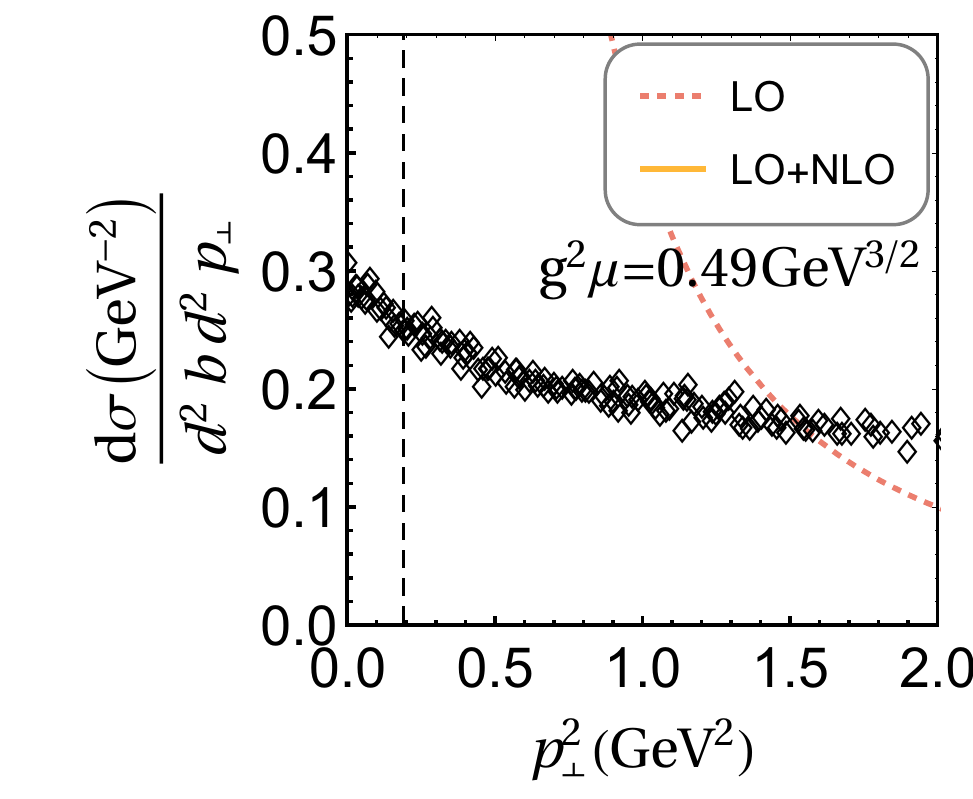}
\hspace{-1mm}
  \includegraphics[width=0.24\textwidth]{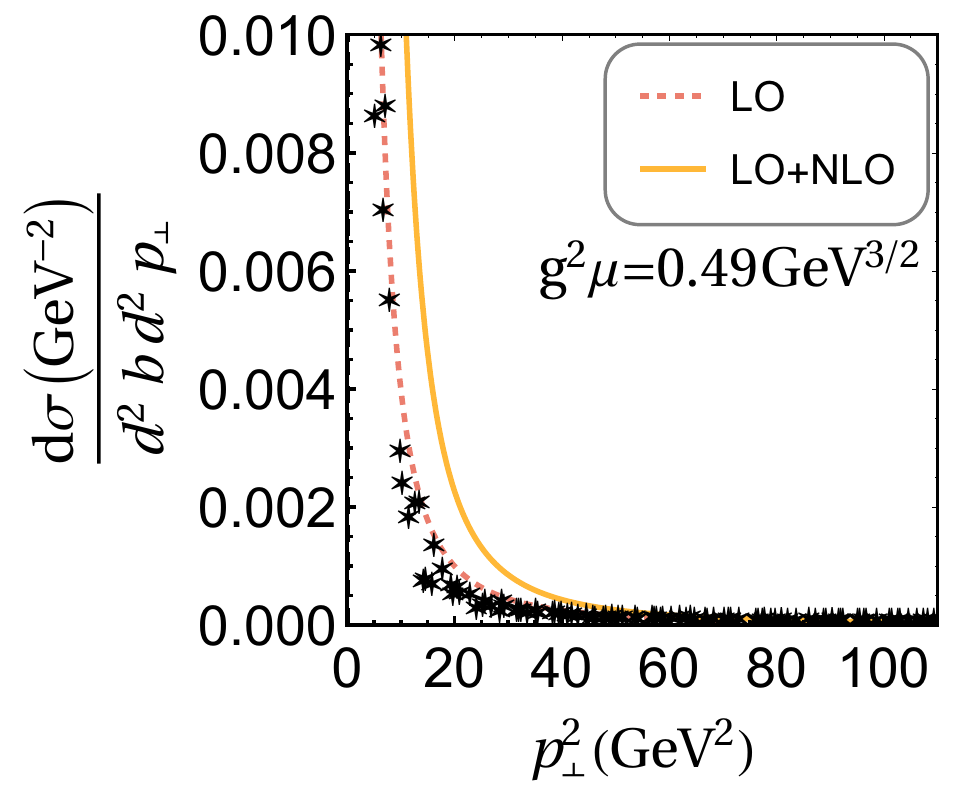}
\vspace{-2mm}

\subfigure[\ $g^2\mu= 0.05~\GeV^{3/2}$]
{\includegraphics[width=0.24\textwidth]{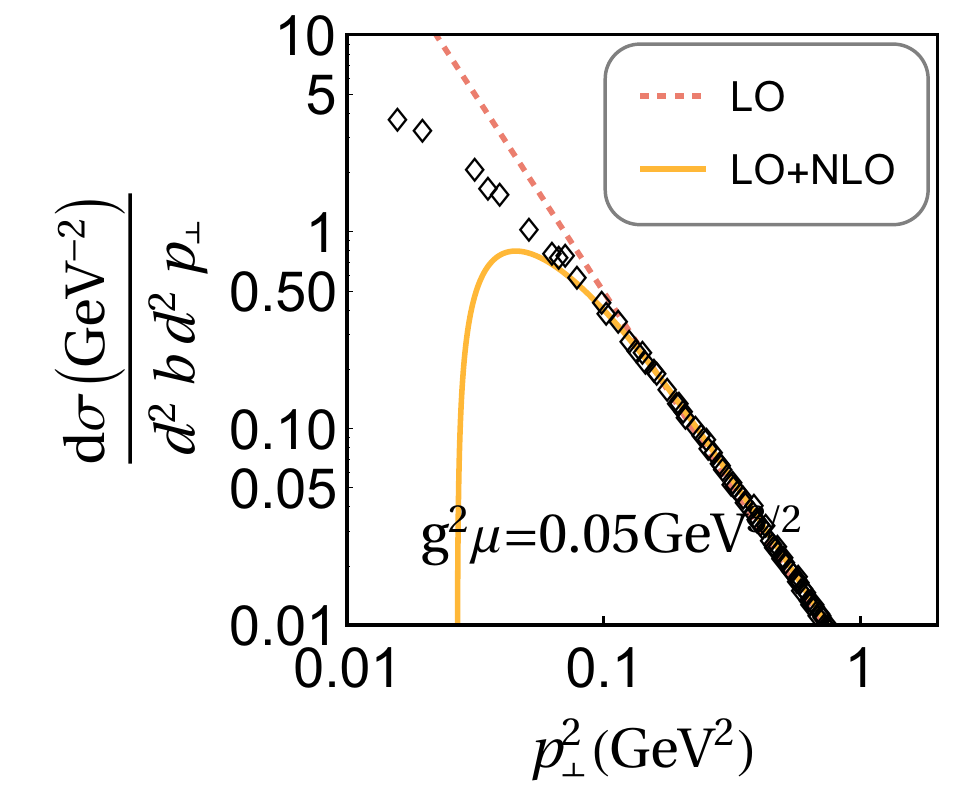}
\label{fig:dp1Log}
} 
\hspace{-2mm}
\subfigure[\ $g^2\mu= 0.14~\GeV^{3/2}$]
{\includegraphics[width=0.24\textwidth]{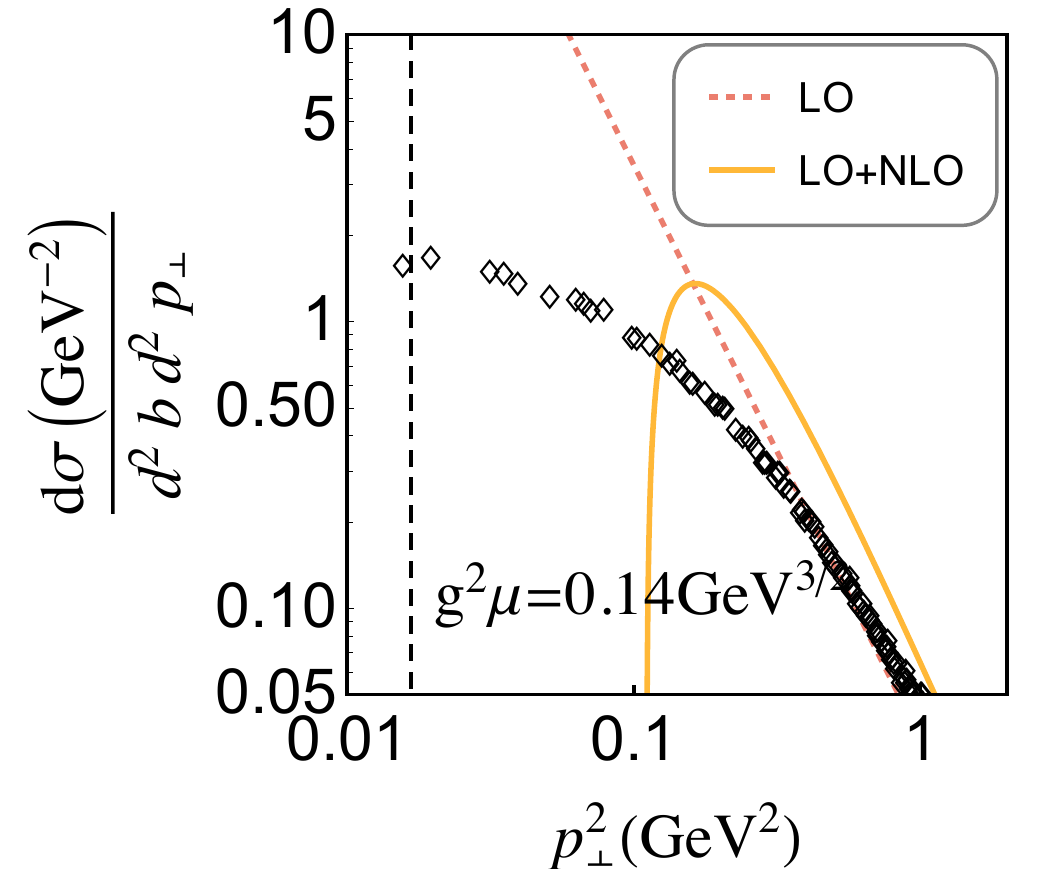}
\label{fig:dp2Log}
} 
\hspace{-2mm}
\subfigure[\ $g^2\mu= 0.49~\GeV^{3/2}$]
{\includegraphics[width=0.24\textwidth]{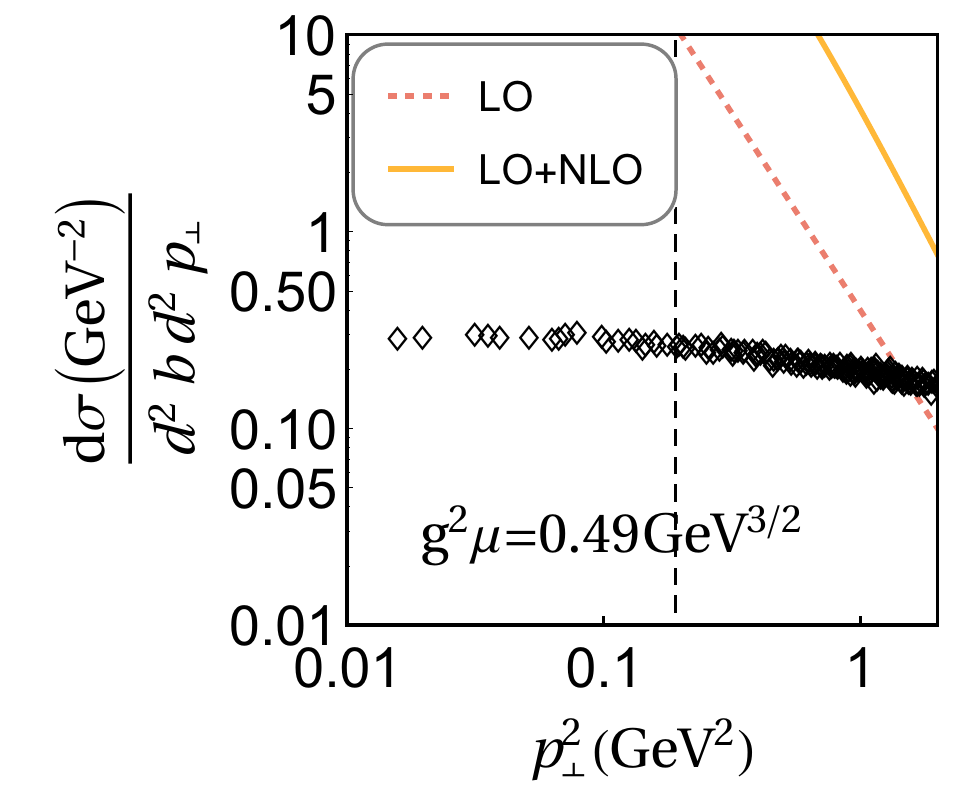}
\label{fig:dp3Log}
} 
\hspace{-2mm}
\subfigure[\ $g^2\mu= 0.49~\GeV^{3/2}$]
{\includegraphics[width=0.24\textwidth]{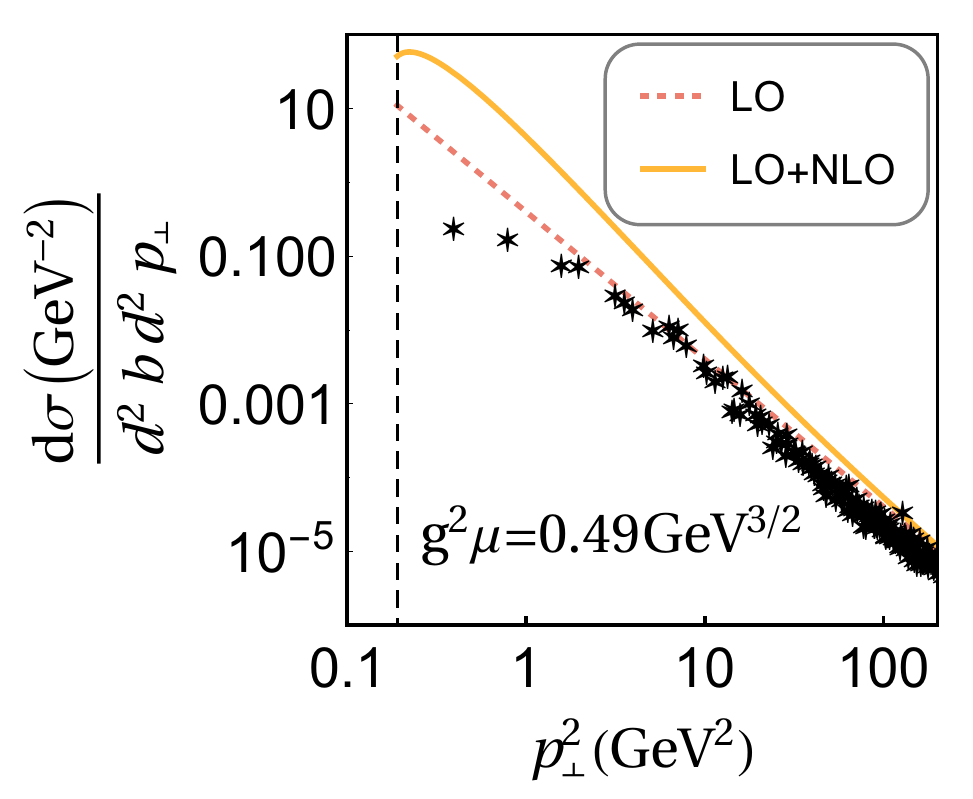}
\label{fig:dp3vLog}
} 
  \caption{The differential cross section of the $qA$ scattering at different $g^2\mu$, (a) $g^2\mu= 0.05~\GeV^{3/2}$, (b) $g^2\mu= 0.14~\GeV^{3/2}$, and (c,d) $g^2\mu=0.49~\GeV^{3/2}$.  
  The top panels are plotted on a linear scale, and the bottom panels are on a log-log scale. 
  The tBLFQ results are plotted as empty diamonds (or black stars), and each data point is averaged over 50 events.
  The vertical dashed line is at the saturation scale $Q_s^2=(g^2\mu)^2L_\eta/(2\pi^2)$. LO (NLO) is the leading (next-to leading) order perturbative approximation (see  Appendix~\ref{app:dcrss}).
  The tBLFQ results in panels (a-c) are calculated on the transverse lattice of  $L=50~\GeV^{-1}$, and that in panel (d) is evaluated on the lattice of $L=5~\GeV^{-1}$ to reveal the large $p_\perp^2$ range at the same $g^2\mu=0.49~\GeV^{3/2}$ as (c).  
  Other parameters for those tBLFQ results: $N=18$, $m_g=0.1 ~\GeV$, $L_\eta=50 ~\GeV^{-1}$ and $N_\eta=4$. 
  }
  \label{fig:dp}
\end{figure*}

In Fig.~\ref{fig:dp}, we present the tBLFQ calculations and compare with perturbative approximations. The differential cross section in the eikonal limit is given in Ref.~\cite{Dumitru:2002qt}, and we derive its perturbative approximations as power series expansions of $Q_s^2$ in Appendix~\ref{app:dcrss}.
In the large $p_\perp$ region, i.e. $p_\perp >> Q_s$, the tBLFQ results agree with the leading order (LO) and next-to-leading order (NLO) perturbative calculations, whereas at small $p_\perp$ region, the perturbation approximation deviates.

\begin{figure}[ht]
  \centering
  \subfigure[\ ]
  {\includegraphics[width=0.24\textwidth]{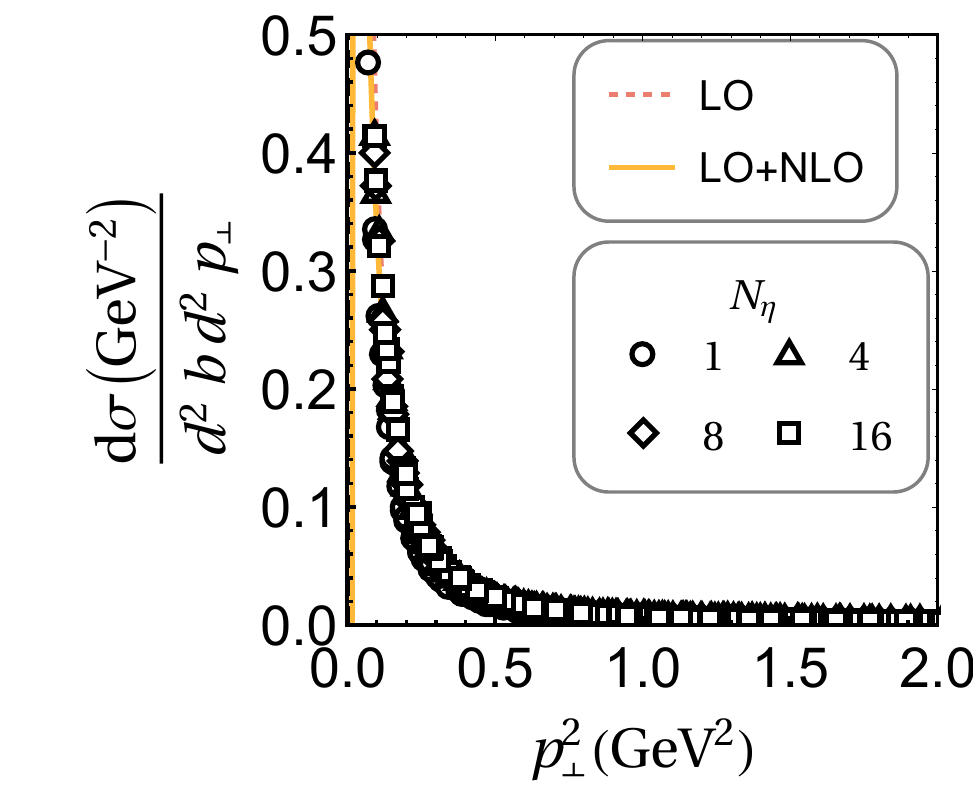}
  } 
  \hspace{-4mm}
  \subfigure[\ ]
  {\includegraphics[width=0.24\textwidth]{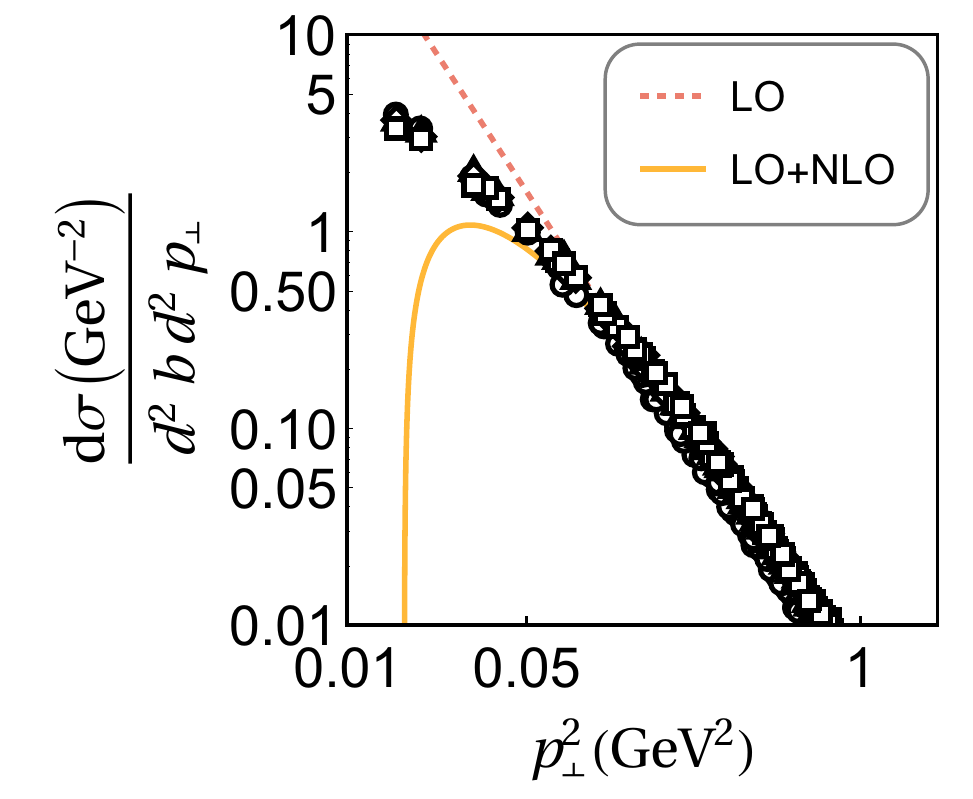}
  } 
  \caption{The differential cross section of the $qA$ scattering with different $N_\eta$ plotted on (a) linear scale and (b) log-log scale. 
  The tBLFQ results are plotted as empty diamonds, and each data point is averaged over 50 events.
  Parameters for those results: $N=18$, $L=50~\GeV^{-1}$, $m_g=0.1 ~\GeV$, $L_\eta=50 ~\GeV^{-1}$ and $g^2\mu= 0.05 ~\GeV^{3/2}$. 
  The vertical dashed line is at $Q_s^2=(g^2\mu)^2L_\eta/(2\pi^2)\approx 0.006~\GeV^2$. 
  LO (NLO) is the leading (next-to leading) order perturbative approximation (see  Appendix~\ref{app:dcrss}).}
  \label{fig:dp_Ny}
\end{figure}

We also check the dependence of the differential cross section on the grid parameters, $N$ and $L$. Like the total and the elastic cross sections, the dependence is not noticeable for grids covering the physical range. 
The result is also not sensitive to the longitudinal resolution, $N_\eta$, as in Fig.~\ref{fig:dp_Ny}. Unlike the cross sections, no ``oscillation'' pattern appears even at $N_\eta = 1$.

\subsection{The evolution of the quark state}
By carrying out the explicit time-evolution of the quark, we are able to access the intermediate information and investigate the
process of the quark-nucleus scattering. In particular, we study how the quark evolves in two aspects, 
the transverse coordinate
space and the color space.

\begin{figure*}[htp!]
  \centering
  \subfigure[\ Evolution of the quark's transverse coordinate distribution at $p^+=\infty$  \label{fig:rt2D_evl_pinf}]
  {\includegraphics[width=.95\textwidth]
  {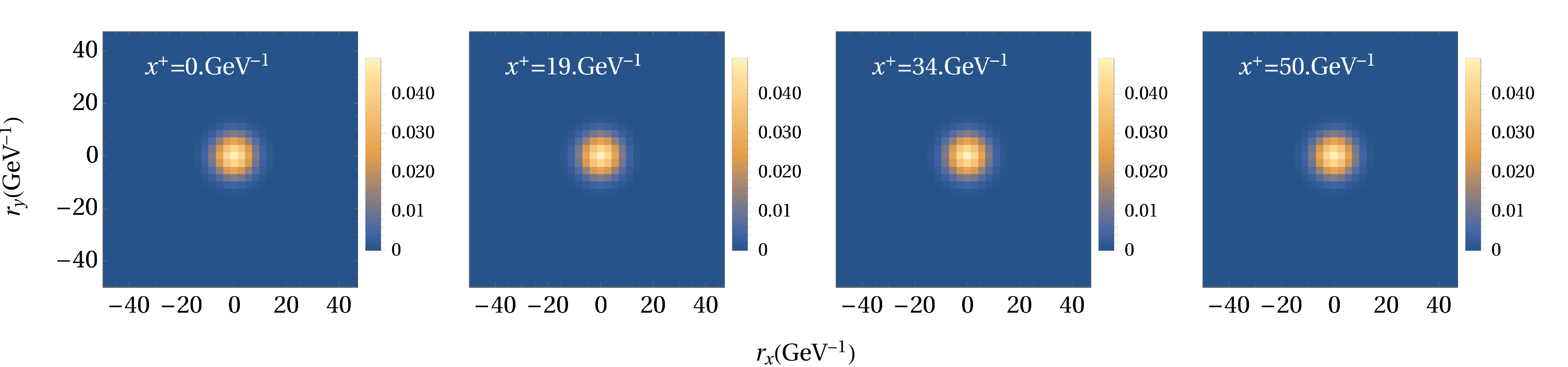}
  }
  \subfigure[\ Single event evolution of the quark's transverse coordinate distribution at $p^+=10~\GeV$  \label{fig:rt2D_evl_p10_1event}]
  {\includegraphics[width=.95\textwidth]{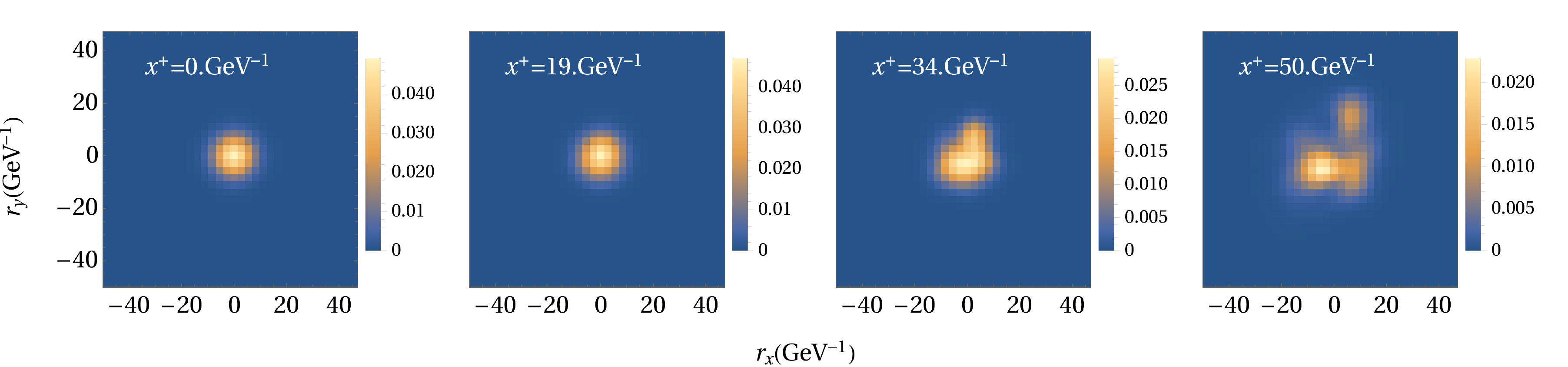}
  }
  \subfigure[\ Evolution of the quark's transverse coordinate distribution at $p^+=10~\GeV$, averaged over 50 events  \label{fig:rt2D_evl_p10_50event}]
  {\includegraphics[width=.95\textwidth]{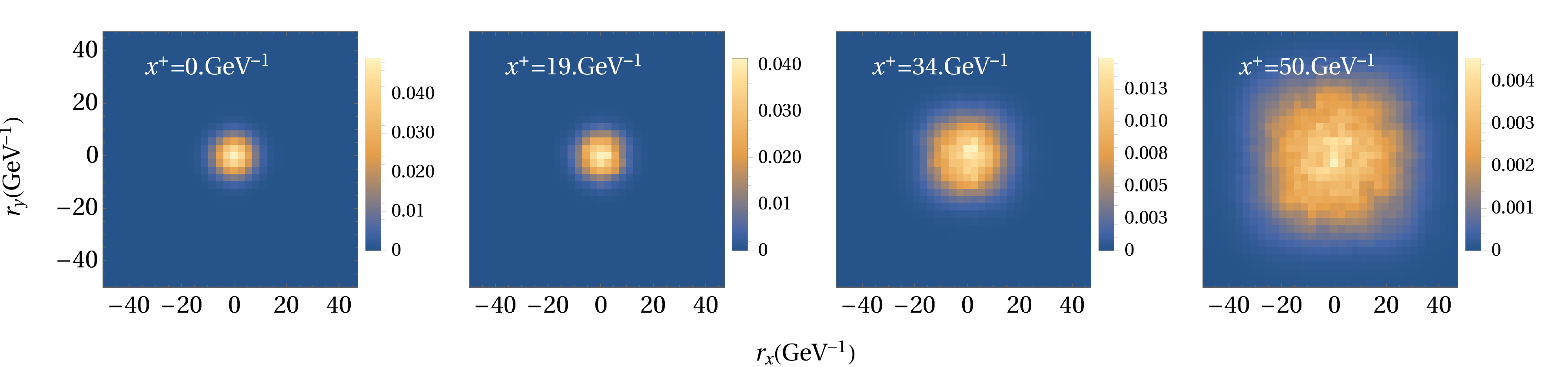}
  }
  \caption{The evolution of the quark's transverse coordinate distribution at different $p^+$, (a) $p^+=\infty$, (b,c) $p^+=10~\GeV$. 
  The initial state of the quark is distributed as $C e^{-|\vec r_\perp|^2/(0.2*50~\GeV^{-1})^2}$, where $C$ is the normalization coefficient. From left to right, the transverse coordinate distributions of the quark are shown at a sequential interaction time calculated by tBLFQ. 
  Parameters in those panels: $L_\eta=50~\GeV^{-1}$, $N_\eta=4$, $m_g=0.1~\GeV$, $N=18$, $L=50~\GeV^{-1}$, $g^2\mu=0.486~\GeV^{-3/2}$. 
  The result in (a) is identical for each event with the same parameters. The result in (b) is an single event and could be different for a different event with the same parameters. The result in (c) is an average of 50 events. }
  \label{fig:rt2D_evl}
\end{figure*}

To explore how the quark state evolves in the transverse coordinate space, we take the initial state of the quark to be a Gaussian packet $C e^{-|\vec r_\perp|^2/r_0^2}$, where $r_0 =0.2L=0.2* 50~\GeV^{-1}=1.97~\text{fm}$ and $C$ is the normalization coefficient. This chosen Gaussian packet has a rather small width, such that the quark is still relatively localized though not point-like.
Snapshots of the quark's transverse coordinate distribution at a sequence of light-front time are presented in Fig.~\ref{fig:rt2D_evl}. In the eikonal limit, i.e. $p^+ =\infty$,  the quark does not change its transverse location. 
But with finite values of $p^+$, the quark undergoes changes in its transverse coordinate distribution. In the plot of a single event as shown in Fig.~\ref{fig:rt2D_evl_p10_1event}, the quark dissipates with a random pattern, which is related to the randomly generated field. In the plot of averaged event as shown in Fig.~\ref{fig:rt2D_evl_p10_50event}, the quark spreads out more evenly, as expected by averaging the field configurations.

\begin{figure*}[t]
  \centering
  \includegraphics[width=.95\textwidth]{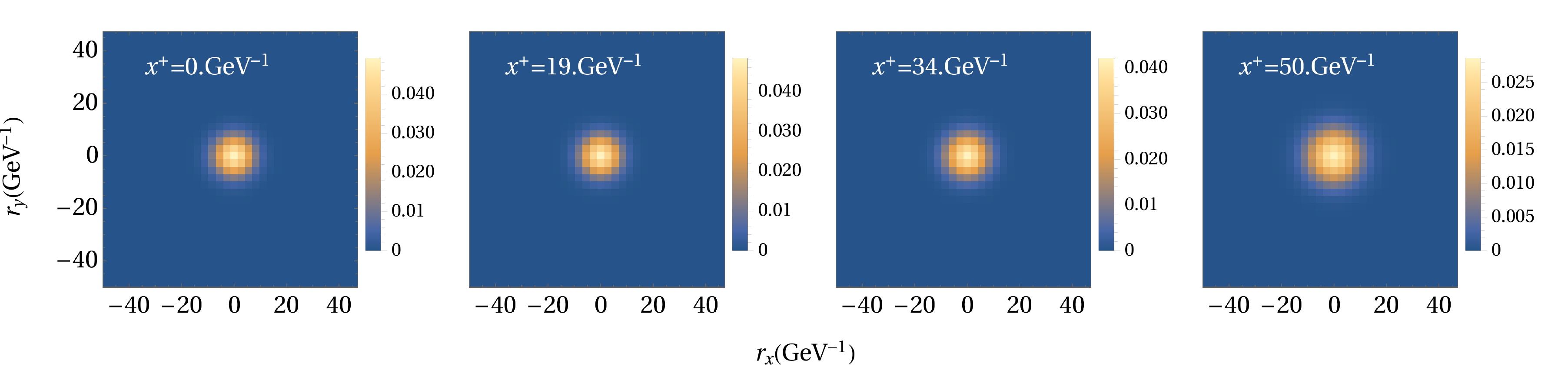}
  \caption{The evolution of the quark's transverse coordinate distribution when no source exists - that is, the light-front time evolution of the wave packet without interactions. The initial state of the quark is distributed as $C e^{-|\vec r_\perp|^2/(0.2*50~\GeV^{-1})^2}$, where $C$ is the normalization coefficient. From left to right, the transverse coordinate distributions of the quark are shown at a sequential interaction time calculated by tBLFQ. Parameters in those panels: $L_\eta=50~\GeV^{-1}$, $N_\eta=4$, $m_g=0.1~\GeV$, $N=18$, $L=50~\GeV^{-1}$, $p^+=10~\GeV$.}
  \label{fig:rt2D_evl_s0}
\end{figure*}

We know that even without an external field, the quark should dissipate in the coordinate space with a finite $p^+$. 
For comparison, Fig.~\ref{fig:rt2D_evl_s0} shows the evolution of the quark's transverse coordinate distribution when no external field exists. The quark spreads out slower with the expected simple isotropic pattern compared with cases where external field exists. 

To study the effect of the external field quantitatively, we calculate the expectation value of the quark's transverse coordinate $|\vec r_\perp|$
in cases with and without external fields.

\begin{figure*}[t]
  \centering
  \includegraphics[width=.95\textwidth]{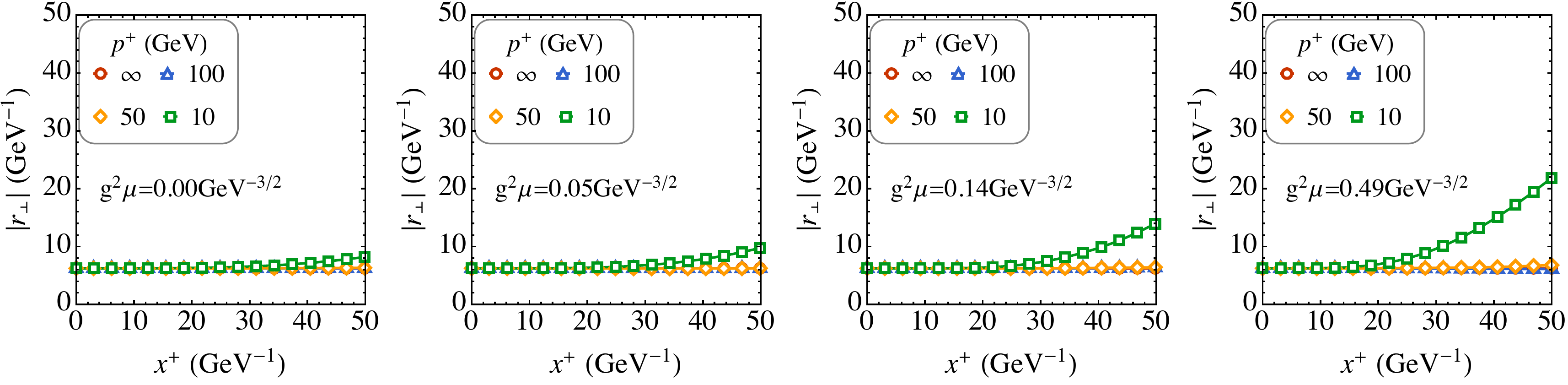}
  \caption{The evolution of the expectation value of the quark's transverse coordinate at different $p^+$ and at different color charge densities. The initial state of the quark is distributed as $C e^{-|\vec r_\perp|^2/(0.2*50~\GeV^{-1})^2}$, where $C$ is the normalization coefficient. From left to right, the first panel is calculated without an external field, the following three panels are calculated with increasing color charge density $g^2\mu$.
  The results are averaged over 10 events. Parameters in those panels:  $L_\eta=50~\GeV^{-1}$, $N_\eta=4$, $m_g=0.1~\GeV$, $N=18$, $L=50~\GeV^{-1}$.}
  \label{fig:rt_evl}
\end{figure*}

We first show how the energy scale of the quark, $p^+$, and the color charge density, $g^2\mu$, affect the evolution. Fig.~\ref{fig:rt_evl} presents the expectation value of the quark's transverse coordinate as a function of light-front time at various $p^+$ and for a range of color charge densities. It shows that the CGC field promotes the quark's dissipation in the transverse plane compared with the no-field case ($g^2\mu=0$). In concert with simple intuition, the quark spreads faster with smaller $p^+$ and larger color charge density.

\begin{figure*}[t]
  \centering
  \includegraphics[width=.95\textwidth]{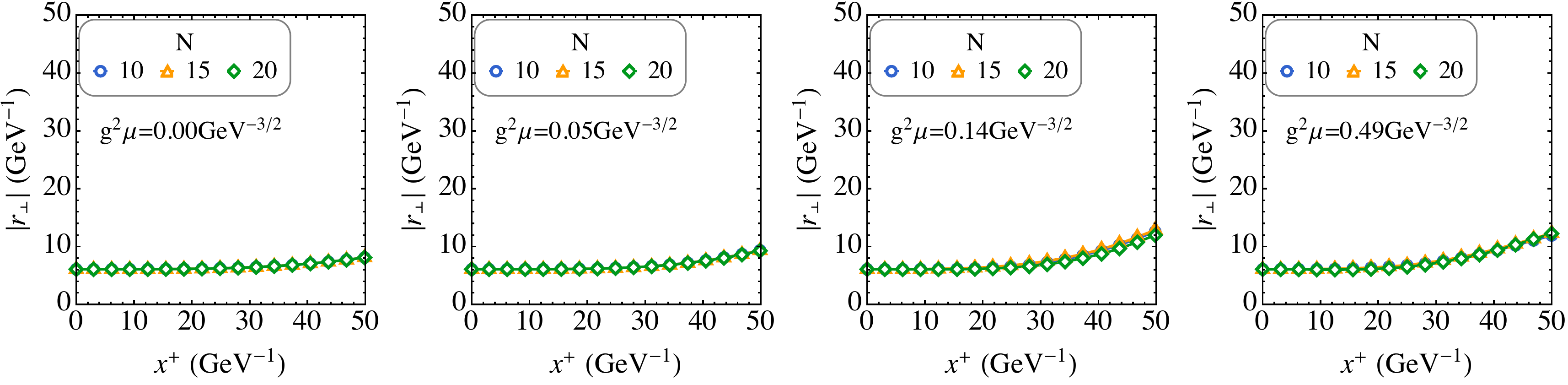}
  \caption{The evolution of the expectation value of the quark's transverse coordinate at different lattice sizes and at different color charge densities with a fixed lattice spacing of $a = L/N= 5~\GeV^{-1}$. Parameters in those panels:  $L_\eta=50~\GeV^{-1}$, $N_\eta=4$, $m_g=0.1
 ~\GeV$, $p^+=10~\GeV$. The initial state of the quark is distributed as $C e^{-|\vec r_\perp|^2/(0.2*50~\GeV^{-1})^2}$, where $C$ is the normalization coefficient.  From left to right, the first panel is calculated without an external field, the following three
  panels are calculated with increasing color charge density $g^2\mu$.
  The results are averaged over 10 events.}
  \label{fig:rt_evl_fix_a}
\end{figure*}

We also check the sensitivity of the quark's evolution to grid parameters at different color charge densities. 
Results at different lattice size $L$ and at different color charge densities with a fixed lattice spacing of $a=L/N=5~\GeV^{-1}=0.99~\text{fm}$ are compared in Fig.~\ref{fig:rt_evl_fix_a}. 
We find that the evolution is not very sensitive to the lattice size described by these parameters over the range of values shown.

\begin{figure*}[t]
  \centering
  \subfigure[\  $\Lambda_{UV}=\infty$  \label{fig:rt_evl_fix_L_noUV}]
  {
  \includegraphics[width=.95\textwidth]{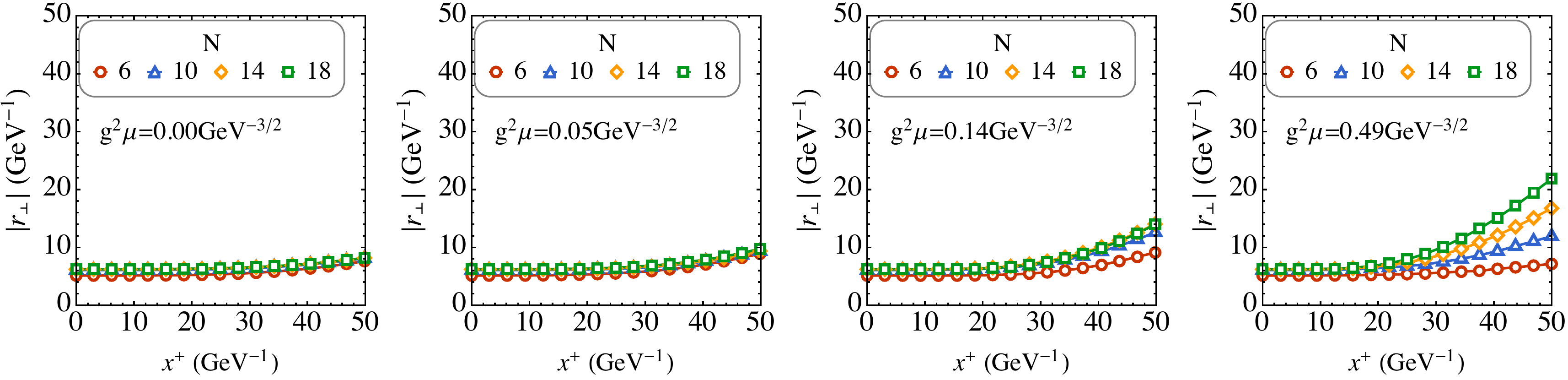}
  }
  \subfigure[\  $\Lambda_{UV}=0.2~\GeV$  \label{fig:rt_evl_fix_L_UVp2}]
  {
  \includegraphics[width=.95\textwidth]{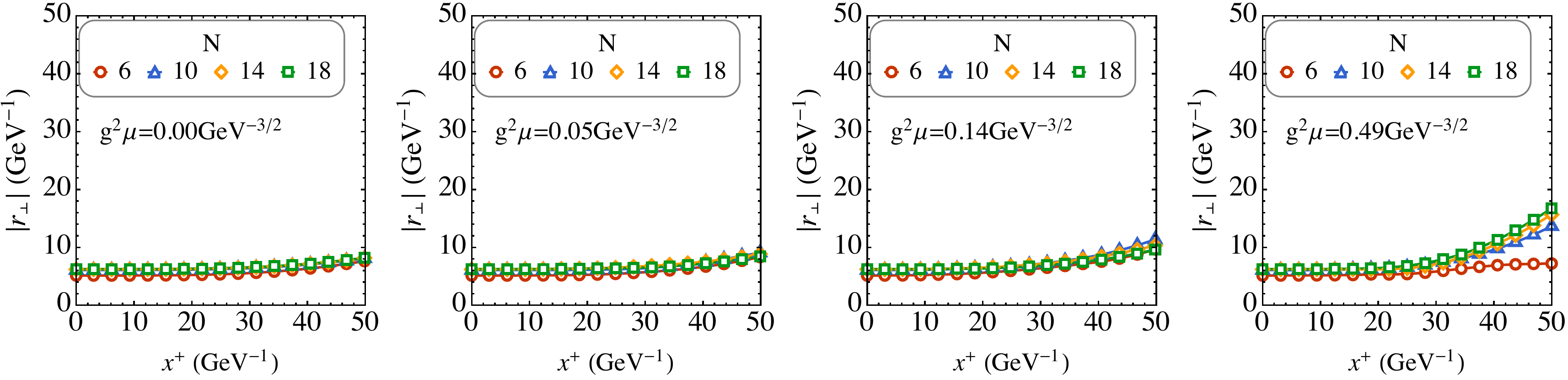}
  }
  \caption{The evolution of the expectation value of the quark's transverse coordinate at four different values of the grid number $N$ and at different color charge densities ($g^2 \mu$) with a fixed lattice size of $L= 50~\GeV^{-1}$.
  The initial state of the quark is $C e^{-|\vec r_\perp|^2/(0.2L)^2}$, where $C$ is the normalization coefficient.
  For both (a) and (b), from left to right, the first panel is calculated without an external field, the following three panels are calculated with increasing color charge density $g^2\mu$. In (b), we impose a UV cutoff when solving the gluon field so that $\tilde{A}(\vec k_\perp)=0$ for $|\vec k_\perp| \ge \Lambda_{UV} = 0.2~\GeV$.
  The results are averaged over 10 events. 
  Parameters in those panels:  $L_\eta=50~\GeV^{-1}$, $N_\eta=4$, $m_g=0.1\GeV$, $p^+=10~\GeV$. }
  \label{fig:rt_evl_fix_L}
\end{figure*}

We study the dependence on the grid number $N$ at a selection of external field strengths in Fig.~\ref{fig:rt_evl_fix_L}. When the external field is absent or weak, the evolution of $|r_\perp|$ agrees among these cases with different $N$. However, with a strong external field, the evolution of $|r_\perp|$ diverges, as seen in Fig.~\ref{fig:rt_evl_fix_L_noUV}. This divergence is expected from the ultravioletly divergent gluon field, as discussed in text associated with Eq.~\eqref{eq:Green}. We verify this source of divergence by imposing an explicit UV cutoff on the gluon field. 
The results become better converged, that is independent of $N$ at the stronger field strength, with the imposed UV cutoff, as presented in Fig.~\ref{fig:rt_evl_fix_L_UVp2}.

\begin{figure*}[htp!]
  \centering
  \includegraphics[width=.95\textwidth]{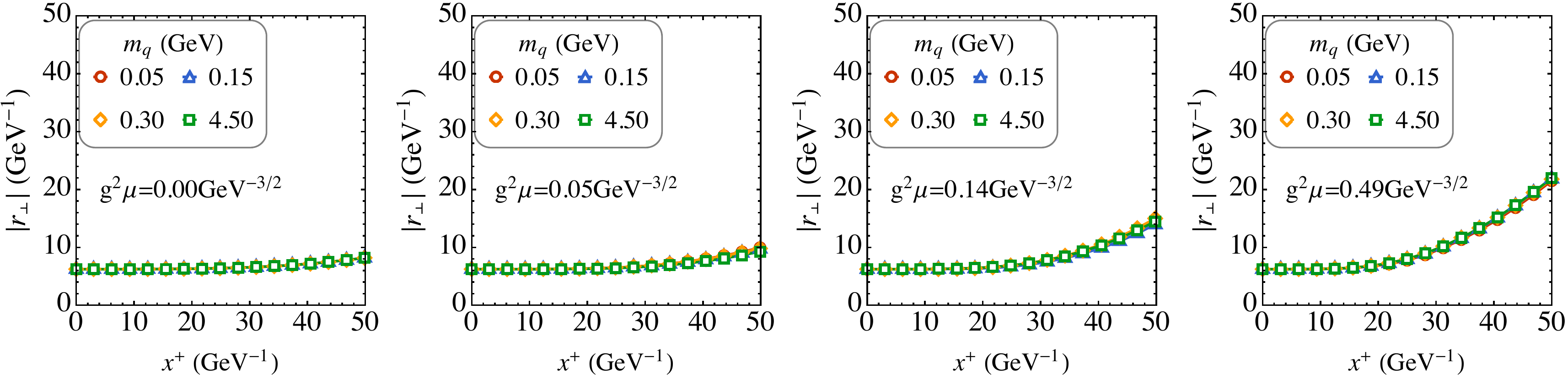}
  \caption{The evolution of the expectation value of the quark's transverse coordinate with different quark masses and at different color charge densities. 
  Parameters in those panels: $L=50~\GeV^{-1}$, $N=18$, $L_\eta=50~\GeV^{-1}$, $N_\eta=4$, $m_g=0.1 ~\GeV$, $p^+=10~\GeV$. 
  The initial state of the quark is distributed as $C e^{-|\vec r_\perp|^2/(0.2*50~\GeV^{-1})^2}$, where $C$ is the normalization coefficient. 
  From left to right, the first panel is calculated without an external field, the following three panels are calculated with increasing color charge density $g^2\mu$.
  The results are averaged over 10 events.}
  \label{fig:rt_evl_mq}
\end{figure*}

The quark admits changes in the transverse coordinate at finite $p^+$, and this is achieved through the phase factor $e^{\pm i\frac{1}{2}p^- x^+}$ with $p^-=(\vec p_\perp^2+m_q^2)/p^+$. 
One might then expect that using different values of the quark mass could influence this effect. However, since $p^+$ does not change through the interaction, $\pm m_q^2/p^+$ is the same for all basis states and therefore, $e^{i\frac{1}{2}m_q^2/p^+ x^+}$ and $e^{-i\frac{1}{2}m_q^2/p^+ x^+}$ cancel out in the phase factors, leaving $V_I(x^+)=e^{i\frac{1}{2}P^-_{QCD}x^+}V(x^+)e^{-i\frac{1}{2}P^-_{QCD}x^+}=e^{i\frac{1}{2}\vec p_\perp^2 x^+/p^+}V(x^+)e^{-i\frac{1}{2}\vec p_\perp^2x^+/p^+}$ (see Eq.~\eqref{eq:ShrodingerEq} and associated text).
We show that the role of the quark mass is indeed minimal at a selection of color charge densities by using $m_q=0.05,0.15,0.3$ and $4.5~\GeV$ in Fig.~\ref{fig:rt_evl_mq}. 

\begin{figure*}[htp!]
  \centering
  \subfigure[\ Evolution of the quark's color distribution at $p^+=\infty $ \label{fig:color_evl_pinf}]
  {
    \includegraphics[width=.95\textwidth]{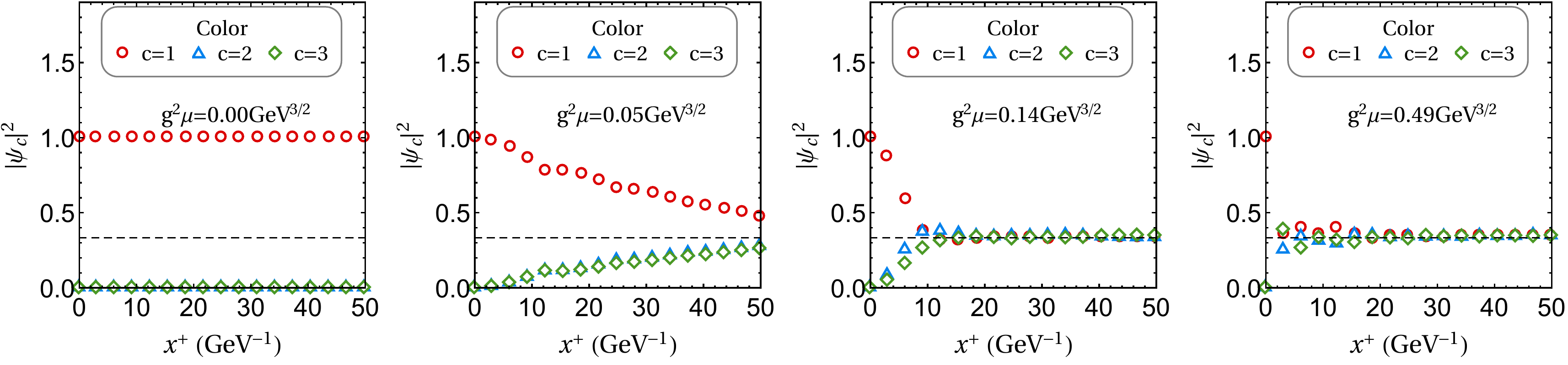}
  }
  \subfigure[\ Evolution of the quark's color distribution at $p^+=10~\GeV$ \label{fig:color_evl_p10}]
  {
    \includegraphics[width=.95\textwidth]{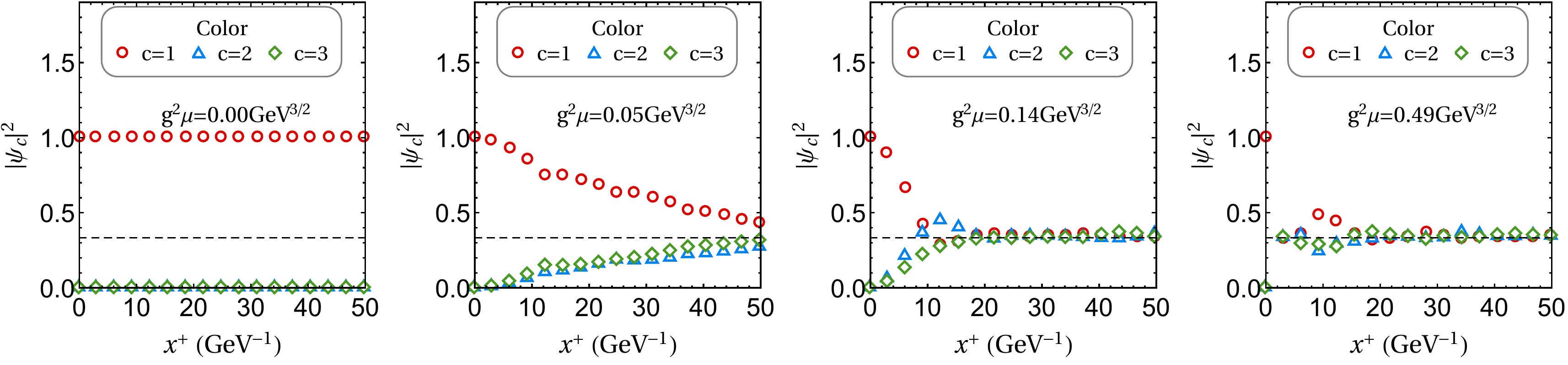}
  }

  \caption{The evolution of the quark's distribution in the color space at a selection of color charge densities with different $p^+$, (a) $p^+=\infty$ and (b) $p^+=10~\GeV$.
  The results are averaged over 50 events. 
  Parameters in those panels: $N=18$, $L=50~\GeV^{-1}$, $L_\eta=50~\GeV^{-1}$, $N_\eta=4$, $m_g=0.1~\GeV$. 
  For both (a) and (b), from left to right, the first panel is calculated without an external field, the following three panels are calculated with increasing color charge density $g^2\mu$. 
 The initial state of the quark is a single color state ($c=1$) with space distribution as $C e^{-|\vec r_\perp|^2/(0.2L)^2}$, where $C$ is the normalization coefficient. 
 The dashed line marks the average probability of the three colors: $0.33$.}
  \label{fig:colorEvl}
\end{figure*}

In the color space, the quark evolves toward a uniformly distributed state, $|\psi_c|^2\to1/3, (c=1,2,3)$. This is shown in Fig.~\ref{fig:colorEvl}. 
The quark evolves faster in the color space with larger color charge density $g^2 \mu$ as may be expected but does not show significant dependence on $p^+$ when comparing the $p^+ = \infty$ and $p^+ = 10~\GeV$ results.

\subsection{Profiled CGC field}
The CGC field we adopt so far is uniform in the transverse plane. In reality, the field generated from a large nucleus should be
stronger at the center than on the edges. We take this into consideration by introducing a Gaussian profile and a Woods-Saxon profile to scale the CGC field in the
transverse coordinate space.

In the Gaussian form, the scale factor reads
\begin{align}
  &f_{\text{Gaussian}}(\vec r_\perp) = e^{-(r_\perp/R_0)^2}
  \;,
\end{align}
where $R_0$ is taken as the nuclear radius. For the gold nucleus, $R_0 = 37 ~\GeV^{-1}$.

In the Woods-Saxon form, the scale factor reads
\begin{align}
  &f_{\text{Woods-Saxon}}(\vec r_\perp) = \frac{1}{1+
  e^{(r_\perp-R_0)/s}}
  \;.
\end{align}
We use the usual parametrization, where $R_0$ is taken as the nuclear radius and $s = 3.2 ~\GeV^{-1}$ is the surface diffuseness~\cite{Suhonen:2007zza}.

\begin{figure*}[htp!]
  \subfigure[\ Uniform profile\label{fig:source}]
  {\includegraphics[width=0.32\textwidth]{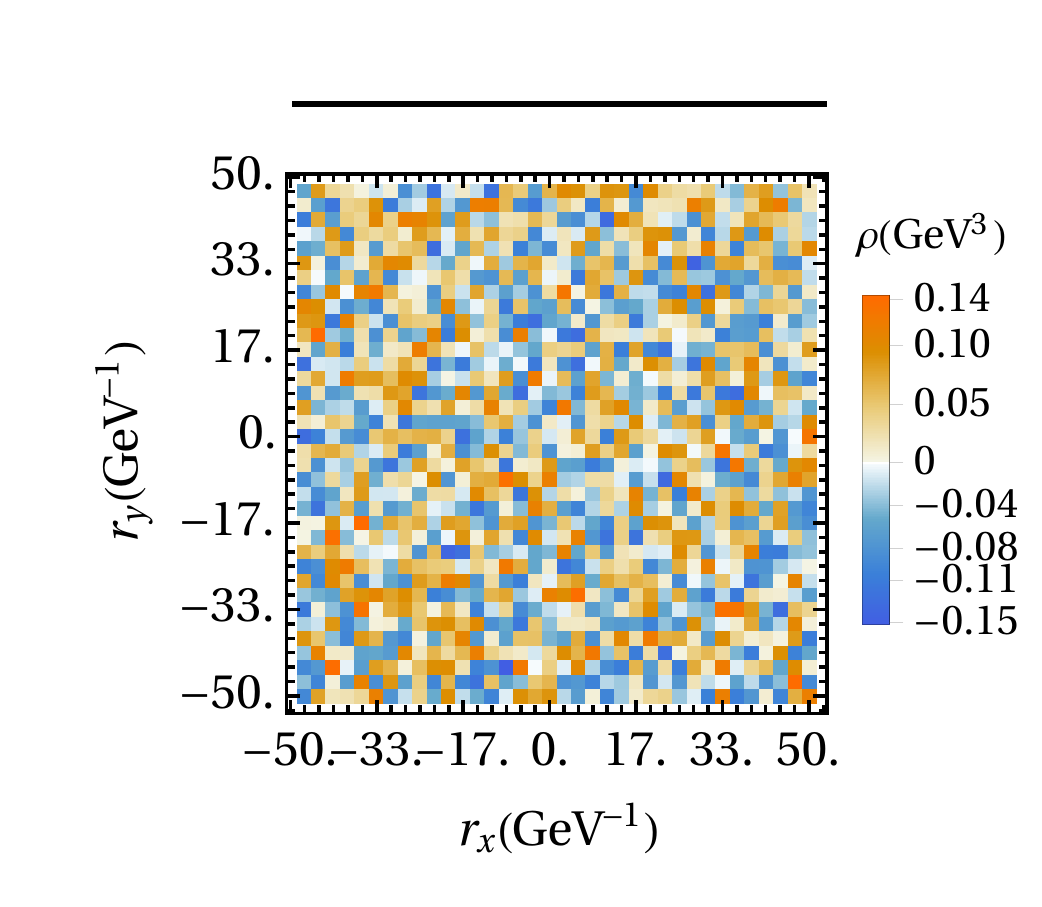}
  }
  \subfigure[\  Gaussion profile \label{fig:source_Gauss}]
  {\includegraphics[width=0.32\textwidth]{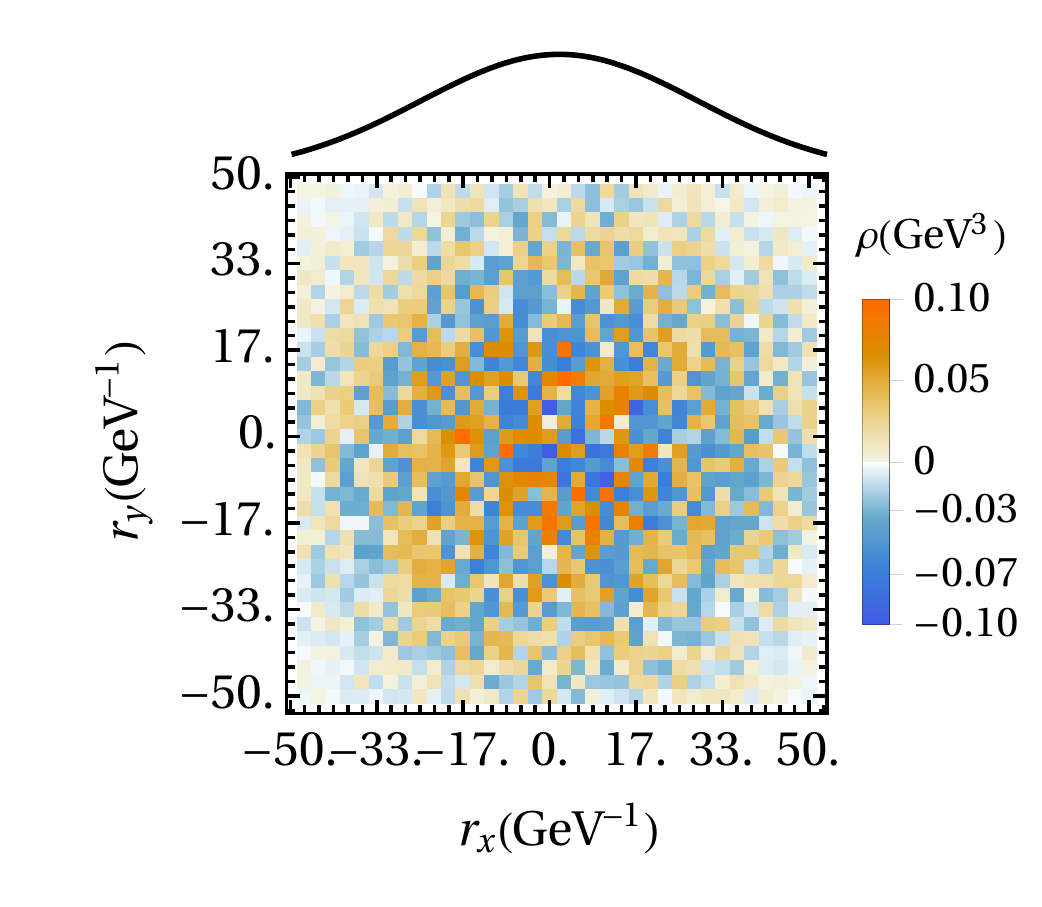}
  }
  \subfigure[\  Woods-Saxon profile \label{fig:source_WS}]
  {\includegraphics[width=0.32\textwidth]{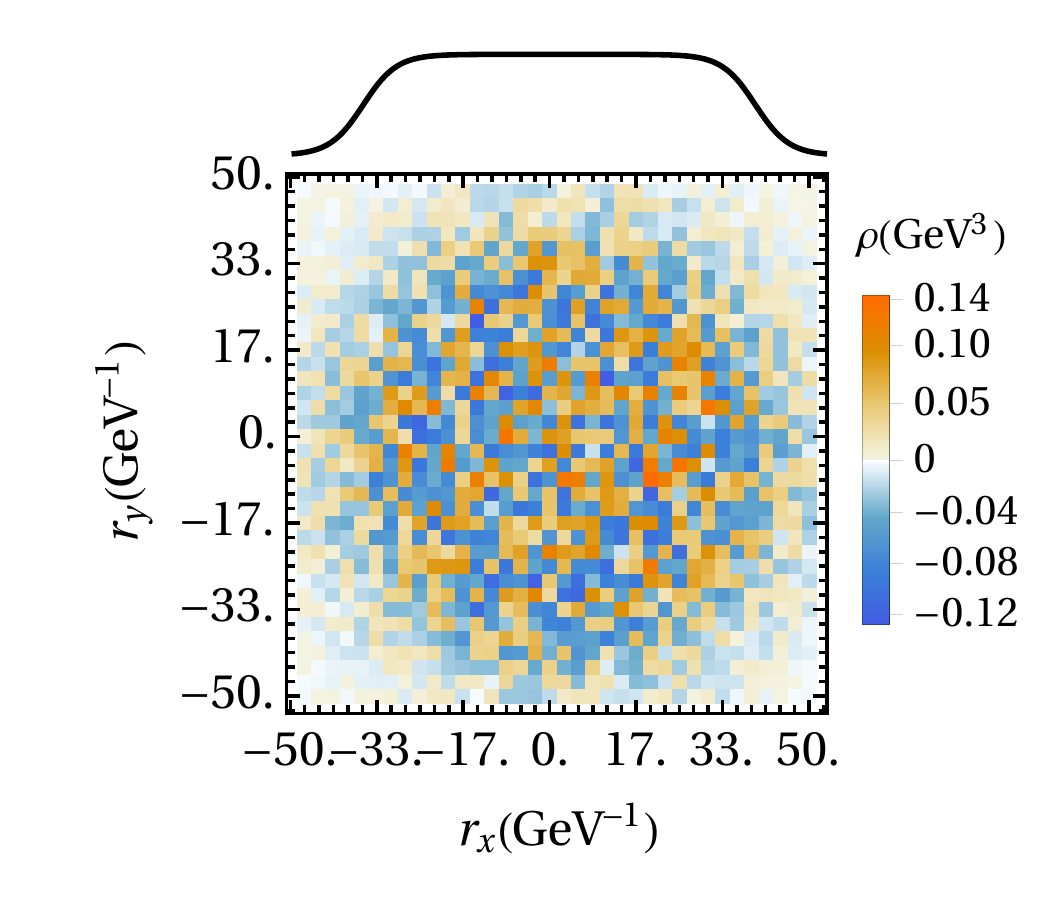}
  }

  \subfigure[\ The total cross section \label{fig:profile_tot}]
  {\includegraphics[width=0.4\textwidth]{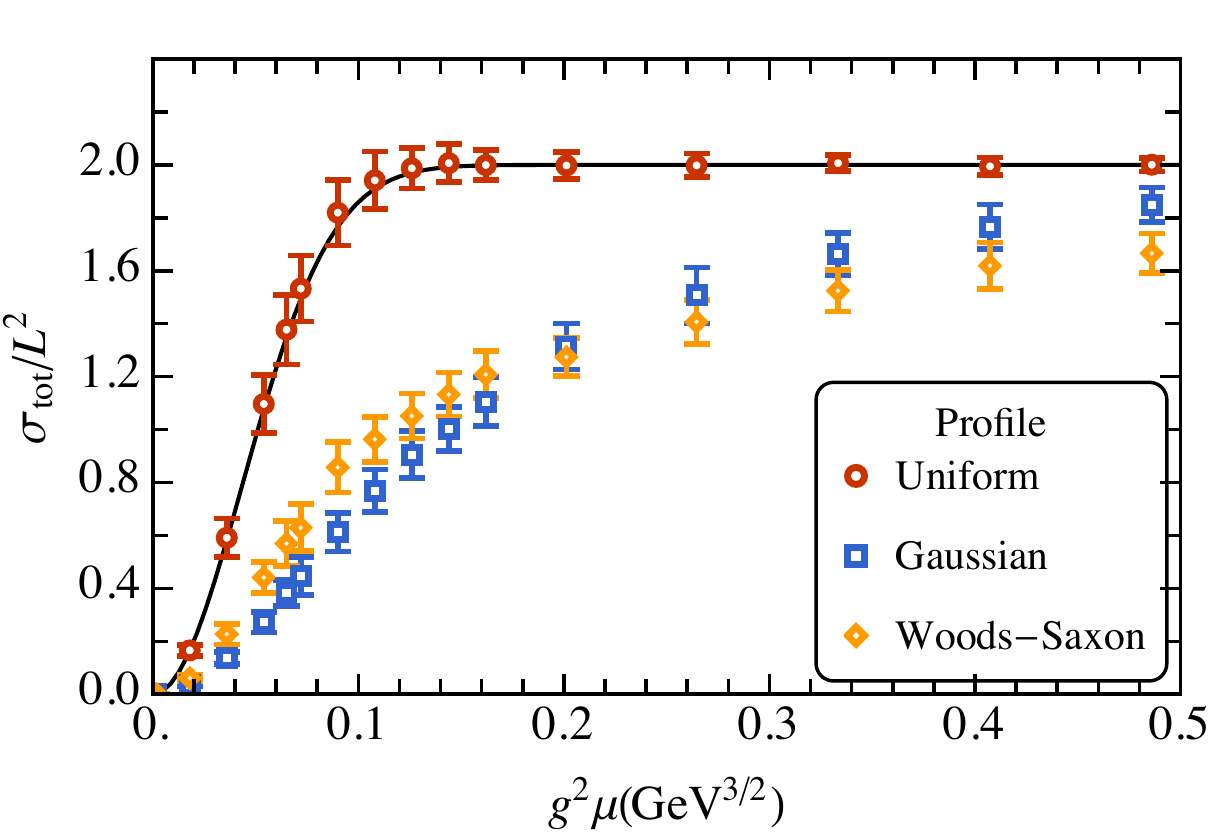}
  }
 \subfigure[\ The differential cross section \label{fig:profile_dcrss}]
 {\includegraphics[width=0.33\textwidth]{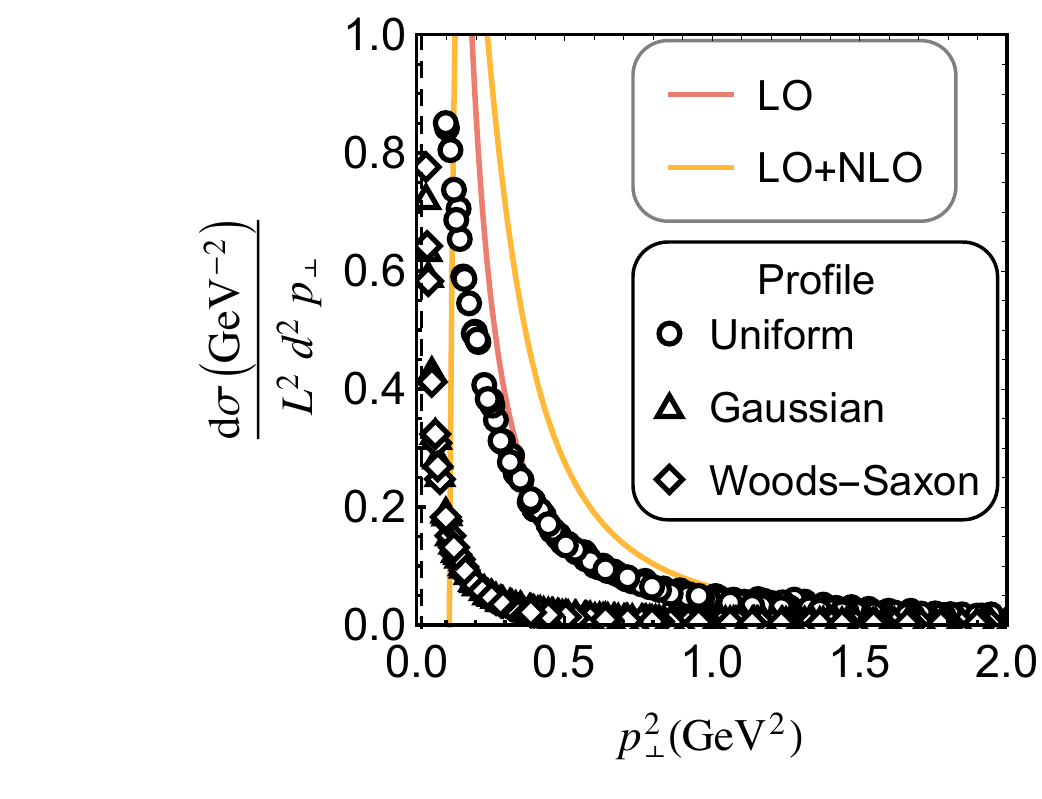}
 }

 \subfigure[\ Evolution of $|r_\perp|$ \label{fig:profile_r_evol}]
 {\includegraphics[width=0.95\textwidth]{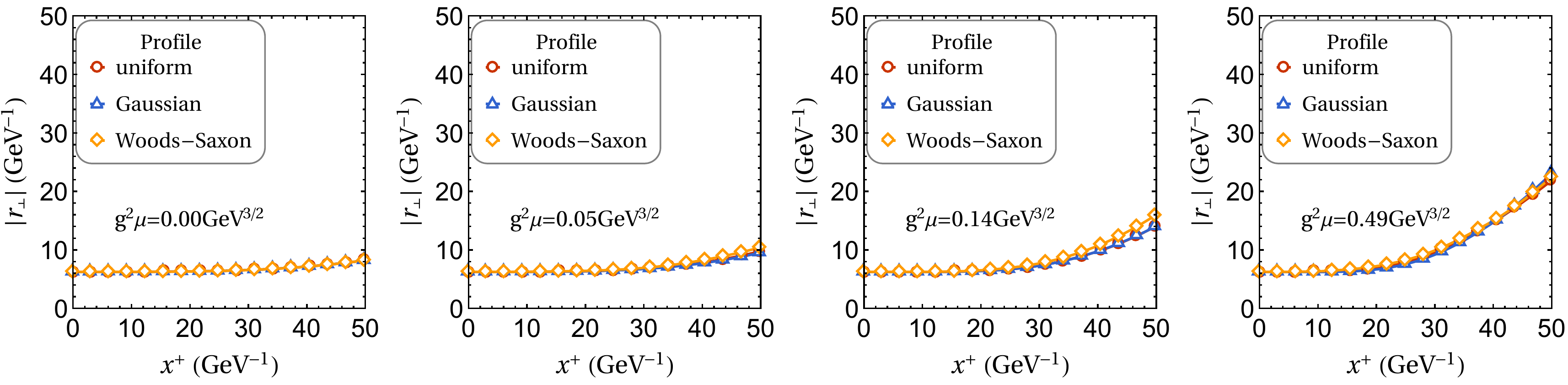}
 }
  \caption{The transverse profiles of the CGC field and related quantities. (a-c): single-event source charges scaled by profile functions plotted on the transverse plane $\vec r_\perp$. The shapes of the adopted profiles are sketched above each panel for reference. 
  (d): the total cross sections using the three different profiles. The initial state of the quark is set as $\vec p_\perp = \vec 0_\perp$. The solid line is the eikonal prediction with the uniform profile according to Eq.~\eqref{eq:crss_eikonal}. (e): the differential cross sections using the three different profiles. The solid lines are the perturbative approximations with the uniform profile, see the caption of Fig.~\ref{fig:dp}. For this result, we choose the initial state of the quark as $\vec p_\perp = \vec 0_\perp$ and the color charge density of value $g^2\mu=0.14~\GeV^{3/2}$. 
  (f): the evolution of the expectation value of the quark's transverse coordinate with different profiles and at a selection of color charge densities. The initial state of the quark is $C e^{-|\vec r_\perp|^2/(0.2L)^2}$, where $C$ is the normalization coefficient. The results are averaged over 100 events. 
  Parameters for those panels: $N=18$, $L=50~\GeV^{-1}$, $L_\eta=50~\GeV^{-1}$, $N_\eta=4$, $m_g=0.1~\GeV$, $p^+=10~\GeV$. }
  \label{fig:profiles}
\end{figure*}

We revisit the quark's evolution, the cross section and the $p_\perp$-dependent differential cross section in Fig.~\ref{fig:profiles}. 

To study the total cross sections with different profiles, we choose the initial state of the quark as $\vec p_\perp =\vec 0_\perp$, such that the quark is distributed on the entire transverse space. We find that the total cross sections are smaller with the Gaussian and Woods-Saxon profiles where the background fields are also smaller overall, as shown in Fig.~\ref{fig:profile_tot}. Note that we use $\sigma_{\text{tot}}/L^2$ as the total cross section, which is equivalent to the average of $\diff \sigma_{\text{tot}}/\diff b$ over the entire transverse space with area $L^2$. For the uniform profile, $\sigma_{\text{tot}}/L^2=\diff \sigma_{\text{tot}}/\diff b$, since the color charge density $g^2\mu$ is a constant on the transverse plane. But for both the Gaussian and the Woods-Saxon profiles, the cross section as a function of the impact parameter $b$ is not constant. 
The tBLFQ result is computed in the transverse momentum space which automatically sums over the contributions in the entire transverse coordinate space, and gives $\sigma_{\text{tot}}/L^2$.

The differential cross section is more peaked around $p_\perp=0$ when the fields are scaled by these two profiles, as shown in Fig.~\ref{fig:profile_dcrss}. Considering that the initial state is $\vec p_\perp =\vec 0_\perp$, the profiled fields which are reduced in strength, make less change to the quark state compared to the uniform field. 

However, the evolution of $|r_\perp|$ does not seem to be sensitive to the profiles, as shown in Fig.~\ref{fig:profile_r_evol}. This is likely for the reason that during the evolution process the quark state is still constrained in the central area where all three profiles have similar strengths of the field.

\section{Conclusions and outlook}\label{sec:summary}

In this work, we applied the tBLFQ formalism to a QCD problem for the first time, the quark-nucleus scattering. We are able to access the wavefunction of the quark at any intermediate time during the evolution. This provides us with an opportunity to carry out detailed studies of the time-dependent process.

Our results of the total and differential cross sections are in good agreement with the analytical expectations under the eikonal condition $p^+=\infty$. In the sub-eikonal case with a finite $p^+$, the cross sections do not show noticeable deviation from the eikonal limit. However, there are clear sub-eikonal effects shown from the distribution of the quark's transverse coordinate. At finite $p^+$, the quark admits changes in its transverse coordinate distribution. We aim to study the implication on transverse momentum distribution in particle production in $p$A collision in the future. 

We used the single quark sector to carry out the calculations as an initial investigation. In general, the single dressed quark state expands in the Fock space,

\begin{align}
  \ket{q}_{\text{dressed}} = a \ket{q}
  + b \ket{qg} + c \ket{qgg} + d \ket{qq\bar q} + \cdots
    \;.
\end{align}
This includes the bare quark as well as its dressed states with gluons and sea quarks. With the parameter sensitivities established in the present work, we are enabled to extend the Fock space to $\ket{q} + \ket{qg}$ and study gluon emission and absorption during the collision process. The QCD Lagrangian will then be restored by including the dynamical gluon in Eq.~\eqref{eq:Lag}. 

In this work, we take the MV model as the background field of the nucleus, and keep the dominant field component ($\mathcal{A}^-$) in our calculation. In future works, we also hope to include the transverse component of the color field ($\mathcal{A}_\perp$) and further investigate its role in the evolution process, for example, the effect on the spin of the quark. 

We foresee more applications of the tBLFQ approach to scattering processes in the near future. The work on quark-nucleus scattering formulated in this paper could be extended to the $q\bar{q}$-nucleus and $gg$-nucleus scatterings, which are the basic ingredients of any cross section at high energies. We also look forward to investigating the particle production and evolution in the glasma field created by heavy-ion collisions. 

\section*{Acknowledgments}
We wish to thank Shaoyang Jia, Wenyang Qian, Shuo Tang, Anji Yu for valuable discussions. M. Li acknowledges the communication with A. Dumitru and discussions with M. Sievert, G. Beuf and T. Lappi.
M. Li would also like to thank the hospitality of the QCD Theory Group of University of Jyväskylä during the progress of this work.
G. Chen acknowledges discussion with R.J. Fries on this project.
X. Zhao is supported by Key Research Program of Frontier Sciences, CAS, Grant No ZDBS-LY-7020. 
This work was supported in part by the US Department of Energy (DOE) under Grant Nos. DE-FG02-87ER40371, DE-SC0018223 (SciDAC-4/NUCLEI), DE-SC0015376 (DOE Topical Collaboration in Nuclear Theory for Double-Beta Decay and Fundamental Symmetries). This research used resources of the National Energy Research Scientific Computing Center (NERSC), a U.S. Department of Energy Office of Science User Facility operated under Contract No. DE-AC02-05CH11231.
This work has been supported in part by the European Research Council (ERC) under the European Union’s Horizon 2020 research and innovation programme (grant agreement No ERC-2015-CoG-681707). The content of this article does not reflect the official opinion of the European Union and responsibility for the information and views expressed therein lies entirely with the authors.
\appendix
\section{Conventions}
The light-front coordinates are defined as \( (x^+, x^-, x^1, x^2) \), where \(x^+=x^0+ x^3\) is the light-front time,  \(x^-=x^0-
x^3\) is the longitudinal coordinate and  \(\vec x_\perp=(x^1, x^2)\) are the transverse coordinates.
Non-vanishing elements of the metric tensor are
\begin{align}
  g^{+-}=g^{-+}=2\;, \
  g_{+-}=g_{-+}=\frac{1}{2}\;,\
  g^{11}=g^{22}=-1\;.
\end{align}
The Dirac matrices are four unitary traceless $4 \times 4$ matrices:
\begin{align}
  \begin{split}
  \gamma^0=\beta=
  \begin{pmatrix}
    0&-i\\
    i&0
  \end{pmatrix},
  \quad
  \gamma^+=
  \begin{pmatrix}
    0 & 0\\
    2i & 0
  \end{pmatrix},\\
  \gamma^-=
  \begin{pmatrix}
    0&-2i\\
    0&0
  \end{pmatrix},
  \quad
  \gamma^i=
  \begin{pmatrix}
    -i\hat{\sigma}^i&0\\
    0&i\hat{\sigma}^i
  \end{pmatrix}
  ,
  \end{split}
\end{align}
where,
\begin{align}
  \hat{\sigma}^1=\sigma^2=
  \begin{pmatrix}
    0&-i\\
    i&0
  \end{pmatrix},
       \quad
       \hat{\sigma}^2=-\sigma^1=
       \begin{pmatrix}
         0&-1\\
         -1&0
       \end{pmatrix}.
\end{align}
We use the following spinor representation,
\begin{equation}
  \begin{split}
    &u(p,\lambda=\frac{1}{2})=\frac{1}{\sqrt{p^+}}{(p^+,0,im_q,ip^x-p^y)}^\intercal \;,\\
    &u(p,\lambda=-\frac{1}{2})=\frac{1}{\sqrt{p^+}}{(0,p^+,-ip^x-p^y,im_q)}^\intercal \;.
  \end{split}
\end{equation}
\section{The interaction time}\label{app:Leta}
To estimate $L_\eta$, consider the quark moving along the positive-z direction with speed $\beta_q$ and the nucleus moving along the negative-z direction with speed $-\beta_A$, as illustrated in Fig.~\ref{fig:dis_zt}. The starting point of the quark-nucleus interaction is at $t_{start}=0, z_{start}=0$, i.e. $x_{start}^+=0$. The end point of their interaction is at $t_{end}=d\sqrt{(1+\beta_{A}\beta_q)^2/(\beta_A+\beta_q)^2-1}, z_{end}=\beta_q t_{end}$. Thereby,
\begin{align}
  \begin{split}
  L_\eta=\Delta x^+ =& (t_{end}+z_{end}) - (t_{start}+z_{start})\\
  =&d(\beta_q+1)\frac{\sqrt{(1-\beta_A^2)(1-\beta_q^2)}}{\beta_A + \beta_q}
  \;.
  \end{split}
\end{align}
$d$ is the width of the nucleus in its rest frame. 
If we consider a gold beam at the RHIC energy of $\sqrt{s} = 100A ~\GeV$, and estimate its velocity according to $\gamma_{A} =1/\sqrt{1-\beta_{A}^2}= \sqrt{s}/m =100$ with $m$ the mass of gold nucleus, thereby $\beta_A=0.9999$. Its rest width for a central impact is $d=14~\text{fm}$. Assuming that the quark has the same speed, i.e. $\beta_q=\beta_A$, we get $\Delta x^+ = 0.0014~\text{fm}$, which is small just as we expected. However the color field generated by the nucleus is identified as the small momentum degrees of freedom in the nucleus, and should admit a smaller longitudinal momentum scaled by Bjorken-x, $p^-=xP^-$. The resulting Lorentz factor is also scaled as $\gamma= x\gamma_A$. 
At $x=0.1$, $\Delta x^+ = 0.14~\text{fm}$, and at $x=0.012$, $\Delta x^+ \approx 14~\text{fm}$. We therefore take $L_\eta=50~\GeV^{-1}\approx 10~\text{fm}$ as the duration of the color field along $x^+$ in our calculations. 
\section{Discretization}\label{app:dis}
  In the numerical calculation, the fields are SU(3) matrices on the sites of a 3-dimensional discrete space.
  The 2-dimensional transverse space is a lattice extending from $-L$ to $L$ for each side. The number of transverse lattice sites is
  $2N$, giving the lattice spacing $a=L/N$.  For any vector in this space, $\vec r_\perp=(r^1,r^2)$,
  \[
    r^i= n_i a (i = 1,2) , \quad n_i=-N,-N+1,\ldots,N-1 .
  \]
  This space satisfies periodic boundary conditions. It follows that in the
  momentum space, for any vector, $\vec p_\perp=(p^1,p^2)$,
  \[
    p^i=k_i d_p (i = 1,2), \quad k_i=-N,-N+1,\ldots,N-1,
  \]
  where $d_p\equiv \pi/L$ is the resolution in momentum space. The momentum space extends from $-\pi/a$ to $\pi/a$.

  The conversion of the integration is
  \begin{align}
    \int\frac{\diff^2 \vec p_\perp}{{(2\pi)}^2} \to \frac{1}{{(2L)}^2}\sum_{k_1, k_2}\;,
    \qquad
    \int \diff^2 \vec r_\perp \to a^2\sum_{n_1, n_2}\;.
  \end{align}
  The Dirac delta is converted to the Kronecker delta as follows
  \begin{align}
    \begin{split}
    &\int \diff^2 \vec r_\perp e^{-i \vec p_\perp\cdot \vec x_\perp}={(2\pi)}^2\delta^2(\vec p_\perp)\ \\
   & \to
    \sum_{n_1, n_2}  a^2 e^{-i (n_1 k_1 + n_2 k_2)\pi/N} = {(2L)}^2 \delta_{k_1,0}\delta_{k_2,0}\;,
    \end{split}
  \end{align}
  and
  \begin{align}
    \begin{split}
      &\int \diff^2 \vec p_\perp e^{i \vec p_\perp\cdot \vec x_\perp}={(2\pi)}^2\delta^2(\vec r_\perp)\ \\
      & \to
      \sum_{k_1, k_2} \frac{1}{{(2L)}^2}  e^{i (n_1 k_1 + n_2 k_2)\pi/N} =  \frac{1}{a^2}\delta_{n_1,0}\delta_{n_2,0}\;.
    \end{split}
  \end{align}
  The (inverse-)Fourier transformation becomes
  \begin{align}
    \begin{split}
    f(n_1, n_2)= & \frac{1}{ {(2 L)}^2} \sum_{k_1, k_2} e^{i(n_1 k_1 + n_2 k_2)\pi/N}\tilde f(k_1, k_2),\\
    \tilde f(k_1, k_2)  = & \sum_{n_1, n_2} a^2 e^{-i(n_1 k_1 + n_2 k_2)\pi/N}  f(n_1, n_2)\;.
    \end{split}
  \end{align}
\section{The eikonal limit of the interaction matrix}\label{app:Wilsonline}
In the eikonal approximation, the momentum of the quark is dominated by $p^+\gg p^-,p_\perp$ and correspondingly for the target $P^-\gg
P^+,P_\perp$. This is usually simplified as $p^\mu=(p^+\approx\sqrt{s},p^-=0,p_\perp=0)$ and
$P^\mu=(P^+=0,P^-\approx\sqrt{s},P_\perp=0)$. In such circumstances, the interaction picture and the Schr\"odinger picture
become equivalent, $V_I(x^+)=V(x^+)$. In our calculation, the field exists during $x^+=[0,L_\eta]$. The evolution of the quark can be written in terms of the Wilson line,
\begin{align}
  U(0,L_\eta; \vec x_\perp)\equiv\mathcal{T}_+\exp\bigg(
  -i g\int_{0}^{L_\eta}\diff x^+\mathcal{A}_a^-(\vec x_\perp, x^+)T_a
  \bigg)\;.
\end{align}
The physical observables such as the cross section could be determined from the configuration average of the Wilson line, as $\braket{U(0,L_\eta; \vec x_\perp)}$. The Taylor expansion of the time-ordered exponential function leads to
  \begin{align}
    \begin{split}
      &\braket{U(0,L_\eta; \vec x_\perp)}\\
      =&\sum_{n=0}^\infty
{(-ig)}^n
      \int\prod_{i=1}^n \diff^2 z_{i\perp}G_0(x_\perp-z_{i\perp})
      \int_{0}^{L_\eta}\diff z_1^+\\
      &\int_{z_1^+}^{L_\eta}\diff  z_2^+ \cdots \int_{z_{n-1}^+}^{L_\eta}\diff z_n^+
      \mathcal{T}_+\langle\rho_{a_1}(z_1^+, z_{1\perp})\rho_{a_2}(z_2^+, z_{2\perp})\\
      &\cdots \rho_{a_n}(z_n^+, z_{n\perp})\rangle
      T_{a_1}T_{a_2}\cdots T_{a_n}\;,
    \end{split}
  \end{align}
  where $G_0$ is given in Eq.~\eqref{eq:Green}.
  According to the correlation function of $\rho_a$ given in Eq.~\eqref{eq:chgcor}, the configuration average for the production of odd number charge densities $\braket{\rho_1\rho_2...\rho_{2j+1}}$ is zero; for even number cases
  $\braket{\rho_1\rho_2...\rho_{2j}}$, only the adjacent contractions survive under the time-ordered integrals.

  \begin{align}
    \begin{split}
     & \braket{\rho_{a_1}(z_1^-, z_{1\perp})\cdots \rho_{a_n}(z_n^-, z_{n\perp})}\\
     \to&\braket{\rho_{a_1}(z_1^-, z_{1\perp})\rho_{a_2}(z_2^-, z_{2\perp})}
     \braket{\rho_{a_3}(z_3^-, z_{3\perp})\rho_{a_4}(z_4^-, z_{4\perp})}
     \\
      &\cdots \braket{\rho_{a_{n-1}}(z_{n-1}^-, z_{n-1 \perp})\rho_{a_n}(z_n^-, z_{n\perp})}
      \;.
    \end{split}
  \end{align}

  The integral on each of the two-point charge correlators can be carried out separately,
and then the Wilson line can be recollected into an exponential,
  \begin{equation}
    \begin{split}
      \braket{U(0,L_\eta; \vec x_\perp)}
      =&\sum_{n=0}^\infty{(-ig^2)}^n
      \int\prod_{i=1}^{n/2} \diff^2 z_{2i\perp}G^2_0(\vec x_\perp- \vec z_{2i\perp})\\
      &\frac{1}{2}\int_{0}^{L_\eta}\diff z_1^- \mu^2(z_1^-)
      \frac{1}{2}\int_{z_1^- }^{L_\eta}\diff z_3^- \mu^2(z_3^-)
      \cdots \\
     &\frac{1}{2}\int_{z_{n-3}^- }^{L_\eta} \diff z_{n-1}^-
      \mu^2(z_{n-1}^-) T^2_{a_1}T^2_{a_3}\cdots T^2_{a_{n-1}}\\
      =&\sum_{n=0}^\infty
      \frac{1}{(n/2)!} 
      \bigg[\frac{-g^4}{2}
      \int \diff^2 z_\perp G^2_0(x_\perp-z_\perp)\\
      &\int_0^{L_\eta}\diff z^-  \mu^2(z^-)
      T^2_a
      \bigg]^{n/2}\\
      =&\exp \bigg[
      \frac{-g^4\mu^2 (N_c^2-1)L_\eta}{16\pi m_g^2 N_c}\bm I_3
      \bigg]
      \;.
    \end{split}
  \end{equation}
  Note that $\bm I_3$ is the identity matrix in the color space.


\section{Perturbative approximations of the differential cross section in the eikonal limit}\label{app:dcrss}
In the eikonal limit, the differential cross section is given by~\cite{Dumitru:2002qt},
\begin{align}\label{eq:dcrss_per}
  \begin{split}
\frac{\diff\sigma_{tot}}{\diff^2 b \diff^2 q_t} 
=&
\frac{1}{(2\pi)^2}
\int\diff^2 r_t
e^{-i q_t r_t}\\
&\times \bigg[ 
e^{-2\pi Q_s^2/N_c\int\diff^2 k_t/k_t^4 (1-e^{ik_t r_t})}\\
&-2 e^{-\pi^2 Q_s^2/N_c m_g^2}
+1
\bigg]
\\
=&
\frac{1}{(2\pi)^2}
\int\diff^2 r_t
e^{-i q_t r_t}
e^{-2\pi Q_s^2/N_c\int\diff^2 k_t/k_t^4 (1-e^{i k_t r_t})}\\
&+\delta^2(q_t)
\left(
-2 e^{-\pi^2 Q_s^2/N_c m_g^2}
+1
\right)
\;.
\end{split}
\end{align}
Note that $q_t$ is the difference between the quark's initial and final transverse momentum. 
Therefore for numerical calculation, it is convenient to study the differential cross section with the quark's initial state as $p_\perp=0$, and this is also what we did in this work.

We implement an infrared cutoff $m_g$ on the integral of the transverse momentum $k_t$, such that
\begin{align}
  \begin{split}
\int\diff^2 k_t \frac{1}{k_t^4} 
= & 2\pi \int_0^\infty\diff k_t \frac{1}{k_t^3}\\
\to & 2\pi \int_{m_g}^\infty\diff k_t \frac{1}{k_t^3}
= \frac{\pi}{m_g^2}\\
\to & 2\pi \int_0^\infty\diff k_t \frac{1}{(k_t+m_g)^3}
= \frac{\pi}{m_g^2}
\;,
\end{split}
\end{align}
and
\begin{align}
  \begin{split}
\int\diff^2 k_t & \frac{1}{k_t^4 (q_t - k_t)^4} \\
\to &  \int_0^\infty\diff k_t \frac{2\pi k_t}{(k_t + m_g)^4 (|q_t - k_t|+ m_g)^4} 
\;.
\end{split}
\end{align}
The first term of Eq.~\eqref{eq:dcrss_per} contributes to $q_t > 0$ region. Expand the exponential over $Q_s^2$ to its second order, 
\begin{align}
  \begin{split}
&
\frac{1}{(2\pi)^2}
\int\diff^2 r_t
e^{-i q_t r_t}
e^{-2\pi Q_s^2/N_c\int\diff^2 k_t/k_t^4 (1-e^{i k_t r_t})}
\\
=&
\frac{1}{(2\pi)^2}
\int\diff^2 r_t
e^{-i q_t r_t}
\bigg\{
1
- \frac{2\pi Q_s^2}{N_c}\int\frac{\diff^2 k_t}{k_t^4} (1-e^{i k_t r_t})\\
&
+\frac{1}{2}
\bigg[-\frac{2\pi Q_s^2}{N_c}\int\frac{\diff^2 k_t}{k_t^4} (1-e^{ik_t r_t})\bigg]^2
+\ldots
\bigg\}
\;.
\end{split}
\end{align}
The second term of Eq.~\eqref{eq:dcrss_per} contributes to $q_t=0$. We also expand the exponential over $Q_s^2$ to its second order, 
\begin{align}
  \begin{split}
    &\delta^2(q_t)
    \left(
-2 e^{-\pi^2 Q_s^2/N_c m_g^2}
+1
\right)\\
=& \delta^2(q_t)
\bigg\{
\bigg[
  -2 \bigg[ 1-\frac{\pi^2 Q_s^2}{N_c m_g^2}+\frac{1}{2}\bigg(\frac{2\pi^2 Q_s^2}{N_c m_g^2}\bigg)^2 +\ldots
\bigg]
+1
\bigg\}
  \end{split}
\end{align}
The approximation up to the leading order (LO) of $Q_s^2$ is
\begin{align}
  \begin{split}
\frac{\diff\sigma_{tot}}{\diff^2 b \diff^2 q_t} \bigg|_{\text{LO}}
=&
\frac{2\pi Q_s^2}{N_c q_t^4}
\;.
\end{split}
\end{align}
The approximation up to the next-to-leading order (NLO) is
\begin{widetext}
\begin{align}
  \begin{split}
\frac{\diff\sigma_{tot}}{\diff^2 b \diff^2 q_t} \bigg|_{\text{LO+NLO}}
=&
\frac{1}{2}
\bigg(
  \frac{2\pi^2 Q_s^2}{N_c m_g^2}
\bigg)^2 \delta^2(q_t)
+
\bigg[
  \frac{2\pi Q_s^2}{N_c }
  - \left(\frac{2\pi Q_s^2}{N_c }\right)^2\frac{\pi}{m_g}
\bigg] \frac{1}{q_t^4}
+\frac{1}{2 (2\pi)^2}
\left(\frac{2\pi Q_s^2}{N_c }\right)^2
\frac{1}{3 m_g^3 q_t^7 (m_g+q_t)^2 (2 m_g+q_t)^7}\\
&\times \pi  \bigg[15360 m_g^{13} \log  \left(\frac{m_g+q_t}{m_g}\right)
-15360 m_g^{12} q_t+76800 m_g^{12} q_t \log \left(\frac{m_g+q_t}{m_g}\right)-69120 m_g^{11} q_t^2\\
 &+161280 m_g^{11} q_t^2 \log \left(\frac{m_g+q_t}{m_g}\right)
 -128000 m_g^{10} q_t^3+180480 m_g^{10} q_t^3 \log \left(\frac{m_g+q_t}{m_g}\right)-121600 m_g^9 q_t^4\\
 &+107520 m_g^9 q_t^4 \log \left(\frac{m_g+q_t}{m_g}\right)-54912 m_g^8 q_t^5+20160 m_g^8 q_t^5 \log \left(\frac{m_g+q_t}{m_g}\right)\\
 &+960 m_g^7 q_t^6-16800 m_g^7 q_t^6 \log \left(\frac{m_g+q_t}{m_g}\right)+14528 m_g^6 q_t^7-14160 m_g^6 q_t^7 \log \left(\frac{m_g+q_t}{m_g}\right)+7808 m_g^5 q_t^8\\
 &-4740 m_g^5 q_t^8 \log \left(\frac{m_g+q_t}{m_g}\right)+2380 m_g^4 q_t^9-600 m_g^4 q_t^9 \log \left(\frac{m_g+q_t}{m_g}\right)+814 m_g^3 q_t^{10}
 \\
 & +60 m_g^3 q_t^{10} \log \left(\frac{m_g+q_t}{m_g}\right)
 +280 m_g^2 q_t^{11}+49 m_g q_t^{12}+4 q_t^{13}\bigg]
\;.
\end{split}
\end{align}
\end{widetext}
\clearpage

\bibliography{qA}

\begin{thebibliography}{47}%
\makeatletter
\providecommand \@ifxundefined [1]{%
 \@ifx{#1\undefined}
}%
\providecommand \@ifnum [1]{%
 \ifnum #1\expandafter \@firstoftwo
 \else \expandafter \@secondoftwo
 \fi
}%
\providecommand \@ifx [1]{%
 \ifx #1\expandafter \@firstoftwo
 \else \expandafter \@secondoftwo
 \fi
}%
\providecommand \natexlab [1]{#1}%
\providecommand \enquote  [1]{``#1''}%
\providecommand \bibnamefont  [1]{#1}%
\providecommand \bibfnamefont [1]{#1}%
\providecommand \citenamefont [1]{#1}%
\providecommand \href@noop [0]{\@secondoftwo}%
\providecommand \href [0]{\begingroup \@sanitize@url \@href}%
\providecommand \@href[1]{\@@startlink{#1}\@@href}%
\providecommand \@@href[1]{\endgroup#1\@@endlink}%
\providecommand \@sanitize@url [0]{\catcode `\\12\catcode `\$12\catcode
  `\&12\catcode `\#12\catcode `\^12\catcode `\_12\catcode `\%12\relax}%
\providecommand \@@startlink[1]{}%
\providecommand \@@endlink[0]{}%
\providecommand \url  [0]{\begingroup\@sanitize@url \@url }%
\providecommand \@url [1]{\endgroup\@href {#1}{\urlprefix }}%
\providecommand \urlprefix  [0]{URL }%
\providecommand \Eprint [0]{\href }%
\providecommand \doibase [0]{http://dx.doi.org/}%
\providecommand \selectlanguage [0]{\@gobble}%
\providecommand \bibinfo  [0]{\@secondoftwo}%
\providecommand \bibfield  [0]{\@secondoftwo}%
\providecommand \translation [1]{[#1]}%
\providecommand \BibitemOpen [0]{}%
\providecommand \bibitemStop [0]{}%
\providecommand \bibitemNoStop [0]{.\EOS\space}%
\providecommand \EOS [0]{\spacefactor3000\relax}%
\providecommand \BibitemShut  [1]{\csname bibitem#1\endcsname}%
\let\auto@bib@innerbib\@empty
\bibitem [{\citenamefont {Dumitru}\ and\ \citenamefont
  {Jalilian-Marian}(2002)}]{Dumitru:2002qt}%
  \BibitemOpen
  \bibfield  {author} {\bibinfo {author} {\bibfnamefont {A.}~\bibnamefont
  {Dumitru}}\ and\ \bibinfo {author} {\bibfnamefont {J.}~\bibnamefont
  {Jalilian-Marian}},\ }\href {\doibase 10.1103/PhysRevLett.89.022301}
  {\bibfield  {journal} {\bibinfo  {journal} {Phys. Rev. Lett.}\ }\textbf
  {\bibinfo {volume} {89}},\ \bibinfo {pages} {022301} (\bibinfo {year}
  {2002})},\ \Eprint {http://arxiv.org/abs/hep-ph/0204028}
  {arXiv:hep-ph/0204028 [hep-ph]} \BibitemShut {NoStop}%
\bibitem [{\citenamefont {Tuchin}(2004)}]{Tuchin:2004rb}%
  \BibitemOpen
  \bibfield  {author} {\bibinfo {author} {\bibfnamefont {K.}~\bibnamefont
  {Tuchin}},\ }\href {\doibase 10.1016/j.physletb.2004.04.057} {\bibfield
  {journal} {\bibinfo  {journal} {Phys. Lett.}\ }\textbf {\bibinfo {volume}
  {B593}},\ \bibinfo {pages} {66} (\bibinfo {year} {2004})},\ \Eprint
  {http://arxiv.org/abs/hep-ph/0401022} {arXiv:hep-ph/0401022 [hep-ph]}
  \BibitemShut {NoStop}%
\bibitem [{\citenamefont {Kovchegov}\ and\ \citenamefont
  {Tuchin}(2006)}]{Kovchegov:2006qn}%
  \BibitemOpen
  \bibfield  {author} {\bibinfo {author} {\bibfnamefont {Y.~V.}\ \bibnamefont
  {Kovchegov}}\ and\ \bibinfo {author} {\bibfnamefont {K.}~\bibnamefont
  {Tuchin}},\ }\href {\doibase 10.1103/PhysRevD.74.054014} {\bibfield
  {journal} {\bibinfo  {journal} {Phys. Rev.}\ }\textbf {\bibinfo {volume}
  {D74}},\ \bibinfo {pages} {054014} (\bibinfo {year} {2006})},\ \Eprint
  {http://arxiv.org/abs/hep-ph/0603055} {arXiv:hep-ph/0603055 [hep-ph]}
  \BibitemShut {NoStop}%
\bibitem [{\citenamefont {Gribov}\ \emph {et~al.}(1983)\citenamefont {Gribov},
  \citenamefont {Levin},\ and\ \citenamefont {Ryskin}}]{Gribov:1984tu}%
  \BibitemOpen
  \bibfield  {author} {\bibinfo {author} {\bibfnamefont {L.~V.}\ \bibnamefont
  {Gribov}}, \bibinfo {author} {\bibfnamefont {E.~M.}\ \bibnamefont {Levin}}, \
  and\ \bibinfo {author} {\bibfnamefont {M.~G.}\ \bibnamefont {Ryskin}},\
  }\href {\doibase 10.1016/0370-1573(83)90022-4} {\bibfield  {journal}
  {\bibinfo  {journal} {Phys. Rept.}\ }\textbf {\bibinfo {volume} {100}},\
  \bibinfo {pages} {1} (\bibinfo {year} {1983})}\BibitemShut {NoStop}%
\bibitem [{\citenamefont {Mueller}\ and\ \citenamefont
  {Qiu}(1986)}]{Mueller:1985wy}%
  \BibitemOpen
  \bibfield  {author} {\bibinfo {author} {\bibfnamefont {A.~H.}\ \bibnamefont
  {Mueller}}\ and\ \bibinfo {author} {\bibfnamefont {J.-w.}\ \bibnamefont
  {Qiu}},\ }\href {\doibase 10.1016/0550-3213(86)90164-1} {\bibfield  {journal}
  {\bibinfo  {journal} {Nucl. Phys.}\ }\textbf {\bibinfo {volume} {B268}},\
  \bibinfo {pages} {427} (\bibinfo {year} {1986})}\BibitemShut {NoStop}%
\bibitem [{\citenamefont {Gelis}\ \emph {et~al.}(2010)\citenamefont {Gelis},
  \citenamefont {Iancu}, \citenamefont {Jalilian-Marian},\ and\ \citenamefont
  {Venugopalan}}]{Gelis:2010nm}%
  \BibitemOpen
  \bibfield  {author} {\bibinfo {author} {\bibfnamefont {F.}~\bibnamefont
  {Gelis}}, \bibinfo {author} {\bibfnamefont {E.}~\bibnamefont {Iancu}},
  \bibinfo {author} {\bibfnamefont {J.}~\bibnamefont {Jalilian-Marian}}, \ and\
  \bibinfo {author} {\bibfnamefont {R.}~\bibnamefont {Venugopalan}},\ }\href
  {\doibase 10.1146/annurev.nucl.010909.083629} {\bibfield  {journal} {\bibinfo
   {journal} {Ann. Rev. Nucl. Part. Sci.}\ }\textbf {\bibinfo {volume} {60}},\
  \bibinfo {pages} {463} (\bibinfo {year} {2010})},\ \Eprint
  {http://arxiv.org/abs/1002.0333} {arXiv:1002.0333 [hep-ph]} \BibitemShut
  {NoStop}%
\bibitem [{\citenamefont {Weigert}(2005)}]{Weigert:2005us}%
  \BibitemOpen
  \bibfield  {author} {\bibinfo {author} {\bibfnamefont {H.}~\bibnamefont
  {Weigert}},\ }\href {\doibase 10.1016/j.ppnp.2005.01.029} {\bibfield
  {journal} {\bibinfo  {journal} {Prog. Part. Nucl. Phys.}\ }\textbf {\bibinfo
  {volume} {55}},\ \bibinfo {pages} {461} (\bibinfo {year} {2005})},\ \Eprint
  {http://arxiv.org/abs/hep-ph/0501087} {arXiv:hep-ph/0501087 [hep-ph]}
  \BibitemShut {NoStop}%
\bibitem [{\citenamefont {Kovchegov}\ and\ \citenamefont
  {Levin}(2012)}]{Kovchegov:2012mbw}%
  \BibitemOpen
  \bibfield  {author} {\bibinfo {author} {\bibfnamefont {Y.~V.}\ \bibnamefont
  {Kovchegov}}\ and\ \bibinfo {author} {\bibfnamefont {E.}~\bibnamefont
  {Levin}},\ }\href {\doibase 10.1017/CBO9781139022187} {\bibfield  {journal}
  {\bibinfo  {journal} {Camb. Monogr. Part. Phys. Nucl. Phys. Cosmol.}\
  }\textbf {\bibinfo {volume} {33}},\ \bibinfo {pages} {1} (\bibinfo {year}
  {2012})}\BibitemShut {NoStop}%
\bibitem [{\citenamefont {Accardi}\ \emph {et~al.}(2016)\citenamefont {Accardi}
  \emph {et~al.}}]{Accardi:2012qut}%
  \BibitemOpen
  \bibfield  {author} {\bibinfo {author} {\bibfnamefont {A.}~\bibnamefont
  {Accardi}} \emph {et~al.},\ }\href {\doibase 10.1140/epja/i2016-16268-9}
  {\bibfield  {journal} {\bibinfo  {journal} {Eur. Phys. J.}\ }\textbf
  {\bibinfo {volume} {A52}},\ \bibinfo {pages} {268} (\bibinfo {year}
  {2016})},\ \Eprint {http://arxiv.org/abs/1212.1701} {arXiv:1212.1701
  [nucl-ex]} \BibitemShut {NoStop}%
\bibitem [{\citenamefont {Zhao}\ \emph {et~al.}(2013)\citenamefont {Zhao},
  \citenamefont {Ilderton}, \citenamefont {Maris},\ and\ \citenamefont
  {Vary}}]{Zhao:2013cma}%
  \BibitemOpen
  \bibfield  {author} {\bibinfo {author} {\bibfnamefont {X.}~\bibnamefont
  {Zhao}}, \bibinfo {author} {\bibfnamefont {A.}~\bibnamefont {Ilderton}},
  \bibinfo {author} {\bibfnamefont {P.}~\bibnamefont {Maris}}, \ and\ \bibinfo
  {author} {\bibfnamefont {J.~P.}\ \bibnamefont {Vary}},\ }\href {\doibase
  10.1103/PhysRevD.88.065014} {\bibfield  {journal} {\bibinfo  {journal} {Phys.
  Rev.}\ }\textbf {\bibinfo {volume} {D88}},\ \bibinfo {pages} {065014}
  (\bibinfo {year} {2013})},\ \Eprint {http://arxiv.org/abs/1303.3273}
  {arXiv:1303.3273 [nucl-th]} \BibitemShut {NoStop}%
\bibitem [{\citenamefont {Vary}\ \emph {et~al.}(2010)\citenamefont {Vary},
  \citenamefont {Honkanen}, \citenamefont {Li}, \citenamefont {Maris},
  \citenamefont {Brodsky}, \citenamefont {Harindranath}, \citenamefont
  {de~Teramond}, \citenamefont {Sternberg}, \citenamefont {Ng},\ and\
  \citenamefont {Yang}}]{1stBLFQ}%
  \BibitemOpen
  \bibfield  {author} {\bibinfo {author} {\bibfnamefont {J.~P.}\ \bibnamefont
  {Vary}}, \bibinfo {author} {\bibfnamefont {H.}~\bibnamefont {Honkanen}},
  \bibinfo {author} {\bibfnamefont {J.}~\bibnamefont {Li}}, \bibinfo {author}
  {\bibfnamefont {P.}~\bibnamefont {Maris}}, \bibinfo {author} {\bibfnamefont
  {S.~J.}\ \bibnamefont {Brodsky}}, \bibinfo {author} {\bibfnamefont
  {A.}~\bibnamefont {Harindranath}}, \bibinfo {author} {\bibfnamefont {G.~F.}\
  \bibnamefont {de~Teramond}}, \bibinfo {author} {\bibfnamefont
  {P.}~\bibnamefont {Sternberg}}, \bibinfo {author} {\bibfnamefont {E.~G.}\
  \bibnamefont {Ng}}, \ and\ \bibinfo {author} {\bibfnamefont {C.}~\bibnamefont
  {Yang}},\ }\href {\doibase 10.1103/PhysRevC.81.035205} {\bibfield  {journal}
  {\bibinfo  {journal} {Phys. Rev.}\ }\textbf {\bibinfo {volume} {C81}},\
  \bibinfo {pages} {035205} (\bibinfo {year} {2010})},\ \Eprint
  {http://arxiv.org/abs/0905.1411} {arXiv:0905.1411 [nucl-th]} \BibitemShut
  {NoStop}%
\bibitem [{\citenamefont {Wiecki}\ \emph {et~al.}(2015)\citenamefont {Wiecki},
  \citenamefont {Li}, \citenamefont {Zhao}, \citenamefont {Maris},\ and\
  \citenamefont {Vary}}]{positronium}%
  \BibitemOpen
  \bibfield  {author} {\bibinfo {author} {\bibfnamefont {P.}~\bibnamefont
  {Wiecki}}, \bibinfo {author} {\bibfnamefont {Y.}~\bibnamefont {Li}}, \bibinfo
  {author} {\bibfnamefont {X.}~\bibnamefont {Zhao}}, \bibinfo {author}
  {\bibfnamefont {P.}~\bibnamefont {Maris}}, \ and\ \bibinfo {author}
  {\bibfnamefont {J.~P.}\ \bibnamefont {Vary}},\ }\href {\doibase
  10.1103/PhysRevD.91.105009} {\bibfield  {journal} {\bibinfo  {journal} {Phys.
  Rev.}\ }\textbf {\bibinfo {volume} {D91}},\ \bibinfo {pages} {105009}
  (\bibinfo {year} {2015})},\ \Eprint {http://arxiv.org/abs/1404.6234}
  {arXiv:1404.6234 [nucl-th]} \BibitemShut {NoStop}%
\bibitem [{\citenamefont {Li}\ \emph {et~al.}(2016)\citenamefont {Li},
  \citenamefont {Maris}, \citenamefont {Zhao},\ and\ \citenamefont
  {Vary}}]{Yang_fix}%
  \BibitemOpen
  \bibfield  {author} {\bibinfo {author} {\bibfnamefont {Y.}~\bibnamefont
  {Li}}, \bibinfo {author} {\bibfnamefont {P.}~\bibnamefont {Maris}}, \bibinfo
  {author} {\bibfnamefont {X.}~\bibnamefont {Zhao}}, \ and\ \bibinfo {author}
  {\bibfnamefont {J.~P.}\ \bibnamefont {Vary}},\ }\href {\doibase
  10.1016/j.physletb.2016.04.065} {\bibfield  {journal} {\bibinfo  {journal}
  {Phys. Lett.}\ }\textbf {\bibinfo {volume} {B758}},\ \bibinfo {pages} {118}
  (\bibinfo {year} {2016})},\ \Eprint {http://arxiv.org/abs/1509.07212}
  {arXiv:1509.07212 [hep-ph]} \BibitemShut {NoStop}%
\bibitem [{\citenamefont {Li}\ \emph {et~al.}(2017)\citenamefont {Li},
  \citenamefont {Maris},\ and\ \citenamefont {Vary}}]{Yang_run}%
  \BibitemOpen
  \bibfield  {author} {\bibinfo {author} {\bibfnamefont {Y.}~\bibnamefont
  {Li}}, \bibinfo {author} {\bibfnamefont {P.}~\bibnamefont {Maris}}, \ and\
  \bibinfo {author} {\bibfnamefont {J.~P.}\ \bibnamefont {Vary}},\ }\href
  {\doibase 10.1103/PhysRevD.96.016022} {\bibfield  {journal} {\bibinfo
  {journal} {Phys. Rev.}\ }\textbf {\bibinfo {volume} {D96}},\ \bibinfo {pages}
  {016022} (\bibinfo {year} {2017})},\ \Eprint
  {http://arxiv.org/abs/1704.06968} {arXiv:1704.06968 [hep-ph]} \BibitemShut
  {NoStop}%
\bibitem [{\citenamefont {Tang}\ \emph {et~al.}(2018)\citenamefont {Tang},
  \citenamefont {Li}, \citenamefont {Maris},\ and\ \citenamefont
  {Vary}}]{Shuo_Bc}%
  \BibitemOpen
  \bibfield  {author} {\bibinfo {author} {\bibfnamefont {S.}~\bibnamefont
  {Tang}}, \bibinfo {author} {\bibfnamefont {Y.}~\bibnamefont {Li}}, \bibinfo
  {author} {\bibfnamefont {P.}~\bibnamefont {Maris}}, \ and\ \bibinfo {author}
  {\bibfnamefont {J.~P.}\ \bibnamefont {Vary}},\ }\href {\doibase
  10.1103/PhysRevD.98.114038} {\bibfield  {journal} {\bibinfo  {journal} {Phys.
  Rev.}\ }\textbf {\bibinfo {volume} {D98}},\ \bibinfo {pages} {114038}
  (\bibinfo {year} {2018})},\ \Eprint {http://arxiv.org/abs/1810.05971}
  {arXiv:1810.05971 [nucl-th]} \BibitemShut {NoStop}%
\bibitem [{\citenamefont {Tang}\ \emph {et~al.}(2019)\citenamefont {Tang},
  \citenamefont {Li}, \citenamefont {Maris},\ and\ \citenamefont
  {Vary}}]{Tang:2019gvn}%
  \BibitemOpen
  \bibfield  {author} {\bibinfo {author} {\bibfnamefont {S.}~\bibnamefont
  {Tang}}, \bibinfo {author} {\bibfnamefont {Y.}~\bibnamefont {Li}}, \bibinfo
  {author} {\bibfnamefont {P.}~\bibnamefont {Maris}}, \ and\ \bibinfo {author}
  {\bibfnamefont {J.~P.}\ \bibnamefont {Vary}},\ }\href@noop {} {\  (\bibinfo
  {year} {2019})},\ \Eprint {http://arxiv.org/abs/1912.02088} {arXiv:1912.02088
  [nucl-th]} \BibitemShut {NoStop}%
\bibitem [{\citenamefont {Jia}\ and\ \citenamefont {Vary}(2019)}]{Jia:2018ary}%
  \BibitemOpen
  \bibfield  {author} {\bibinfo {author} {\bibfnamefont {S.}~\bibnamefont
  {Jia}}\ and\ \bibinfo {author} {\bibfnamefont {J.~P.}\ \bibnamefont {Vary}},\
  }\href {\doibase 10.1103/PhysRevC.99.035206} {\bibfield  {journal} {\bibinfo
  {journal} {Phys. Rev.}\ }\textbf {\bibinfo {volume} {C99}},\ \bibinfo {pages}
  {035206} (\bibinfo {year} {2019})},\ \Eprint
  {http://arxiv.org/abs/1811.08512} {arXiv:1811.08512 [nucl-th]} \BibitemShut
  {NoStop}%
\bibitem [{\citenamefont {Du}\ \emph {et~al.}(2019)\citenamefont {Du},
  \citenamefont {Li}, \citenamefont {Zhao}, \citenamefont {Miller},\ and\
  \citenamefont {Vary}}]{Du:2019ips}%
  \BibitemOpen
  \bibfield  {author} {\bibinfo {author} {\bibfnamefont {W.}~\bibnamefont
  {Du}}, \bibinfo {author} {\bibfnamefont {Y.}~\bibnamefont {Li}}, \bibinfo
  {author} {\bibfnamefont {X.}~\bibnamefont {Zhao}}, \bibinfo {author}
  {\bibfnamefont {G.~A.}\ \bibnamefont {Miller}}, \ and\ \bibinfo {author}
  {\bibfnamefont {J.~P.}\ \bibnamefont {Vary}},\ }\href@noop {} {\  (\bibinfo
  {year} {2019})},\ \Eprint {http://arxiv.org/abs/1911.10762} {arXiv:1911.10762
  [nucl-th]} \BibitemShut {NoStop}%
\bibitem [{\citenamefont {Hu}\ \emph {et~al.}(2019)\citenamefont {Hu},
  \citenamefont {Ilderton},\ and\ \citenamefont {Zhao}}]{Hu:2019hjx}%
  \BibitemOpen
  \bibfield  {author} {\bibinfo {author} {\bibfnamefont {B.}~\bibnamefont
  {Hu}}, \bibinfo {author} {\bibfnamefont {A.}~\bibnamefont {Ilderton}}, \ and\
  \bibinfo {author} {\bibfnamefont {X.}~\bibnamefont {Zhao}},\ }\href@noop {}
  {\  (\bibinfo {year} {2019})},\ \Eprint {http://arxiv.org/abs/1911.12307}
  {arXiv:1911.12307 [nucl-th]} \BibitemShut {NoStop}%
\bibitem [{\citenamefont {Chen}\ \emph {et~al.}(2017)\citenamefont {Chen},
  \citenamefont {Zhao}, \citenamefont {Li}, \citenamefont {Tuchin},\ and\
  \citenamefont {Vary}}]{Chen:2017uuq}%
  \BibitemOpen
  \bibfield  {author} {\bibinfo {author} {\bibfnamefont {G.}~\bibnamefont
  {Chen}}, \bibinfo {author} {\bibfnamefont {X.}~\bibnamefont {Zhao}}, \bibinfo
  {author} {\bibfnamefont {Y.}~\bibnamefont {Li}}, \bibinfo {author}
  {\bibfnamefont {K.}~\bibnamefont {Tuchin}}, \ and\ \bibinfo {author}
  {\bibfnamefont {J.~P.}\ \bibnamefont {Vary}},\ }\href {\doibase
  10.1103/PhysRevD.95.096012} {\bibfield  {journal} {\bibinfo  {journal} {Phys.
  Rev.}\ }\textbf {\bibinfo {volume} {D95}},\ \bibinfo {pages} {096012}
  (\bibinfo {year} {2017})},\ \Eprint {http://arxiv.org/abs/1702.06932}
  {arXiv:1702.06932 [nucl-th]} \BibitemShut {NoStop}%
\bibitem [{\citenamefont {Du}\ \emph {et~al.}(2018)\citenamefont {Du},
  \citenamefont {Yin}, \citenamefont {Li}, \citenamefont {Chen}, \citenamefont
  {Zuo}, \citenamefont {Zhao},\ and\ \citenamefont {Vary}}]{Du:2018tce}%
  \BibitemOpen
  \bibfield  {author} {\bibinfo {author} {\bibfnamefont {W.}~\bibnamefont
  {Du}}, \bibinfo {author} {\bibfnamefont {P.}~\bibnamefont {Yin}}, \bibinfo
  {author} {\bibfnamefont {Y.}~\bibnamefont {Li}}, \bibinfo {author}
  {\bibfnamefont {G.}~\bibnamefont {Chen}}, \bibinfo {author} {\bibfnamefont
  {W.}~\bibnamefont {Zuo}}, \bibinfo {author} {\bibfnamefont {X.}~\bibnamefont
  {Zhao}}, \ and\ \bibinfo {author} {\bibfnamefont {J.~P.}\ \bibnamefont
  {Vary}},\ }\href {\doibase 10.1103/PhysRevC.97.064620} {\bibfield  {journal}
  {\bibinfo  {journal} {Phys. Rev.}\ }\textbf {\bibinfo {volume} {C97}},\
  \bibinfo {pages} {064620} (\bibinfo {year} {2018})},\ \Eprint
  {http://arxiv.org/abs/1804.01156} {arXiv:1804.01156 [nucl-th]} \BibitemShut
  {NoStop}%
\bibitem [{\citenamefont {Yin}\ \emph {et~al.}(2019)\citenamefont {Yin},
  \citenamefont {Du}, \citenamefont {Zuo}, \citenamefont {Zhao},\ and\
  \citenamefont {Vary}}]{Yin:2019kqv}%
  \BibitemOpen
  \bibfield  {author} {\bibinfo {author} {\bibfnamefont {P.}~\bibnamefont
  {Yin}}, \bibinfo {author} {\bibfnamefont {W.}~\bibnamefont {Du}}, \bibinfo
  {author} {\bibfnamefont {W.}~\bibnamefont {Zuo}}, \bibinfo {author}
  {\bibfnamefont {X.}~\bibnamefont {Zhao}}, \ and\ \bibinfo {author}
  {\bibfnamefont {J.~P.}\ \bibnamefont {Vary}},\ }\href@noop {} {\  (\bibinfo
  {year} {2019})},\ \Eprint {http://arxiv.org/abs/1910.10586} {arXiv:1910.10586
  [nucl-th]} \BibitemShut {NoStop}%
\bibitem [{\citenamefont {McLerran}\ and\ \citenamefont
  {Venugopalan}(1994{\natexlab{a}})}]{McLerran:1993ni}%
  \BibitemOpen
  \bibfield  {author} {\bibinfo {author} {\bibfnamefont {L.~D.}\ \bibnamefont
  {McLerran}}\ and\ \bibinfo {author} {\bibfnamefont {R.}~\bibnamefont
  {Venugopalan}},\ }\href {\doibase 10.1103/PhysRevD.49.2233} {\bibfield
  {journal} {\bibinfo  {journal} {Phys. Rev.}\ }\textbf {\bibinfo {volume}
  {D49}},\ \bibinfo {pages} {2233} (\bibinfo {year} {1994}{\natexlab{a}})},\
  \Eprint {http://arxiv.org/abs/hep-ph/9309289} {arXiv:hep-ph/9309289 [hep-ph]}
  \BibitemShut {NoStop}%
\bibitem [{\citenamefont {McLerran}\ and\ \citenamefont
  {Venugopalan}(1994{\natexlab{b}})}]{McLerran:1993ka}%
  \BibitemOpen
  \bibfield  {author} {\bibinfo {author} {\bibfnamefont {L.~D.}\ \bibnamefont
  {McLerran}}\ and\ \bibinfo {author} {\bibfnamefont {R.}~\bibnamefont
  {Venugopalan}},\ }\href {\doibase 10.1103/PhysRevD.49.3352} {\bibfield
  {journal} {\bibinfo  {journal} {Phys. Rev.}\ }\textbf {\bibinfo {volume}
  {D49}},\ \bibinfo {pages} {3352} (\bibinfo {year} {1994}{\natexlab{b}})},\
  \Eprint {http://arxiv.org/abs/hep-ph/9311205} {arXiv:hep-ph/9311205 [hep-ph]}
  \BibitemShut {NoStop}%
\bibitem [{\citenamefont {McLerran}\ and\ \citenamefont
  {Venugopalan}(1994{\natexlab{c}})}]{McLerran:1994vd}%
  \BibitemOpen
  \bibfield  {author} {\bibinfo {author} {\bibfnamefont {L.~D.}\ \bibnamefont
  {McLerran}}\ and\ \bibinfo {author} {\bibfnamefont {R.}~\bibnamefont
  {Venugopalan}},\ }\href {\doibase 10.1103/PhysRevD.50.2225} {\bibfield
  {journal} {\bibinfo  {journal} {Phys. Rev.}\ }\textbf {\bibinfo {volume}
  {D50}},\ \bibinfo {pages} {2225} (\bibinfo {year} {1994}{\natexlab{c}})},\
  \Eprint {http://arxiv.org/abs/hep-ph/9402335} {arXiv:hep-ph/9402335 [hep-ph]}
  \BibitemShut {NoStop}%
\bibitem [{\citenamefont {Jalilian-Marian}(2017)}]{Jalilian-Marian:2017ttv}%
  \BibitemOpen
  \bibfield  {author} {\bibinfo {author} {\bibfnamefont {J.}~\bibnamefont
  {Jalilian-Marian}},\ }\href {\doibase 10.1103/PhysRevD.96.074020} {\bibfield
  {journal} {\bibinfo  {journal} {Phys. Rev.}\ }\textbf {\bibinfo {volume}
  {D96}},\ \bibinfo {pages} {074020} (\bibinfo {year} {2017})},\ \Eprint
  {http://arxiv.org/abs/1708.07533} {arXiv:1708.07533 [hep-ph]} \BibitemShut
  {NoStop}%
\bibitem [{\citenamefont
  {Jalilian-Marian}(2019{\natexlab{a}})}]{Jalilian-Marian:2019kaf}%
  \BibitemOpen
  \bibfield  {author} {\bibinfo {author} {\bibfnamefont {J.}~\bibnamefont
  {Jalilian-Marian}},\ }\href@noop {} {\  (\bibinfo {year}
  {2019}{\natexlab{a}})},\ \Eprint {http://arxiv.org/abs/1912.08878}
  {arXiv:1912.08878 [hep-ph]} \BibitemShut {NoStop}%
\bibitem [{\citenamefont
  {Jalilian-Marian}(2019{\natexlab{b}})}]{Jalilian-Marian:2018iui}%
  \BibitemOpen
  \bibfield  {author} {\bibinfo {author} {\bibfnamefont {J.}~\bibnamefont
  {Jalilian-Marian}},\ }\href {\doibase 10.1103/PhysRevD.99.014043} {\bibfield
  {journal} {\bibinfo  {journal} {Phys. Rev.}\ }\textbf {\bibinfo {volume}
  {D99}},\ \bibinfo {pages} {014043} (\bibinfo {year} {2019}{\natexlab{b}})},\
  \Eprint {http://arxiv.org/abs/1809.04625} {arXiv:1809.04625 [hep-ph]}
  \BibitemShut {NoStop}%
\bibitem [{\citenamefont {Kovchegov}\ and\ \citenamefont
  {Sievert}(2019{\natexlab{a}})}]{Kovchegov:2018znm}%
  \BibitemOpen
  \bibfield  {author} {\bibinfo {author} {\bibfnamefont {Y.~V.}\ \bibnamefont
  {Kovchegov}}\ and\ \bibinfo {author} {\bibfnamefont {M.~D.}\ \bibnamefont
  {Sievert}},\ }\href {\doibase 10.1103/PhysRevD.99.054032} {\bibfield
  {journal} {\bibinfo  {journal} {Phys. Rev.}\ }\textbf {\bibinfo {volume}
  {D99}},\ \bibinfo {pages} {054032} (\bibinfo {year} {2019}{\natexlab{a}})},\
  \Eprint {http://arxiv.org/abs/1808.09010} {arXiv:1808.09010 [hep-ph]}
  \BibitemShut {NoStop}%
\bibitem [{\citenamefont {Kovchegov}\ and\ \citenamefont
  {Sievert}(2019{\natexlab{b}})}]{Kovchegov:2018zeq}%
  \BibitemOpen
  \bibfield  {author} {\bibinfo {author} {\bibfnamefont {Y.~V.}\ \bibnamefont
  {Kovchegov}}\ and\ \bibinfo {author} {\bibfnamefont {M.~D.}\ \bibnamefont
  {Sievert}},\ }\href {\doibase 10.1103/PhysRevD.99.054033} {\bibfield
  {journal} {\bibinfo  {journal} {Phys. Rev.}\ }\textbf {\bibinfo {volume}
  {D99}},\ \bibinfo {pages} {054033} (\bibinfo {year} {2019}{\natexlab{b}})},\
  \Eprint {http://arxiv.org/abs/1808.10354} {arXiv:1808.10354 [hep-ph]}
  \BibitemShut {NoStop}%
\bibitem [{\citenamefont {Altinoluk}\ \emph {et~al.}(2016)\citenamefont
  {Altinoluk}, \citenamefont {Armesto}, \citenamefont {Beuf},\ and\
  \citenamefont {Moscoso}}]{Altinoluk:2015gia}%
  \BibitemOpen
  \bibfield  {author} {\bibinfo {author} {\bibfnamefont {T.}~\bibnamefont
  {Altinoluk}}, \bibinfo {author} {\bibfnamefont {N.}~\bibnamefont {Armesto}},
  \bibinfo {author} {\bibfnamefont {G.}~\bibnamefont {Beuf}}, \ and\ \bibinfo
  {author} {\bibfnamefont {A.}~\bibnamefont {Moscoso}},\ }\href {\doibase
  10.1007/JHEP01(2016)114} {\bibfield  {journal} {\bibinfo  {journal} {JHEP}\
  }\textbf {\bibinfo {volume} {01}},\ \bibinfo {pages} {114} (\bibinfo {year}
  {2016})},\ \Eprint {http://arxiv.org/abs/1505.01400} {arXiv:1505.01400
  [hep-ph]} \BibitemShut {NoStop}%
\bibitem [{\citenamefont {Chirilli}(2019)}]{Chirilli:2018kkw}%
  \BibitemOpen
  \bibfield  {author} {\bibinfo {author} {\bibfnamefont {G.~A.}\ \bibnamefont
  {Chirilli}},\ }\href {\doibase 10.1007/JHEP01(2019)118} {\bibfield  {journal}
  {\bibinfo  {journal} {JHEP}\ }\textbf {\bibinfo {volume} {01}},\ \bibinfo
  {pages} {118} (\bibinfo {year} {2019})},\ \Eprint
  {http://arxiv.org/abs/1807.11435} {arXiv:1807.11435 [hep-ph]} \BibitemShut
  {NoStop}%
\bibitem [{\citenamefont {Chen}\ \emph {et~al.}(2015)\citenamefont {Chen},
  \citenamefont {Fries}, \citenamefont {Kapusta},\ and\ \citenamefont
  {Li}}]{Chen:2015wia}%
  \BibitemOpen
  \bibfield  {author} {\bibinfo {author} {\bibfnamefont {G.}~\bibnamefont
  {Chen}}, \bibinfo {author} {\bibfnamefont {R.~J.}\ \bibnamefont {Fries}},
  \bibinfo {author} {\bibfnamefont {J.~I.}\ \bibnamefont {Kapusta}}, \ and\
  \bibinfo {author} {\bibfnamefont {Y.}~\bibnamefont {Li}},\ }\href {\doibase
  10.1103/PhysRevC.92.064912} {\bibfield  {journal} {\bibinfo  {journal} {Phys.
  Rev.}\ }\textbf {\bibinfo {volume} {C92}},\ \bibinfo {pages} {064912}
  (\bibinfo {year} {2015})},\ \Eprint {http://arxiv.org/abs/1507.03524}
  {arXiv:1507.03524 [nucl-th]} \BibitemShut {NoStop}%
\bibitem [{\citenamefont {Brodsky}\ \emph {et~al.}(1998)\citenamefont
  {Brodsky}, \citenamefont {Pauli},\ and\ \citenamefont
  {Pinsky}}]{Brodsky:1997de}%
  \BibitemOpen
  \bibfield  {author} {\bibinfo {author} {\bibfnamefont {S.~J.}\ \bibnamefont
  {Brodsky}}, \bibinfo {author} {\bibfnamefont {H.-C.}\ \bibnamefont {Pauli}},
  \ and\ \bibinfo {author} {\bibfnamefont {S.~S.}\ \bibnamefont {Pinsky}},\
  }\href {\doibase 10.1016/S0370-1573(97)00089-6} {\bibfield  {journal}
  {\bibinfo  {journal} {Phys. Rept.}\ }\textbf {\bibinfo {volume} {301}},\
  \bibinfo {pages} {299} (\bibinfo {year} {1998})},\ \Eprint
  {http://arxiv.org/abs/hep-ph/9705477} {arXiv:hep-ph/9705477 [hep-ph]}
  \BibitemShut {NoStop}%
\bibitem [{\citenamefont {McLerran}\ and\ \citenamefont
  {Venugopalan}(1999)}]{McLerran:1998nk}%
  \BibitemOpen
  \bibfield  {author} {\bibinfo {author} {\bibfnamefont {L.~D.}\ \bibnamefont
  {McLerran}}\ and\ \bibinfo {author} {\bibfnamefont {R.}~\bibnamefont
  {Venugopalan}},\ }\href {\doibase 10.1103/PhysRevD.59.094002} {\bibfield
  {journal} {\bibinfo  {journal} {Phys. Rev.}\ }\textbf {\bibinfo {volume}
  {D59}},\ \bibinfo {pages} {094002} (\bibinfo {year} {1999})},\ \Eprint
  {http://arxiv.org/abs/hep-ph/9809427} {arXiv:hep-ph/9809427 [hep-ph]}
  \BibitemShut {NoStop}%
\bibitem [{\citenamefont {Jalilian-Marian}\ \emph {et~al.}(1997)\citenamefont
  {Jalilian-Marian}, \citenamefont {Kovner}, \citenamefont {McLerran},\ and\
  \citenamefont {Weigert}}]{JalilianMarian:1996xn}%
  \BibitemOpen
  \bibfield  {author} {\bibinfo {author} {\bibfnamefont {J.}~\bibnamefont
  {Jalilian-Marian}}, \bibinfo {author} {\bibfnamefont {A.}~\bibnamefont
  {Kovner}}, \bibinfo {author} {\bibfnamefont {L.~D.}\ \bibnamefont
  {McLerran}}, \ and\ \bibinfo {author} {\bibfnamefont {H.}~\bibnamefont
  {Weigert}},\ }\href {\doibase 10.1103/PhysRevD.55.5414} {\bibfield  {journal}
  {\bibinfo  {journal} {Phys. Rev.}\ }\textbf {\bibinfo {volume} {D55}},\
  \bibinfo {pages} {5414} (\bibinfo {year} {1997})},\ \Eprint
  {http://arxiv.org/abs/hep-ph/9606337} {arXiv:hep-ph/9606337 [hep-ph]}
  \BibitemShut {NoStop}%
\bibitem [{\citenamefont {Krasnitz}\ \emph {et~al.}(2003)\citenamefont
  {Krasnitz}, \citenamefont {Nara},\ and\ \citenamefont
  {Venugopalan}}]{krasnitz2003gluon}%
  \BibitemOpen
  \bibfield  {author} {\bibinfo {author} {\bibfnamefont {A.}~\bibnamefont
  {Krasnitz}}, \bibinfo {author} {\bibfnamefont {Y.}~\bibnamefont {Nara}}, \
  and\ \bibinfo {author} {\bibfnamefont {R.}~\bibnamefont {Venugopalan}},\
  }\href {\doibase 10.1016/S0375-9474(03)00636-5} {\bibfield  {journal}
  {\bibinfo  {journal} {Nucl. Phys.}\ }\textbf {\bibinfo {volume} {A717}},\
  \bibinfo {pages} {268} (\bibinfo {year} {2003})},\ \Eprint
  {http://arxiv.org/abs/hep-ph/0209269} {arXiv:hep-ph/0209269 [hep-ph]}
  \BibitemShut {NoStop}%
\bibitem [{\citenamefont {Lappi}(2006)}]{Lappi:2006hq}%
  \BibitemOpen
  \bibfield  {author} {\bibinfo {author} {\bibfnamefont {T.}~\bibnamefont
  {Lappi}},\ }\href {\doibase 10.1016/j.physletb.2006.10.017} {\bibfield
  {journal} {\bibinfo  {journal} {Phys. Lett.}\ }\textbf {\bibinfo {volume}
  {B643}},\ \bibinfo {pages} {11} (\bibinfo {year} {2006})},\ \Eprint
  {http://arxiv.org/abs/hep-ph/0606207} {arXiv:hep-ph/0606207 [hep-ph]}
  \BibitemShut {NoStop}%
\bibitem [{\citenamefont {Müller}(2019)}]{Muller:2019bwd}%
  \BibitemOpen
  \bibfield  {author} {\bibinfo {author} {\bibfnamefont {D.}~\bibnamefont
  {Müller}},\ }\emph {\bibinfo {title} {{Simulations of the Glasma in
  3+1D}}},\ \href@noop {} {Ph.D. thesis},\ \bibinfo  {school} {TU Vienna}
  (\bibinfo {year} {2019}),\ \Eprint {http://arxiv.org/abs/1904.04267}
  {arXiv:1904.04267 [hep-ph]} \BibitemShut {NoStop}%
\bibitem [{\citenamefont {Fukushima}\ and\ \citenamefont
  {Hidaka}(2007)}]{fukushima2007light}%
  \BibitemOpen
  \bibfield  {author} {\bibinfo {author} {\bibfnamefont {K.}~\bibnamefont
  {Fukushima}}\ and\ \bibinfo {author} {\bibfnamefont {Y.}~\bibnamefont
  {Hidaka}},\ }\href {\doibase 10.1088/1126-6708/2007/06/040} {\bibfield
  {journal} {\bibinfo  {journal} {JHEP}\ }\textbf {\bibinfo {volume} {06}},\
  \bibinfo {pages} {040} (\bibinfo {year} {2007})},\ \Eprint
  {http://arxiv.org/abs/0704.2806} {arXiv:0704.2806 [hep-ph]} \BibitemShut
  {NoStop}%
\bibitem [{\citenamefont {Pauli}\ and\ \citenamefont
  {Brodsky}(1985)}]{Pauli:1985pv}%
  \BibitemOpen
  \bibfield  {author} {\bibinfo {author} {\bibfnamefont {H.~C.}\ \bibnamefont
  {Pauli}}\ and\ \bibinfo {author} {\bibfnamefont {S.~J.}\ \bibnamefont
  {Brodsky}},\ }\href {\doibase 10.1103/PhysRevD.32.1993} {\bibfield  {journal}
  {\bibinfo  {journal} {Phys. Rev.}\ }\textbf {\bibinfo {volume} {D32}},\
  \bibinfo {pages} {1993} (\bibinfo {year} {1985})}\BibitemShut {NoStop}%
\bibitem [{\citenamefont {Eller}\ \emph {et~al.}(1987)\citenamefont {Eller},
  \citenamefont {Pauli},\ and\ \citenamefont {Brodsky}}]{Eller:1986nt}%
  \BibitemOpen
  \bibfield  {author} {\bibinfo {author} {\bibfnamefont {T.}~\bibnamefont
  {Eller}}, \bibinfo {author} {\bibfnamefont {H.~C.}\ \bibnamefont {Pauli}}, \
  and\ \bibinfo {author} {\bibfnamefont {S.~J.}\ \bibnamefont {Brodsky}},\
  }\href {\doibase 10.1103/PhysRevD.35.1493} {\bibfield  {journal} {\bibinfo
  {journal} {Phys. Rev.}\ }\textbf {\bibinfo {volume} {D35}},\ \bibinfo {pages}
  {1493} (\bibinfo {year} {1987})}\BibitemShut {NoStop}%
\bibitem [{\citenamefont {Lappi}(2008)}]{lappi2008wilson}%
  \BibitemOpen
  \bibfield  {author} {\bibinfo {author} {\bibfnamefont {T.}~\bibnamefont
  {Lappi}},\ }\href {\doibase 10.1140/epjc/s10052-008-0588-4} {\bibfield
  {journal} {\bibinfo  {journal} {Eur. Phys. J.}\ }\textbf {\bibinfo {volume}
  {C55}},\ \bibinfo {pages} {285} (\bibinfo {year} {2008})},\ \Eprint
  {http://arxiv.org/abs/0711.3039} {arXiv:0711.3039 [hep-ph]} \BibitemShut
  {NoStop}%
\bibitem [{\citenamefont {Peskin}\ and\ \citenamefont
  {Schroeder}(1995)}]{Peskin:1995ev}%
  \BibitemOpen
  \bibfield  {author} {\bibinfo {author} {\bibfnamefont {M.~E.}\ \bibnamefont
  {Peskin}}\ and\ \bibinfo {author} {\bibfnamefont {D.~V.}\ \bibnamefont
  {Schroeder}},\ }\href {http://www.slac.stanford.edu/~mpeskin/QFT.html} {\emph
  {\bibinfo {title} {{An Introduction to quantum field theory}}}}\ (\bibinfo
  {publisher} {Addison-Wesley},\ \bibinfo {address} {Reading, USA},\ \bibinfo
  {year} {1995})\BibitemShut {NoStop}%
\bibitem [{\citenamefont {Mueller}(1998)}]{Mueller:1997ik}%
  \BibitemOpen
  \bibfield  {author} {\bibinfo {author} {\bibfnamefont {A.~H.}\ \bibnamefont
  {Mueller}},\ }\bibfield  {booktitle} {\emph {\bibinfo {booktitle} {{Workshop
  on Interplay between Soft and Hard Interactions in Deep Inelastic Scattering
  Heidelberg, Germany, September 29-October 1, 1997}}},\ }\href {\doibase
  10.1007/s100500050027} {\bibfield  {journal} {\bibinfo  {journal} {Eur. Phys.
  J.}\ }\textbf {\bibinfo {volume} {A1}},\ \bibinfo {pages} {19} (\bibinfo
  {year} {1998})},\ \Eprint {http://arxiv.org/abs/hep-ph/9710531}
  {arXiv:hep-ph/9710531 [hep-ph]} \BibitemShut {NoStop}%
\bibitem [{\citenamefont {Kovchegov}\ and\ \citenamefont
  {McLerran}(1999)}]{Kovchegov:1999kx}%
  \BibitemOpen
  \bibfield  {author} {\bibinfo {author} {\bibfnamefont {Y.~V.}\ \bibnamefont
  {Kovchegov}}\ and\ \bibinfo {author} {\bibfnamefont {L.~D.}\ \bibnamefont
  {McLerran}},\ }\href {\doibase 10.1103/PhysRevD.62.019901,
  10.1103/PhysRevD.60.054025} {\bibfield  {journal} {\bibinfo  {journal} {Phys.
  Rev.}\ }\textbf {\bibinfo {volume} {D60}},\ \bibinfo {pages} {054025}
  (\bibinfo {year} {1999})},\ \bibinfo {note} {[Erratum: Phys.
  Rev.D62,019901(2000)]},\ \Eprint {http://arxiv.org/abs/hep-ph/9903246}
  {arXiv:hep-ph/9903246 [hep-ph]} \BibitemShut {NoStop}%
\bibitem [{\citenamefont {Suhonen}(2007)}]{Suhonen:2007zza}%
  \BibitemOpen
  \bibfield  {author} {\bibinfo {author} {\bibfnamefont {J.}~\bibnamefont
  {Suhonen}},\ }\href {\doibase 10.1007/978-3-540-48861-3} {\emph {\bibinfo
  {title} {{From Nucleons to Nucleus}}}},\ Theoretical and Mathematical
  Physics\ (\bibinfo  {publisher} {Springer},\ \bibinfo {address} {Berlin,
  Germany},\ \bibinfo {year} {2007})\BibitemShut {NoStop}%
\end{thebibliography}%
\end{document}